%% file: 2021_wiecha_pygdm_update.tex
\documentclass[final, 5p, twocolumn]{elsarticle}

\usepackage[utf8]{inputenc}
\usepackage[T1]{fontenc}
\usepackage[english]{babel}
\usepackage[none]{hyphenat}

\usepackage{amsmath,amssymb,amsfonts}
\usepackage{amstext, mathrsfs, mathtools, textcomp}
\usepackage{empheq}
\usepackage{array}
\usepackage{bbm}
\usepackage{cuted}
\usepackage{nicefrac}
\usepackage{multirow}

\usepackage[most]{tcolorbox}
\usepackage{booktabs}
\usepackage{makecell}

\usepackage{gensymb}
\usepackage{graphicx}
\usepackage{hyperref}

\usepackage{xcolor}
\usepackage[export]{adjustbox}

\usepackage{listings}
\definecolor{codegreen}{rgb}{0,0.6,0}
\definecolor{codegray}{rgb}{0.5,0.5,0.5}
\definecolor{codepurple}{rgb}{0.58,0,0.82}
\definecolor{backcolour}{rgb}{0.95,0.95,0.92}
 
\lstdefinestyle{mystyle}{
    backgroundcolor=\color{backcolour},   
    commentstyle=\color{codegreen},
    keywordstyle=\color{blue},
    numberstyle=\tiny\color{codegray},
    stringstyle=\color{codepurple},
    basicstyle=\ttfamily\scriptsize,
    breakatwhitespace=false,         
    breaklines=true,                 
    captionpos=b,                    
    keepspaces=true,                 
    numbers=left,                    
    numbersep=5pt,                  
    showspaces=false,                
    showstringspaces=false,
    showtabs=false,                  
    tabsize=2
}
\DeclareFontFamily{U}{mathb}{\hyphenchar\font45}
\DeclareFontShape{U}{mathb}{m}{n}{<5> <6> <7> <8> <9> <10> gen * mathb
<10.95> mathb10 <12> <14.4> <17.28> <20.74> <24.88> mathb12}{}
\DeclareSymbolFont{mathb}{U}{mathb}{m}{n}
\DeclareMathSymbol{\rcirclearrow}{0}{mathb}{'367}

\renewcommand{\imath}{\text{i}}
\renewcommand{\Im}{\text{Im}}

\usepackage{changes}

\graphicspath{{Figures/}}

\newcommand{\pygdm}{pyGDM}
\newcommand{\pyGDM}{pyGDM}

\definecolor{pansypurple}{rgb}{0.47, 0.09, 0.29}
\definecolor{lincolngreen}{rgb}{0.11, 0.35, 0.02}
\definecolor{internationalkleinblue}{rgb}{0.0, 0.18, 0.65}

\newcommand{\function}[1]{\textcolor{internationalkleinblue}{\textbf{#1}}}
\newcommand{\object}[1]{\textcolor{lincolngreen}{\textbf{#1}}}

\renewcommand{\Re}{\text{Re}}
\newcommand{\iu}{\mathrm{i}}
\newcommand{\e}{\mathrm{e}}
\newcommand{\stepsize}{d}
\newcommand{\pdiff}[3][\empty]{\ifx\empty#1
		\frac{\partial\,#2}{\partial #3}
	\else
		\frac{\partial^{#1}\,#2}{\partial #3^{#1}}
	\fi}		

\newcommand*{\functiondescription}[4]{%
  \vspace{0.5\baselineskip}
  
  \noindent\hspace{0.035\linewidth}
  \fbox{
   \begin{minipage}{0.85\linewidth}
    \vspace{0.25\baselineskip}
    \begin{center}
      \ifthenelse{\equal{#1}{f}}
        {\function{#2}\\\small(function)}
        {\object{#2}\\\small(class)}
    \end{center}
    \if\relax\detokenize{#4}\relax
    \else
	\relax
	\vspace{-0.5\baselineskip}\noindent
	\ifthenelse{\equal{#1}{f}}
	    {\textit{arguments:}}
	    {\textit{constructor arguments:}}
	\relax
	\functiondescr{#4}
    \fi
   \end{minipage}
  }
  \vspace{\baselineskip}
  
  \par\noindent\relax
}
\newcommand*{\functiondescr}[1]{%
  \begin{itemize}
   \funcparameter#1\relax
  \end{itemize}
}
\newcommand{\funcparameter}[1]{%
  \ifx\relax#1\empty
  \else
     \vspace{-0.5\baselineskip}
     \item #1
    \relax
    \expandafter\funcparameter
  \fi
}

\newcommand{\TITLE}{``pyGDM'' - new functionalities and major improvements to the python toolkit for nano-optics full-field simulations}

\hypersetup{
    colorlinks,       
    linkcolor=blue,          
    citecolor=blue,         
    filecolor=blue,       
    urlcolor=blue,           
    pdfauthor     = {Wiecha et al.},
    pdftitle      = {\TITLE},
    pdfsubject    = {pyGDM},
    pdfcreator    = {pdflatex},
    pdfkeywords   = {}
    }

\usepackage{etoolbox}
\newtoggle{revtex}
\togglefalse{revtex}

\iftoggle{revtex}
{
	\bibliographystyle{apsrev4-1}}
{
	\bibliographystyle{naturemag}
	
	\newcounter{bla}

	\journal{Computer Physics Communications}
	
}

\begin{document}

\iftoggle{revtex}
{
	\title{\TITLE}
	
	\author{\firstname{Peter R.} \surname{Wiecha}}
	\email[e-mail~: ]{pwiecha@laas.fr}
	\affiliation{LAAS, Universit\'e de Toulouse, CNRS, Toulouse, France}
	
	\author{\firstname{Clément} \surname{Majorel}}
	\affiliation{CEMES-CNRS, Universit\'e de Toulouse, CNRS, UPS, Toulouse, France}
	
	\author{\firstname{Arnaud} \surname{Arbouet}}
	\affiliation{CEMES-CNRS, Universit\'e de Toulouse, CNRS, UPS, Toulouse, France}
	
	\author{\firstname{Adelin} \surname{Patoux}}
	\affiliation{CEMES-CNRS, Universit\'e de Toulouse, CNRS, UPS, Toulouse, France}
	
	\author{\firstname{Yoann} \surname{Brûlé}}
	\affiliation{ICB, UMR 6303 CNRS - Universit\'e Bourgogne-Franche Comt\'e, Dijon, France}
	
	\author{\firstname{Gérard} \surname{Colas des Francs}}
	\affiliation{ICB, UMR 6303 CNRS - Universit\'e Bourgogne-Franche Comt\'e, Dijon, France}
	
	\author{\firstname{Christian} \surname{Girard}}
	\affiliation{CEMES-CNRS, Universit\'e de Toulouse, CNRS, UPS, Toulouse, France}
}
{
	\begin{frontmatter}
		
		
		
		\title{\TITLE}
		
		
		\author[a]{Peter R. Wiecha\corref{author}}
		\author[b]{Clément Majorel}
		\author[b]{Arnaud Arbouet}
		\author[a,b,c]{Adelin Patoux}
		\author[d]{Yoann Brûlé}
		\author[d]{Gérard {Colas des Francs}}
		\author[b]{Christian Girard}
		\cortext[author] {Corresponding author.\\\textit{E-mail address:} pwiecha@laas.fr}
		\address[a]{LAAS, Universit\'e de Toulouse, CNRS, Toulouse, France}
		\address[b]{CEMES-CNRS, Universit\'e de Toulouse, CNRS, UPS, Toulouse, France}
		\address[c]{Airbus Defence and Space, Toulouse, France}
		\address[d]{ICB, UMR 6303 CNRS - Universit\'e Bourgogne-Franche Comt\'e, Dijon, France}
}
\begin{abstract}
\pygdm\ is a python toolkit for electro-dynamical simulations of individual nano-structures, based on the Green Dyadic Method (GDM).
\pygdm\ uses the concept of a generalized propagator, which allows to solve cost-efficiently monochromatic problems with a large number of varying illumination conditions such as incident angle scans or focused beam raster-scan simulations.
We provide an overview of new features added since the initial publication \href{https://doi.org/10.1016/j.cpc.2018.06.017}{[Wiecha, Computer Physics Communications 233, pp.167-192 (2018)]}.
The updated version of \pyGDM\ is implemented in pure python, removing the former dependency on fortran-based binaries.
In the course of this re-write, the toolkit's internal architecture has been completely redesigned to offer a much wider range of possibilities to the user such as the choice of the dyadic Green's functions describing the environment.
A new class of dyads allows to perform 2D simulations of infinitely long nanostructures.
While the Green's dyads in \pygdm\ are based on a quasistatic description for interfaces, we also provide as new external python package ``\textsf{pyGDM2\_retard}'' a module with retarded Green's tensors for an environment with two interfaces.
We have furthermore added functionalities for simulations using fast-electron excitation, namely electron energy loss spectroscopy and cathodoluminescence.
Along with several further new tools and improvements, the update includes also the possibility to calculate the magnetic field and the magnetic LDOS inside nanostructures, field-gradients in- and outside a nanoparticle, optical forces or the chirality of nearfields. All new functionalities remain compatible with the evolutionary optimization module of pyGDM for nano-photonics inverse design.
\end{abstract}
\iftoggle{revtex}
{\maketitle}
{
	
\end{frontmatter}

\textbf{PROGRAM SUMMARY}

\begin{small}
	\noindent
	\textit{Program Title:} pyGDM2                               \\
	\textit{Licensing provisions:} GPLv3                         \\
	\textit{Programming language:} python                        \\
	\textit{Nature of problem:}\\
	Full-field electrodynamical simulations of photonic nanostructures. This includes calculations of the optical extinction, scattering and absorption, as well as the near-field distribution or the interaction of quantum emitters with nanostructures as well as fast electron beam simulations. 
	The toolkit includes a module for automated evolutionary optimization of nanostructure geometries to obtain a user-defined optical response. \\
	\textit{Solution method:}\\
	The optical response of photonic nanostructures is calculated using field susceptibilities (``Green Dyadic Method'', GDM) via a volume discretization. The approach is formally similar to the coupled dipole approximation.\\
	\textit{Additional comments including Restrictions and Unusual features:}\\
	2D and 3D nanostructures. On typical office PCs (8-16GB RAM) the discretization is limited to about 10000-15000 meshpoints, therefore it applies best to single, small nanostructures.
	\\
\end{small}
}




\input{00_intro}

\input{01_general_theory_magnetic_fields}

\input{03_theory_new_functionalities}

\input{06_bugfixes}

\section{Outlook}

In the future, from the physical modeling point of view we plan to implement Green's tensors for periodic structures \cite{chaumetGeneralizationCoupledDipole2003}, retarded tensors for 2D simulations in layered environments and possibly for 3D systems with arbitrary numbers of interfaces \cite{paulusAccurateEfficientComputation2000, paulusGreenTensorTechnique2001}, functions to calculate the non-radiative part of the LDOS \cite{baffouMolecularQuenchingRelaxation2008, cucheNearfieldHyperspectralQuantum2017} and the possibility to model femto-second laser pulses for ultrafast optics applications \cite{arbouetInteractionUltrashortOptical2012}.
We also plan to implement nonlinear nano-optics functions from earlier work into \pygdm\ \cite{blackTailoringSecondHarmonicGeneration2015, wiechaOriginSecondharmonicGeneration2016}.
On the long term we want to add support for meta-material unit-cells having a direct magnetic response  \cite{chaumetCoupleddipoleMethodMagnetic2009}.
We are constantly implementing further improvements. 
For instance we work on including transmission and reflection coefficients at an interface for the new focused vector-beams. 
In the \textsf{pyGDM2\_retard} module for the retarded description of a layered environment we are working on including field calculations in other zones than the one hosting the nanostructure.

From the technical side, we plan to implement a conjugate gradients solver to avoid full inversion of systems with large amounts of mesh-cells \cite{draineDiscretedipoleApproximationScattering1994}. Furthermore, such iterative solver could be strongly accelerated using GPU-based fast Fourier transforms \cite{govindarajuHighPerformanceDiscrete2008, huntemannDiscreteDipoleApproximation2011}.

More flexible meshing is also a feature that we plan to implement in the future \cite{kottmannAccurateSolutionVolume2000, ouldaghaNearFieldPropertiesPlasmonic2014, smunevRectangularDipolesDiscrete2015}.
Furthermore, we are beginning to conceptualize a module to evaluate complex problems with pre-trained deep learning models, and we think about how to construct an interface which allows the easy creation of such deep learning accelerated models for specific applications \cite{wiechaDeepLearningMeets2020}.

\section*{Acknowledgments}
We gratefully thank Aurélien Cuche, Otto Muskens and Frank Mersch for many inspiring discussions.
This work was supported by the German Research Foundation (DFG) through a research fellowship (WI 5261/1-1), and by the computing facility center CALMIP of the University of Toulouse under grant P20010.
of Toulouse under grant P20010.
YB and GCF acknowledge support from the French National Research Agency (ANR) project HiLight (ANR-19-CE24-0026), the French ``Investissements d'Avenir'' program (EUR-EIPHI 17-EURE-0002) and access to the HPC facilities of the University of Bourgogne (ccuB).

\section*{Conflicts of interests}

The authors declare no competing financial interests.


\appendix
\input{Appendix_Nearfield_Farfield_retardation}
\input{Appendix_2D_tensors}

\input{Appendix_fast_electrons}
\input{Appendix_technical_details}

\input{bibliography.bbl}
\end{document}

%% file: 00_intro.tex
Solving Maxwell's equations~\cite{maxwellDynamicalTheoryElectromagnetic1865} for arbitrary nano-structure geometries in complex environments is a key challenge in modern nano-optics. 
In such nano-optics problems, effects of light-matter interaction can be computed only numerically.
Different methods can be used to this end, which have distinct strengths and drawbacks, depending on the respective configuration.
To name a few, popular methods are for example the finite difference time domain method (FDTD) \cite{oskooiMEEPFlexibleFreesoftware2010}, the coupled dipole approximation (CDA) \cite{draineDiscretedipoleApproximationScattering1994, chaumetCoupleddipoleMethodMagnetic2009} or the boundary element method (BEM, also called surface integral equation method) \cite{hohenesterMNPBEMMatlabToolbox2012, garciadeabajoOpticalExcitationsElectron2010} or the finite element method (FEM) \cite{demesyAllpurposeFiniteElement2010, hoffmannComparisonElectromagneticField2009}.
Detailed comparisons between the methods can be found in literature \cite{barchiesiComputingOpticalNearfield1996, parsonsComparisonTechniquesUsed2010, gallinetNumericalMethodsNanophotonics2015}.

We describe here \pyGDM, a python toolkit implementing a volume integral approach based on field-susceptibilities, namely the Green's Dyadic Method (GDM).
It is well suited for the numerical description of nano-particles deposited on a substrate. 
The full optical response of both dielectric \cite{kuznetsovOpticallyResonantDielectric2016} or plasmonic nanostructures is accessible  \cite{muhlschlegelResonantOpticalAntennas2005, bharadwajOpticalAntennas2009}. 
The GDM can describe well nano-plasmonics in general \cite{wiechaPolarizationConversionPlasmonic2017, girardDesigningThermoplasmonicProperties2018}, but in particular it is an almost ideal method to describe ``flat plasmonics'', hence large planar nanostructures of thicknesses smaller than the skin depth \cite{girardShapingManipulationLight2008, viarbitskayaTailoringImagingPlasmonic2013, viarbitskayaMorphologyinducedRedistributionSurface2013, ouldaghaNearFieldPropertiesPlasmonic2014, cucheDipolarRegimeHighorder2017, cucheNearfieldHyperspectralQuantum2017}.
Concerning dielectric materials, low-index dielectric nanostructures can be described very accurately \cite{barchiesiComputingOpticalNearfield1996}, but also high-index dielectric nanostructures are within the scope of the GDM \cite{wiechaEvolutionaryMultiobjectiveOptimization2017, wiechaStronglyDirectionalScattering2017, wiechaOriginSecondharmonicGeneration2016}.
Beyond linear optics it is also capable to describe non-linear effects such as multi-photon photoluminescence or harmonics generation \cite{wiechaOriginSecondharmonicGeneration2016, blackTailoringSecondHarmonicGeneration2015, wiechaLinearNonlinearOptical2016}.

One of the major modifications in the new version is the entirely rewritten internal API.
While the initial version was based on fortran routines for a fast computation of Green's tensors, the new \pygdm\ is fully written in python and accelerated using the \textit{numba} package \cite{lamNumbaLLVMbasedPython2015a}.
In addition to easier installation and better platform independence, the python re-write offers an enormous gain in flexibility and renders \pygdm\ fully modular, which means that external packages can be easily written and fully integrate in the toolkit without technical barriers.
A huge advantage is that the environment is no longer hard-coded with the nano-structure class, but is now fully described through a separate \textit{dyads} class, in which the set of Green's tensors is implemented. 
This means that the environment can be flexibly modified in the new \pygdm.
With the new version, we currently provide an alternative dyads class for 2D simulations based on a quasistatic approximation for a layered environment. 
As an external package we provide another dyads class offering a fully retarded description of layered environments with one or two interfaces.
In the future we plan to implement further sets of Green's dyads, for instance for periodic structures.

Further new features, among others, are several focused vectorbeams, the generalization of the plane wave incident field, supporting arbitrary illumination angles and polarizations in environments containing up to 3 interfaces. 
We added the computation of optical chirality, a fast electron illumination for electron energy loss spectroscopy (EELS) and cathodoluminescence (CL) within the new \texttt{electron} submodule, or routines for field gradient calculations and optical forces.
In the appendix we provide a full description of the Green's tensors used in \pyGDM\ for the different simulation functions.
In a further appendix section we also provide more technical documentation as well as instructions for the installation and the use of \pygdm.

\section{Summary of \pygdm\ functionalities}\label{sec:summary_functionality}

We provide here a non-exhaustive list of the functionalities available in the \pyGDM\ toolkit:

\begin{itemize}
	\setlength\itemsep{.1em}
	\item Easy to use, easy to install, open source. Depends exclusively on open-source python libraries available via \href{https://pypi.org/}{pip} (numpy, numba, scipy, matplotlib)
	\item Fast: Performance-critical parts are accelerated by just-in-time compilation via numba and are parallelized. Efficient and parallelized scipy libraries are used whenever possible. An MPI-parallelized routine for the calculation of spectra is available
	\item Electro-dynamical simulations including a substrate, either in a computationally fast quasistatic approximation, or computationally slower including retardation
	\item 2D and 3D simulations
	\item Various illumination sources such as plane waves, tightly focused beams with linear, radial or azimuthal polarization, dipolar emitters
	\item Very efficient calculation of large monochromatic problems (raster-scans, incident angle scans, ...)
	\item Fast electron simulations (EELS, ca\-thodo\-lumi\-nescence)
	\item Includes various tools to rapidly post-process the simulations and derive physical quantities such as
	\begin{itemize}
		\item extinction, absorption and scattering cross-sections
		\item optical near-field inside and around nanostructures
		\item polarization- and spatially resolved far-field scattering
		\item optical chirality of the near-field
		\item heat generation, temperature distribution around nano-objects
		\item photonic local density of states (LDOS)
		\item field gradients and optical forces
		\item multi-photon photoluminescence
		\item electric / magnetic dipole decomposition of local fields and the extinction cross section
	\end{itemize}
	\item Evolutionary optimization of the nano-particle geometry with regards to specific optical properties
	\item Easy to use visualization tools including animations of the electro-magnetic fields
\end{itemize}

%% file: 01_general_theory_magnetic_fields.tex
\section{The Green's Dyadic Method}\label{section_1}

\subsection{From Maxwell's equations to Lippmann-Schwinger equation}

\begin{figure}[t]
	\includegraphics[width=\columnwidth]{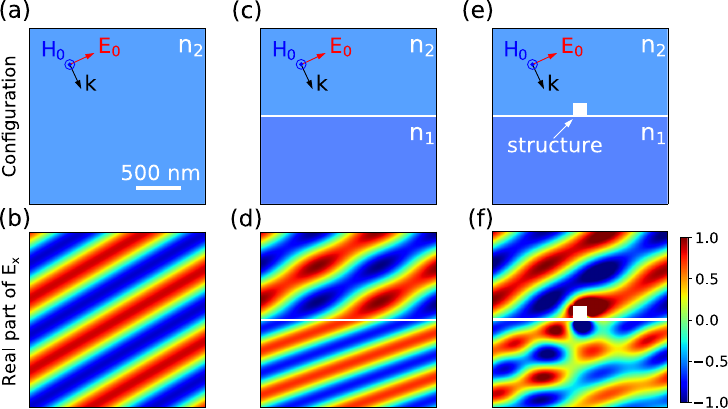}
	\caption{
		Different environment configurations, through which an oblique incident plane wave passes. Angle of incidence $\theta_{\text{in}}=30^{\circ}$, wavelength $\lambda_0 = 710\,$nm.
		(a-b) air ($n_2=1$).
		(c-d) plane wave incident from air ($n_2=1$) onto a dielectric substrate ($n_1=1.5$).
		(e-f) same as in (c-d) with a nanostructure deposited on top of the interface (silicon nano-cube of size $160\times 160\times 160\,$nm$^3$).
		(a,c,e) illustration of the respective simulated configuration.
		(b,d,f) color plot of the real part of $E_x$. 
		Panels show $2\times 2\,$\textmu m$^2$ large areas.
		In contrast to most other electrodynamics solvers, in (e-f) only the silicon cube needs to be discretized.
	}\label{fig:oblique_incidence_vac_subst_struct}
\end{figure}

We will first recapitulate the underlying theory behind the Green's Dyadic Method (GDM) \cite{martinGeneralizedFieldPropagator1995}. 
Compared to the initial \pyGDM\ paper \cite{wiechaPyGDMPythonToolkit2018}, we present here the formalism in a more general way, including the self-consistent equations describing both, electric and magnetic optical fields, since one new feature of the toolkit is the calculation of the magnetic field inside nanostructures.

The GDM allows to treat light-matter interaction problems such as depicted in figure~\ref{fig:oblique_incidence_vac_subst_struct}, where the progressive increase in complexity of the scenario is illustrated by the example of the perturbation of a plane wave.
Fig.~\ref{fig:oblique_incidence_vac_subst_struct}a shows the field amplitude of a plane wave passing through a homogeneous environment, in \ref{fig:oblique_incidence_vac_subst_struct}b an interface is added and in \ref{fig:oblique_incidence_vac_subst_struct}c an additional small nano-structure.
Problems such as the one shown in figure~\ref{fig:oblique_incidence_vac_subst_struct} require to solve Maxwell's equations, the GDM does this for monochromatic fields (fixed frequency $\omega$). 
In cgs units (centimeter, gram, second), the frequency domain Maxwell's equations are
\begin{align}\label{eq:Eq_Maxwell_Faraday}
&\nabla\times\mathbf{E}(\mathbf{r},\omega)=ik_0[\mu_{\text{env}}(\omega)\mathbf{H}(\mathbf{r},\omega) +4\pi \mathbf{M}(\mathbf{r},\omega)]\\
&\nabla\cdot\mathbf{H}(\mathbf{r},\omega)=-\frac{4\pi}{\mu_{\text{env}}(\omega)}\nabla\cdot\mathbf{M}(\mathbf{r},\omega)\\\label{eq:Eq_Maxwell_Ampere}
&\nabla\times\mathbf{H}(\mathbf{r},\omega)=-ik_0[\epsilon_{\text{env}}(\omega)\mathbf{E}(\mathbf{r},\omega) +4\pi\mathbf{P}(\mathbf{r},\omega)]\\
&\nabla\cdot\mathbf{E}(\mathbf{r},\omega)=-\frac{4\pi}{\epsilon_{\text{env}}(\omega)}\nabla\cdot\mathbf{P}(\mathbf{r},\omega)
\end{align}
where $\mathbf{E}(\mathbf{r},\omega)$ (respectively $\mathbf{H}(\mathbf{r},\omega)$) is the electric (respectively magnetic) field at the position $\mathbf{r}$. 
$\mathbf{P}(\mathbf{r},\omega)$ and $\mathbf{M}(\mathbf{r},\omega)$ are the electric and magnetic polarizations, $\epsilon_{\text{env}}$ and $\mu_{\text{env}}$ are respectively the dielectric constant and the permeability of the surrounding medium and $k_0=\omega/c$ is the wavenumber of the light in vacuum.

By applying the rotational operator to the Maxwell-Faraday (Eq. (\ref{eq:Eq_Maxwell_Faraday})) and Maxwell-Ampere (Eq. (\ref{eq:Eq_Maxwell_Ampere})) equations and by reorganizing the obtained equations, we can define two wave-equations \cite{agarwalQuantumElectrodynamicsPresence1975, jacksonClassicalElectrodynamics1999}. One for $\mathbf{E}(\mathbf{r}, \omega)$ and one for $\mathbf{H}(\mathbf{r}, \omega)$: 
\begin{subequations}\label{eq:wave-equations}
	\begin{align}\label{wave-eq-E}
		&(\Delta+k^2)\mathbf{E}(\mathbf{r},\omega)=\\
		&-4\pi\Big[\frac{1}{\epsilon_{\text{env}}}(\nabla\nabla+k^{2}) \mathbf{P}(\mathbf{r},\omega)+ ik_{0}\nabla\times\mathbf{M}(\mathbf{r},\omega)\Big]\nonumber\\
	\label{wave-eq-H}
		&(\Delta+k^2)\mathbf{H}(\mathbf{r},\omega)=\\
		&-4\pi\Big[\frac{1}{\mu_{\text{env}}}(\nabla\nabla+k^{2}) \mathbf{M}(\mathbf{r},\omega)+ ik_{0}\nabla\times\mathbf{P}(\mathbf{r},\omega)\Big]\nonumber
	\end{align}
\end{subequations}
with $k=\sqrt{\epsilon_{\text{env}}\mu_{\text{env}}}k_0$ the wavenumber of light in the environment medium.
In the approximation of a local and linear optical response, the electric and magnetic polarizations are proportional to the electric and magnetic field as \(\mathbf{P} = \boldsymbol{\chi}_{\text{e}} \cdot \mathbf{E}\) and \(\mathbf{M} = \boldsymbol{\chi}_{\text{m}} \cdot \mathbf{H}\). 
The electric (respectively magnetic) susceptibility is a tensor of rank 2, corresponding to the difference between the relative permittivity (respectively permeability) of the structure and the environment $\boldsymbol{\chi}_{\text{e}}(\mathbf{r},\omega)=(\boldsymbol\epsilon_{\text{r}} - \boldsymbol\epsilon_{\text{env}})/4\pi$ (respectively $\boldsymbol{\chi}_{\text{m}}(\mathbf{r},\omega)=(\boldsymbol\mu_{\text{r}} - \boldsymbol\mu_{\text{env}})/4\pi$).

If we replace $\mathbf{P}$ and $\mathbf{M}$ by their expressions in Eq. (\ref{wave-eq-E}) and Eq. (\ref{wave-eq-H}), we obtain a system of coupled equations relating $\mathbf{E}(\mathbf{r},\omega)$ and $\mathbf{H}(\mathbf{r},\omega)$.
To lighten the notation, we define the super vector $\mathbf{F}(\mathbf{r},\omega)=\lbrace\mathbf{E}(\mathbf{r},\omega),\mathbf{H}(\mathbf{r},\omega) \rbrace$.
We can then write the solution of Eq. (\ref{eq:wave-equations}), as a single vectorial Lippman-Schwinger equation \cite{patouxPolarizabilitiesComplexIndividual2020, sersicMagnetoelectricPointScattering2011, chaumetCoupleddipoleMethodMagnetic2009, schroterModellingMagneticEffects2003}
\begin{equation}\label{LPS-superpropag}
\begin{multlined}
	\mathbf{F}(\mathbf{r},\omega)=\mathbf{F}_{0}(\mathbf{r},\omega) + \\
	 \int_{V}\pmb{\mathbbm{G}}(\mathbf{r},\,\mathbf{r}',\omega)\cdot\boldsymbol\chi(\mathbf{r}', \omega)\cdot\mathbf{F}(\mathbf{r}',\omega)\mathrm{d}\mathbf{r}'.
\end{multlined}
\end{equation}
This equation establishes a self-consistent relation between the local fields $\mathbf{F}(\mathbf{r},\omega)$ and the illumination fields $\mathbf{F}_0(\mathbf{r},\omega)$ at the position $\mathbf{r}$ inside the structure.
The integral is performed over the nanostructure volume $V$. 
$\boldsymbol{\chi}(\mathbf{r}, \omega)$ is a $(6\times 6)$ tensor comprising the electric and magnetic susceptibilities:
\begin{equation}
\boldsymbol{\chi}(\mathbf{r}, \omega)=\left( \begin{matrix}
\boldsymbol{\chi}_{e}(\mathbf{r}, \omega) & 0 \\
0 & \boldsymbol{\chi}_{m}(\mathbf{r}, \omega)
\end{matrix}\right) \, .
\end{equation}
The $(6\times6)$ ``super-propagator'' $\pmb{\mathbbm{G}}$ describes light propagation in the environment:
\begin{equation}
\pmb{\mathbbm{G}}(\mathbf{r},\,\mathbf{r}',\omega)=
\left( \begin{matrix}
\mathbf{G}_{\text{tot}}^{\text{EE}}(\mathbf{r}, \mathbf{r'}, \omega) & \mathbf{G}_{\text{tot}}^{\text{EH}}(\mathbf{r}, \mathbf{r'}, \omega)\\[6pt]
\mathbf{G}_{\text{tot}}^{\text{HE}}(\mathbf{r}, \mathbf{r'}, \omega) & \mathbf{G}_{\text{tot}}^{\text{HH}}(\mathbf{r}, \mathbf{r'}, \omega)
\end{matrix}\right).
\label{superprop_G}
\end{equation}
It is composed of four Green's dyads (called field susceptibilities in the context of electrodynamics). 
The two ``direct'' field susceptibilities $\mathbf{G}_{\text{tot}}^{\text{EE}}(\mathbf{r}, \mathbf{r'}, \omega)$ and $\mathbf{G}_{\text{tot}}^{\text{HH}}(\mathbf{r}, \mathbf{r'}, \omega)$ describe, respectively, electric-electric and magnetic-magnetic coupling.
The terms $\mathbf{G}_{\text{tot}}^{\text{EH}}(\mathbf{r}, \mathbf{r'}, \omega)$ and $\mathbf{G}_{\text{tot}}^{\text{HE}}(\mathbf{r}, \mathbf{r'}, \omega)$ on the other hand are the mixed field susceptibilities which describe coupling between the electric and magnetic fields.

Since all materials in nature interact only marginally with the optical magnetic field, the implementation in \pyGDM\ is so far limited to treat only non-magnetic media, hence $\boldsymbol{\chi}_{\text{e}}\neq 0, \boldsymbol{\chi}_{\text{m}}=0$.
In consequence, equation (\ref{LPS-superpropag}) involves only the field susceptibilities $\mathbf{G}_{\text{tot}}^{\text{EE}}(\mathbf{r}, \mathbf{r'}, \omega)$ and $\mathbf{G}_{\text{tot}}^{\text{HE}}(\mathbf{r}, \mathbf{r'}, \omega)$. 
The local electric and magnetic fields are then defined by the two following Lippmann-Schwinger equations
\begin{equation}\label{LPS-only-chiE}
\begin{split}
	\begin{multlined}
		\mathbf{E}(\mathbf{r}, \omega)  = 
		\mathbf{E}_0(\mathbf{r}, \omega) + \\
		\int \mathbf{G}_{\text{tot}}^{\text{EE}}(\mathbf{r}, \mathbf{r'}, \omega) \cdot 
		\boldsymbol{\chi}_{\text{e}}(\mathbf{r}', \omega) \cdot \mathbf{E}(\mathbf{r'}, \omega) \text{d} \mathbf{r'}
	\end{multlined}\\
	\begin{multlined}
		\mathbf{H}(\mathbf{r}, \omega)  = 
		\mathbf{H}_0(\mathbf{r}, \omega) + \\
		\int \mathbf{G}_{\text{tot}}^{\text{HE}}(\mathbf{r}, \mathbf{r'}, \omega) \cdot 
		\boldsymbol{\chi}_{\text{e}}(\mathbf{r}', \omega) \cdot \mathbf{E}(\mathbf{r'}, \omega) \text{d} \mathbf{r'}
	\end{multlined}
\end{split}
\end{equation}
By default, \pyGDM\ uses a Green's tensor $\mathbf{G}_{\text{tot}}^{\text{EE}} =  \mathbf{G}_0^{\text{EE}} + \mathbf{G}_{\text{3-layer}}$, composed of a field susceptibility for the homogeneous medium $\mathbf{G}_0^{\text{EE}}$ and a field susceptibility $\mathbf{G}_{\text{3-layer}}$ associated with a substrate and/or cladding layer \cite{wiechaPyGDMPythonToolkit2018} (for computational efficiency, the latter is based on a quasi-static approximation).
Setting $\mu=1$ everywhere, $\mathbf{G}_{\text{tot}}^{\text{HE}} =  \mathbf{G}_0^{\text{HE}}$ (no contribution from the layered environment), which writes explicitly \cite{girardOpticalMagneticNearfield1997}
\begin{equation}
\mathbf{G}_0^{\text{HE}}(\mathbf{r}, \mathbf{r}', \omega) = -ik_{0}\nabla\times G_0(\mathbf{r}, \mathbf{r}', \omega) 
\end{equation}
with  \(G_0(\mathbf{r}, \mathbf{r}', \omega)=\e^{\iu k\, | \mathbf{r} - \mathbf{r'} |}/ | \mathbf{r} - \mathbf{r'} | \) being the scalar Green's function.

\section{Numerical resolution of the Lippmann-Schwinger equation -- Volume discretization}

In the following we will briefly describe how the above theory can be numerically solved for structures of arbitrary shape.
In such case, the set of equations (\ref{LPS-only-chiE}) can in general not be solved analytically and needs to be approached with numerical methods. 
Here, we solve Eqs. (\ref{LPS-only-chiE}) via a volume discretization of the structure on a regular mesh of $N$ unit cells centered at positions $\mathbf{r}_{\text{i}}$.
Consequently, the integrals become discrete sums over the ensemble of mesh-cells and the differential term $\text{d} \mathbf{r'}$ is replaced by the volume of the unit cell $V_{\text{cell}}$. 
Note that we omit the dependency on $\omega$ in the following for better readability.

\begin{multline}\label{eq:discretized_sums_H_E}
	\begin{aligned}
    \mathbf{E}(\mathbf{r}_{\text{i}}) =& 
     \mathbf{E}_0(\mathbf{r}_{\text{i}}) \\
&+ 
         \sum_{\text{j}=1}^{\text{N}} \mathbf{G}_{\text{tot}}^{\text{EE}}(\mathbf{r}_{\text{i}}, \mathbf{r}_{\text{j}}) \cdot 
              \boldsymbol{\chi}_{\text{e}}(\mathbf{r}_{\text{j}}) \cdot \mathbf{E}(\mathbf{r}_{\text{j}})V_{\text{cell}}\\
    \mathbf{H}(\mathbf{r}_{\text{i}}) =& 
     \mathbf{H}_0(\mathbf{r}_{\text{i}}) \\
&+ 
         \sum_{\text{j}=1}^{\text{N}} \mathbf{G}_{\text{tot}}^{\text{HE}}(\mathbf{r}_{\text{i}}, \mathbf{r}_{\text{j}}) \cdot 
              \boldsymbol{\chi}_{\text{e}}(\mathbf{r}_{\text{j}}) \cdot \mathbf{E}(\mathbf{r}_{\text{j}})V_{\text{cell}}.
	\end{aligned}
\end{multline}
It is worth emphasizing that the incident electromagnetic field ($\mathbf{E}_0$, $\mathbf{H}_0$) can be \textit{e.g.} a plane wave, a highly focused beam with linear, radial or azimuthal polarization, as well as a dipolar field.
A key advantage of the GDM is that different excitation conditions can be considered by computationally highly efficient post-processing, once the Green's dyadic tensor (also ``field susceptibility tensor'') of the full structure has been determined.

We now define $6N$-dimensional super-vectors $\mathbf{F}_{\text{obj}}$ and $\mathbf{F}_{0,\text{obj}}$, which are composed of the electromagnetic fields at each unit cell's position $\mathbf{r}_{\text{i}}$ as
\begin{multline*}
	\mathbf{F}_{0, \text{obj.}} = 
	\Big( 
	\mathbf{E}_0(\mathbf{r}_1),\mathbf{E}_0(\mathbf{r}_2),\ldots,\quad 
	\ldots,\mathbf{E}_0(\mathbf{r}_{\text{N}}), \\
	\mathbf{H}_0(\mathbf{r}_1),\mathbf{H}_0(\mathbf{r}_2),\ldots,\quad 
	\ldots,\mathbf{H}_0(\mathbf{r}_{\text{N}})
	\Big) 
\end{multline*}
\begin{multline*}
	\mathbf{F}_{\text{obj.}} = 
	\Big( 
	\mathbf{E}(\mathbf{r}_1),\mathbf{E}(\mathbf{r}_2),\ldots,\quad 
	\ldots,\mathbf{E}(\mathbf{r}_{\text{N}}), \\
	\mathbf{H}(\mathbf{r}_1),\mathbf{H}(\mathbf{r}_2),\ldots,\quad 
	\ldots,\mathbf{H}(\mathbf{r}_{\text{N}})
	\Big). 
\end{multline*}

Using those super-vectors, the system of equations defined by Eq.~(\ref{eq:discretized_sums_H_E}) can be written in matrix form :
\begin{equation}\label{eq:F-MF0}
\mathbf{F}_{0,\text{obj}} =
\left( \begin{matrix}  
\mathbf{M}^{\text{EE}} & 0 \\ 
\mathbf{M}^{\text{HE}} &  \mathbf{I}_{3\text{N}}
\end{matrix} \right)
\cdot
\mathbf{F}_{\text{obj}} =
\pmb{\mathbbm{M}}\cdot\mathbf{F}_{\text{obj}}
\end{equation}
with the identity matrix $\mathbf{I}_{3\text{N}}$ of size $3N$. 
$\mathbf{M}^{\text{EE}}$ and $\mathbf{M}^{\text{HE}}$ are ($3N\times 3N$) matrices composed of $N$ (3$\times$3) matrices defined by equation~(\ref{eq:discretized_sums_H_E}):
\begin{multline}\label{eq:MEE_MHE_for_inversion}
	\begin{aligned}
    &\mathbf{M}^{\text{EE}}_{ij}(\mathbf{r}_{\text{i}},\,\mathbf{r}_{\text{j}})  = 
     \mathbf{I}\delta_{\text{ij}} - \sum_{\text{j}=1}^{\text{N}} \mathbf{G}_{\text{tot}}^{\text{EE}}(\mathbf{r}_{\text{i}}, \mathbf{r}_{\text{j}}) \cdot 
              \boldsymbol{\chi}_{\text{e}}(\mathbf{r}_{\text{j}}) \cdot V_{\text{cell}}\\
&\mathbf{M}^{\text{HE}}_{ij}(\mathbf{r}_{\text{i}},\,\mathbf{r}_{\text{j}})  = 
     - \sum_{\text{j}=1}^{\text{N}} \mathbf{G}_{\text{tot}}^{\text{HE}}(\mathbf{r}_{\text{i}}, \mathbf{r}_{\text{j}}) \cdot 
              \boldsymbol{\chi}_{\text{e}}(\mathbf{r}_{\text{j}}) \cdot V_{\text{cell}}.
	\end{aligned}
\end{multline}
$\mathbf{I}$ is the unitary tensor.
Now, by inverting the matrix $\pmb{\mathbbm{M}}$, we can deduce the local electromagnetic field $\mathbf{F}_{\text{obj}}$ inside the structure from the incident fields $\mathbf{F}_{0,\text{obj}}$ from the inverse $\pmb{\mathbbm{K}}=\pmb{\mathbbm{M}}^{-1}$ via a single matrix-vector product:
\begin{equation}\label{Field_inside}
\mathbf{F}_{\text{obj}} = \pmb{\mathbbm{K}}\cdot\mathbf{F}_{0,\text{obj}}.
\end{equation}
\pygdm\ supports conventional CPU based inversion algorithms as well as LU decomposition, both through \texttt{scipy}. A new functionality is graphics processing unit (GPU) accelerated inversion with CUDA compatible GPUs through the optional \texttt{cupy} package. On high-end GPUs, up to 10-fold speed-up can be reached.

\subsection{Interfaces: near-field, far-field and retardation}

The Lippmann-Schwinger equation (\ref{LPS-superpropag}) allows to determine the electromagnetic field everywhere in space.
Indeed, as mentioned previously, we determine the electromagnetic field inside the structure by solving self-consistently the equation (\ref{LPS-superpropag}).
Then, to determine the electric field outside the structure, we reuse the same equation to repropagate the local internal field.
The superpropagator $\pmb{\mathbbm{G}}$ introduced in the last section (Eq. (\ref{superprop_G})) contains the four field susceptibilities ($\mathbf{G}_{\text{tot}}^{\text{EE}}(\mathbf{r}, \mathbf{r'}, \omega)$, $\mathbf{G}_{\text{tot}}^{\text{EH}}(\mathbf{r}, \mathbf{r'}, \omega)$, 
$\mathbf{G}_{\text{tot}}^{\text{HE}}(\mathbf{r}, \mathbf{r'}, \omega)$, $\mathbf{G}_{\text{tot}}^{\text{HH}}(\mathbf{r}, \mathbf{r'}, \omega)$) describing the propagation of light emitted by an electric or a magnetic dipole.

For a dipole in the vicinity of a plane interface these susceptibilities can be written as a sum of a free space contribution $\mathbf{G}_{0}^{\alpha\beta}(\mathbf{r}, \mathbf{r'}, \omega)$ and a surface dyad $\mathbf{G}_{\text{2-layer}}^{\alpha\beta}(\mathbf{r}, \mathbf{r'}, \omega)$. 
The superscript index \(\alpha\) represents  the nature of the induced field at $\mathbf{r}$ (electric ``E'' or magnetic ``H'') and \(\beta\) represents the nature of the dipole at position $\mathbf{r}'$.
The free space susceptibilities, solutions of the wave-equations Eq.~(\ref{eq:wave-equations}), are proportional to the scalar Green's function $G_0(\mathbf{r}, \mathbf{r}', \omega)$ (see appendix Eq.~\ref{eq:props_1}).
These susceptibilities can be decomposed in a sum of dyadic tensors representing the near field, intermediate and far field contributions (Eq. (\ref{eq:props_2})) \cite{colasdesfrancsOptiqueSublongueurOnde2002, girardFieldsNanostructures2005}.

For a single plane interface, the susceptibilities $\mathbf{G}^{\alpha\beta}_{\text{2-layer}}(\mathbf{r}, \mathbf{r'}, \omega)$, associated with the surface, can be obtained by expanding the electric and magnetic fields using plane waves while applying the boundary conditions for the tangential component of the field at the interface  \cite{agarwalQuantumElectrodynamicsPresence1975, colasdesfrancsOptiqueSublongueurOnde2002, novotnyPrinciplesNanooptics2006}. 
We summarize the main results of the calculation in \ref{sec:weyl_representation}.

\subsubsection{Near-field in the electrostatic region: mirror-charge approximation}

\begin{figure}[t!]
	\centering
	\includegraphics[width=0.95\columnwidth]{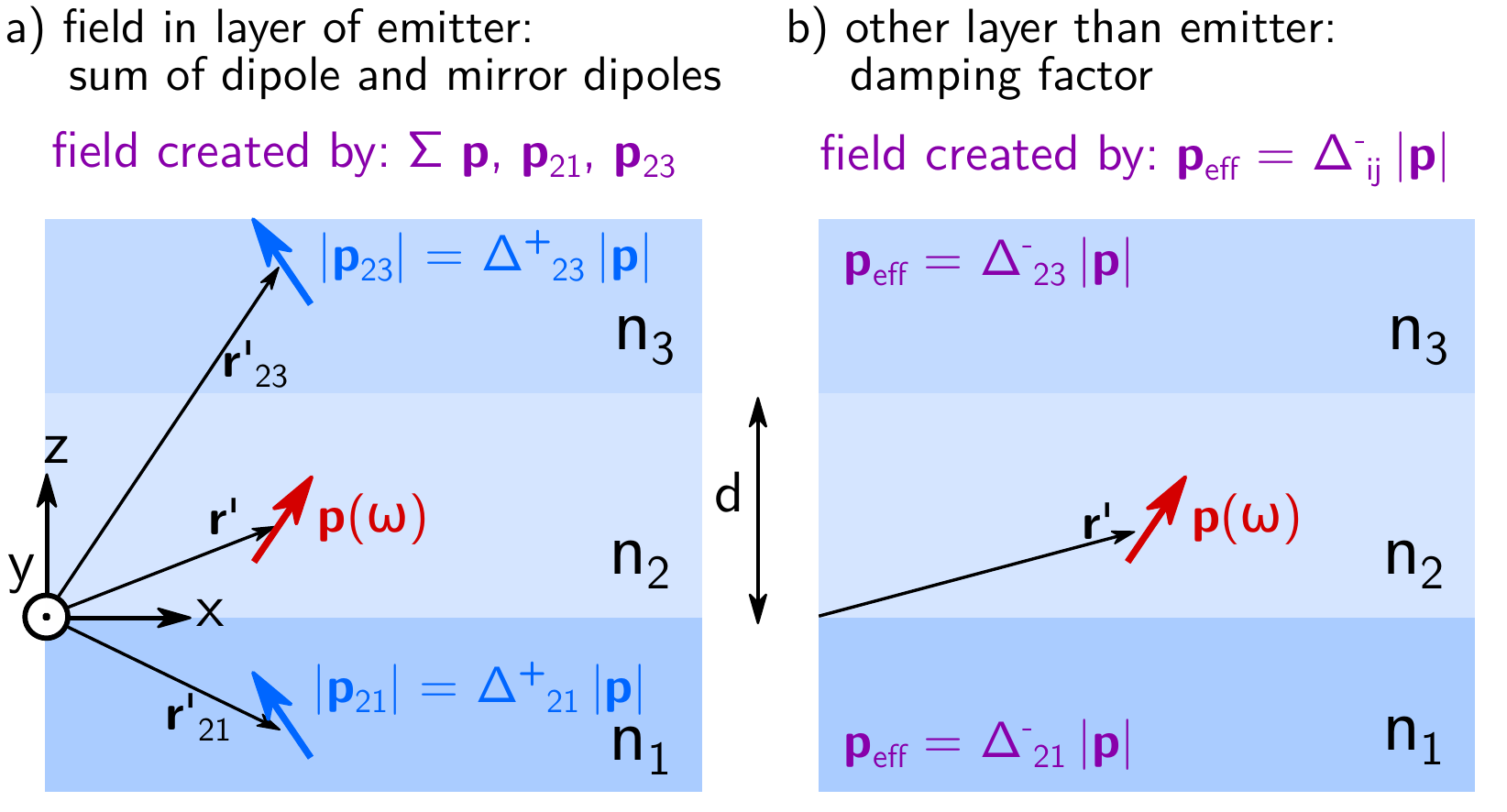}
	\caption{
		Illustration of the image dipole approximation.
		Schematic representation of an electric dipole $\mathbf{p}(\omega)$ (red) at the position $\mathbf{r}'$ in a layer of refractive index n$_2$ and thickness $d$. The layer is sandwiched between two semi-infinite layers of indices $n_1$ (below) and $n_3$ (above).
		a) $\mathbf{p}_{21}(\omega)$ and $\mathbf{p}_{23}(\omega)$ are the image dipoles in the lower and upper layers, used to approximate the field inside the layer of the emitting physical dipole.
		Their orientation is along $\mathbf{p}_{21} = \mathbf{p}_{23} = (-p_x, -p_y, +p_z)$, their $z$-distance to $\mathbf{p}$ is twice the distance between the physical dipole and the respective interface.
		b) in order to calculate the fields in the layers where no physical emitter is present, no mirror dipoles are required. The field corresponds to the original dipole's field, attenuated by a screening factor $\Delta^-$.
	}\label{fig:dipole_image}
\end{figure}

In case of a low index dielectric layer or if the emitted field is evaluated at distances to the dipole, significantly smaller than the wavelength (\textit{i.e.} in the electrostatic limit with no retardation), we can use an approximation based on the method of mirror charges to describe the effect of the interface.
As illustrated in figure \ref{fig:dipole_image}a for the presence of two interfaces, for an evaluation inside the layer of the emitter dipole a virtual dipole is added to the configuration, with an amplitude vector of opposite sign and which is placed at identical distance behind the interface. 
This ensures the continuity of the electromagnetic field at the interface.
Inside the empty medium, the emitted field corresponds to the single dipole, however attenuated by a screening factor (see Fig.~\ref{fig:dipole_image}).
This leads to the expressions (\ref{G_2layer_EE_elec+}) and (\ref{G_2layer_EE_elec-}) for the electric-electric contribution. 
The magnetic-electric susceptibility of the interface is zero  (Eq. (\ref{G_2layer_HE_elec})).
The mirror dipole method provides a good approximation for the contribution of the interface if retardation effects are negligible \cite{gay-balmazValidityDomainLimitation2000}.

In \pygdm\ the environment can consist of at most three layers of refractive indexes $n_1$, $n_2$ and $n_3$ for the bottom, center and top layer, respectively. 
The lower and upper layers in Fig.~\ref{fig:dipole_image} are semi-infinite and the intermediate layer has a thickness \textit{d} (also called ``spacing'').
In the shown example with an electric dipole $\mathbf{p}(\omega)$ in the intermediate layer, the two mirror dipoles are represented by the blue arrows $\mathbf{p}_{21}$ and $\mathbf{p}_{23}$.

\subsubsection{Asymptotic far-field approximation}

Conversely if the evaluation position is at a large distance from the emitting dipole, we are in the far-field region.
In this case the surface contributions above and below the interface have also explicit analytical expressions. 
They can be obtained using the asymptotic limit, defined in equation (\ref{eq:lim_asymp}).
For the electric-electric surface term $\mathbf{G}^{\text{EE}}_{\text{2-layer}}(\mathbf{r}, \mathbf{r'}, \omega)$, the analytical asymptotic expressions implemented in pyGDM are (\ref{eq:Ss_+inf}) and (\ref{eq:Ss_-inf}).
In the far-field the electromagnetic wave is transverse, therefore it is possible to deduce the magnetic field $\mathbf{H}$ via the cross product of $\mathbf{k}$ and $\mathbf{E}$.

\subsubsection{Intermediate region - retardation}

If the field is to be evaluated at intermediate distances to the dipole (roughly at distances of less than a few wavelengths), no general approximation exists and the integral in Eq.~(\ref{eq:G_2_layer}) must be performed numerically \cite{novotnyInterferenceLocallyExcited1997}, requiring integration in the complex plane to avoid the singularities on the real axis. 
This can be interpreted as an integration over propagating (real valued wavevectors) and evanescent modes (complex valued wavevectors) \cite{rahmaniFieldPropagatorDressed1997}.
Reference~\cite{paulusAccurateEfficientComputation2000} provides a detailed discussion about how to choose the best integration path in the complex plane, and extends the procedure to layered media with arbitrary numbers of interfaces.
Due to the necessity of the complex integration, the evaluation of the retarded Green's tensors is computationally relatively expensive. 
It is usually necessary to include retardation effects when considering large structures deposited on high index dielectric, thin film or plasmonic substrates, in order to take into account the effect of, for example, propagating modes or surface plasmons. 

\begin{figure}[t!]
	\centering
	\includegraphics[width=0.95\columnwidth]{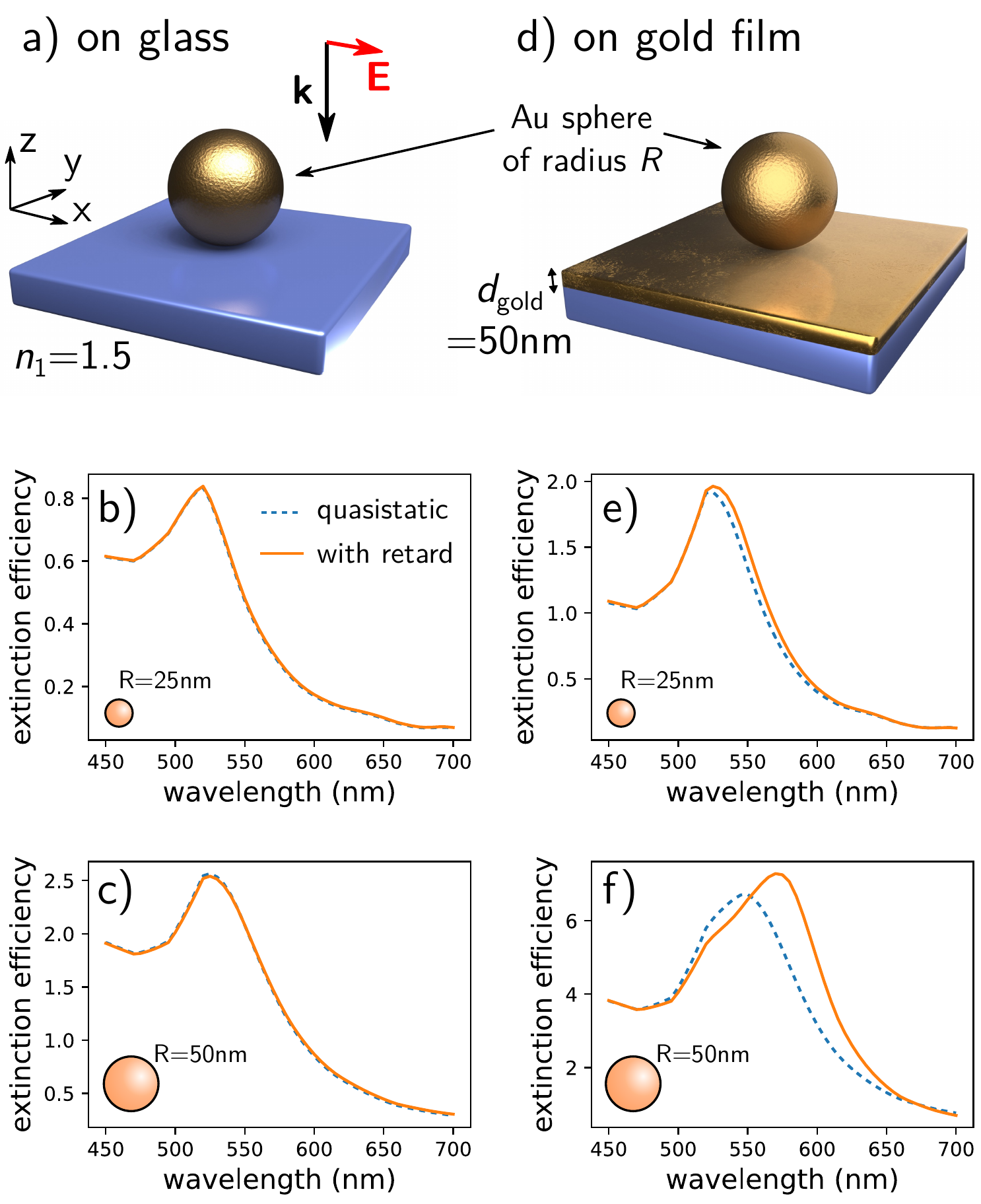}
	\caption{
		Comparison between quasistatic mirror-dipole approximation (dashed blue lines) and Green's dyads with retardation (solid orange lines) to describe the substrate.
		(a-c) Extinction simulations of a gold nanosphere on a semi-infinite dielectric substrate ($n_1=1.5$) as depicted in (a). The sphere is illuminated by a plane wave from the top, with linear polarization along $X$.
		(b) extinction efficiency for a gold sphere of radius $R=25\,$nm and (c) for $R=50\,$nm.
		(d-f) same as (a-c) but the substrate is composed of an additional $d_{\text{gold}}=50\,$nm thin gold layer on top of the dielectric substrate.
	}\label{fig:quasistatic_vs_retard}
\end{figure}

A comparison between the mirror dipole approximation and a fully retarded description of the substrate is given in figure~\ref{fig:quasistatic_vs_retard}. Panels~\ref{fig:quasistatic_vs_retard}a-c show simulation results of a gold nanosphere lying on a glass substrate ($n_1=1.5$), illuminated by a plane wave polarized along $X$. 
A small nanosphere ($R=25\,$nm in Fig.~\ref{fig:dipole_image}b) can be very accurately described in the quasistatic approximation, and even for larger particles (Fig.~\ref{fig:dipole_image}c, $R=50\,$nm), only small deviations occur in comparison with the retarded simulation. 
Figure~\ref{fig:quasistatic_vs_retard}d-f show simulations of an identical gold sphere as in \ref{fig:quasistatic_vs_retard}a-c, but now lying on a thin gold film ($d_{\text{gold}}=50\,$nm). 
Even in this case, small particles ($R=25\,$nm, Fig.~\ref{fig:quasistatic_vs_retard}e) can be described accurately using quasistatic Green's dyads.
But when the radius of the sphere is increased ($R=50\,$nm, Fig.~\ref{fig:quasistatic_vs_retard}f), significant deviations occur between the non-retarded and retarded descriptions.
For a script to reproduce the results in figure~\ref{fig:quasistatic_vs_retard} see \href{https://wiechapeter.gitlab.io/pyGDM2-doc/examples/exampleRetard_goldsphere_on_substrate.html}{this link}.

Because of poor performance of pure python code and to avoid a dependency of \pygdm\ on external binaries, the retarded Green's tensors are not implemented in the main \pygdm\ module.
However, we provide a separate python module based on a fortran implementation of the retarded Green's tensors for the 2-interface environment, used in \pyGDM . Please note that this extension of \pygdm\ is still in a development version, and not all aspects have been implemented yet.
The ``\texttt{pyGDM2\_retard}'' module requires compilation of the fortran part. 
Via the python package index \textit{pypi}, we provide pre-compiled binaries for windows, on other platforms the module needs to be compiled from the source code (see \ref{sec:installation}).
For technical details on the numerical calculation of the retarded Green's tensors, we refer the interested reader to Ref.~\cite{paulusAccurateEfficientComputation2000}.
We note that the ``\texttt{pyGDM2\_retard}'' package requires the nanostructure to be placed in the top-most layer (layer ``3'' in figure~\ref{fig:dipole_image}). Moreover, field calculations are only possible in the layer of the nanostructure. We plan to extend the retarded module in the future.

If for reasons of computation speed, the mirror dipole approximation is used to take into account a layered environment, the above described limitations should be kept in mind.

\subsection{Mesh-type, self-terms}

We will recall here briefly the meshing in \pygdm. For details, please refer to Ref. \cite{wiechaPyGDMPythonToolkit2018}.
\pygdm\ can currently handle discretizations on regular cubic and hexagonal compact meshes in 3D and on square meshes in 2D. 
The divergence of the Green's function at $\mathbf{r}_i = \mathbf{r}_j$ needs to be accounted for by analytically evaluating the so-called ``self-term'' at that location, which depends on the geometry of the mesh-cells. 

For the cubic, respectively hexagonal mesh, in 3D, we find (see \cite{girardShapingManipulationLight2008}, section~3.1)
\begin{equation}\label{eq:renormalization_cube}
\mathbf{G}_{0,\text{cube}}^{\text{EE}}(\mathbf{r}_i, \mathbf{r}_i) = 
- \frac{4\pi}{3 \epsilon_{\text{env}} \stepsize^3} \, \mathbf{I}
\end{equation}
\begin{equation}\label{eq:renormalization_hex}
\mathbf{G}_{0,\text{hex}}^{\text{EE}}(\mathbf{r}_i, \mathbf{r}_i) = 
-\frac{4\pi \sqrt{2}}{3 \epsilon_{\text{env}} \stepsize^3}  \, \mathbf{I}
\, .
\end{equation}
For the self-terms of a 2D square mesh, see \ref{sec:2D_selfterms}.

Self-terms for more complex mesh-cell geometries like cuboids or tetrahedrons can be calculated numerically \cite{kottmannAccurateSolutionVolume2000, ouldaghaNearFieldPropertiesPlasmonic2014, smunevRectangularDipolesDiscrete2015}. 
This is currently not implemented but might be added in a future version of \pygdm.

We note here that the structure needs to be discretized with a sufficiently small step to resolve all fields and their gradients inside the nanostructure. For best convergence the step should be in the order of $d\lesssim \lambda_0 / 10 n_{\text{struct}}$, but significantly finer meshes may be required for certain configurations, for instance in cases of strong field gradients.

\subsection{Outlook: Magnetic media}

As mentioned before, the current version of \pygdm\ assumes media with no direct magnetic response ($\boldsymbol{\chi}_{\text{m}}=0$), but it might be implemented in a future version of \pyGDM. 
It is technically straightforward to include the $\boldsymbol{\chi}_{\text{m}}\neq 0$ terms in the calculation. 
While natural materials don't interact with the optical magnetic field, this may be interesting to model meta-materials with a macroscopic \textit{effective} magnetic response \cite{pendryNegativeRefractionMakes2000}.
Every mesh-cell would correspond to a meta-atom of the meta-material.
In such case the upper and lower right sub-matrices in $\pmb{\mathbbm{M}}$ would be filled with \textit{electric-magnetic} and \textit{magnetic-magnetic} coupling terms $\mathbf{M}^{\text{EH}}_{ij}$ and  $\mathbf{M}^{\text{HH}}_{ij}$ \cite{chaumetCoupleddipoleMethodMagnetic2009}.

\subsection{pyGDM simulation description}

\begin{figure}[t]
	\centering
	\includegraphics[width=0.9\columnwidth]{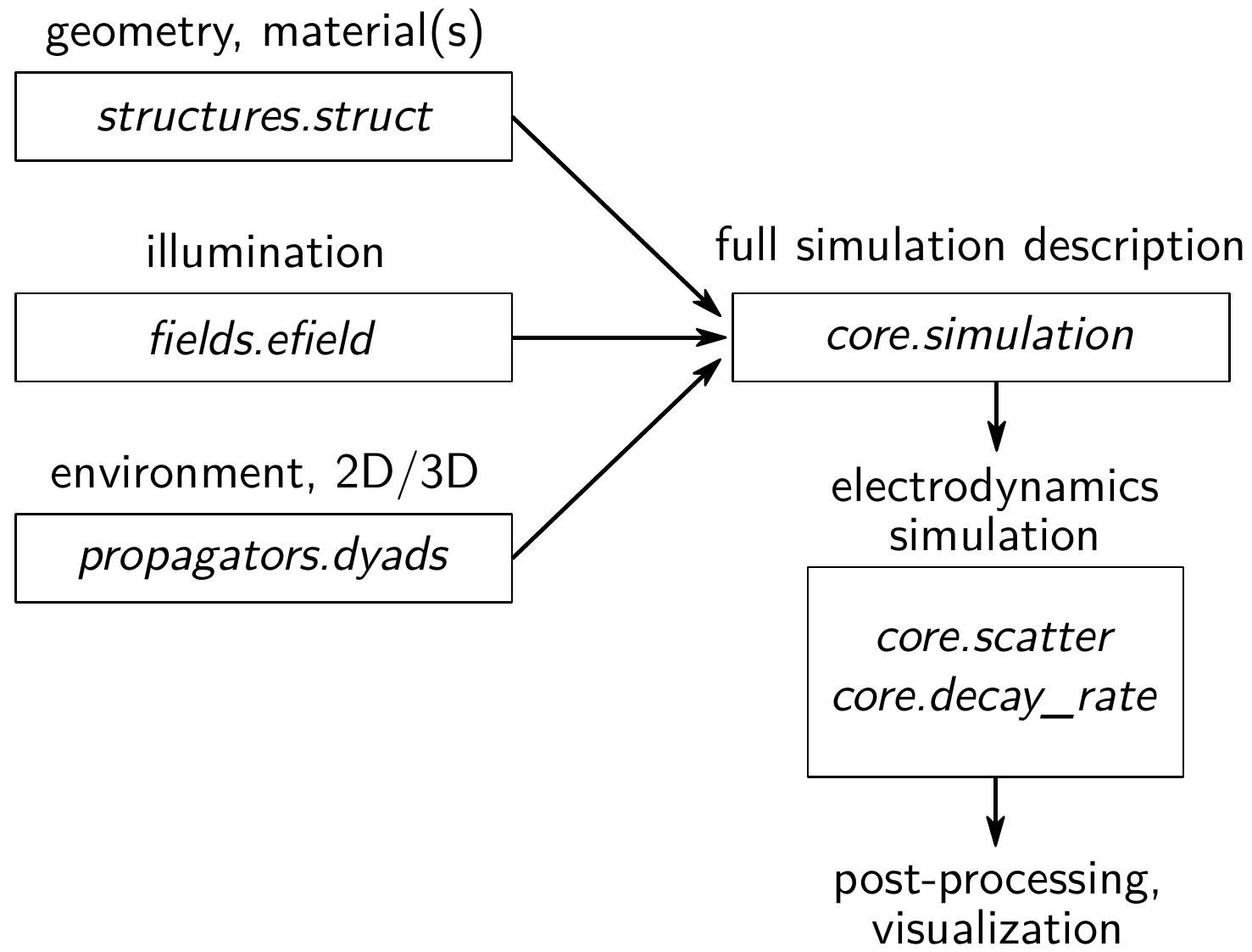}
	\caption{
		Structure of a simulation in pyGDM and the typical workflow.
		The geometry and material(s) of  the nano-object are defined in the \textit{struct}-class, the illumination field by the \textit{efield}-class and the simulation environment (environment refractive index, substrate, 2D/3D, ...) is configured in a \textit{dyads} object. 
		The \textit{simulation} class bundles all information and serves also as a container for the simulation results, which is then used for further evaluation and post-processing or visualizations.
	}\label{fig:pygdm_structure}
\end{figure}

A \pygdm\ simulation is organized as depicted in figure~\ref{fig:pygdm_structure}. 
A first class describes the geometry and material(s) of the nanostructure (class \textit{struct}). 
A second class describes the incident field (class \textit{efield}), and a third one the environment and simulation type (new class \textit{dyads}).
These three objects are assembled in a \textit{simulation} class, which also serves as a container to store the results of the GDM simulation.
The simulation object is also used by all further post-processing and visualization functions provided by \pygdm.

%% file: 03_theory_new_functionalities.tex
\section{New functionalities}

\subsection{Decomposition in electric and magnetic dipole modes}

We have implemented in the updated version the decomposition of the internal field in electric and magnetic multipoles, following the work of Evlyukhin \textit{et al.} \cite{evlyukhinMultipoleLightScattering2011}. 
By default, the new functions \textsf{linear.multipole\_decomp} and \textsf{linear.multipole\_decomp\_extinct} return only the electric dipole (ED) and magnetic dipole (MD) contributions, those are stable and extensively tested. 
The quadrupoles can be returned by passing the optional keyword argument \textsf{quadrupoles=True}, but those are still experimental and have not been sufficiently tested.
The effective dipole moments $\mathbf{p}_{\text{tot}}$ (ED) and $\mathbf{m}_{\text{tot}}$ (MD) are obtained from the internal field distribution as
\begin{equation}\label{eq:ED_moment}
\mathbf{p}_{\text{tot}} = \sum_{i}^{N} \mathbf{p}_i = \sum_{i}^{N}  V_{\text{cell}}\, \boldsymbol{\chi}_{e}(\mathbf{r}_i) \cdot \mathbf{E}(\mathbf{r}_{i})
\end{equation}
and
\begin{equation}\label{eq:MD_moment}
\mathbf{m}_{\text{tot}} = - \frac{\text{i} k_0}{2} \sum_{i}^{N} \big( \mathbf{r}_i - \mathbf{r}_0 \big) \times \mathbf{p}_i \, .
\end{equation}
We note that the choice of the position $\mathbf{r}_0$ of the total dipole moment is crucial in the calculation of the MD moment. By default, \pygdm\ uses the center of mass of the nanostructure, but $\mathbf{r}_0$ can be specified manually using the optional keyword argument ``\textsf{r0}''.

\begin{figure}[t!]
	\centering
	\includegraphics[width=0.85\columnwidth]{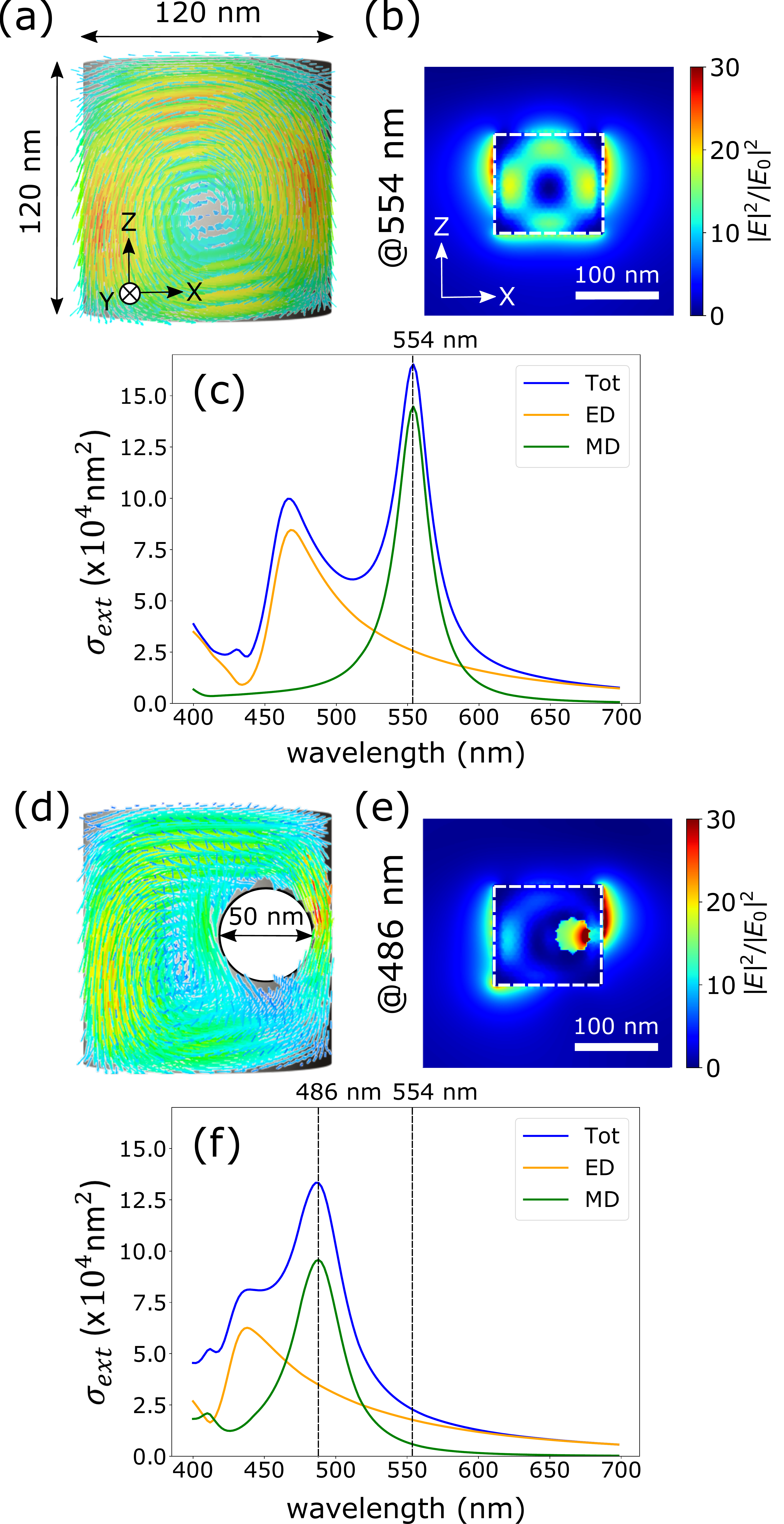}
	\caption{
		Influence of a hole inside a silicon nanodisc on the electric and magnetic dipole resonances. 
		(a-c) Full nanodisc of diameter $D=120\,$nm and height $H=120\,$nm.
		(a) Field vectors inside (real part) and (b) near-field intensity inside and around the disc at the magnetic resonance position $\lambda_0=554\,$nm.
		(c) Total extinction cross section (blue line) as well as the electric (orange line) and magnetic (green line) dipole contributions. 
		The illumination is a plane wave traveling along $Z$, polarized along $X$, the environment is vacuum ($n_{\text{env}}=1$).
		(d-f) The same nanodisc but with a hole of diameter $D_{\text{hole}}=50\,$nm along the (Oy) direction, cutting through the field vortex which induces the magnetic dipole moment. 
		(d-e) show the near-field as in (a-b), here at the resonance wavelength $\lambda_0 = 486$\,nm. 
		(f) extinction cross sections similar to (c). 
	}\label{fig:multipole_decomp_extinct}
\end{figure}

The respective contributions to the extinction cross section can be calculated as \cite{evlyukhinMultipoleLightScattering2011}
\begin{equation}\label{eq:ED_extinction}
\begin{aligned}
\sigma_{\text{ext,ED}} & = \frac{4 \pi k_0}{|\mathbf{E}_0|^2 \, n_{\text{env}}}  \, \text{Im} \Big(  \sum \mathbf{E}_0^* \cdot \mathbf{p}_{\text{tot}} \Big) \\
\sigma_{\text{ext,MD}} & = \frac{4 \pi k_0}{|\mathbf{E}_0|^2 \, n_{\text{env}}}  \, \text{Im} \Big(  \sum \mathbf{H}_0^* \cdot \mathbf{m}_{\text{tot}} \Big) \, ,
\end{aligned}
\end{equation}
where $n_{\text{env}}$ is the refractive index of the environment.

We demonstrate the decomposition of the extinction cross section on a silicon nano-disc, as illustrated in figure~\ref{fig:multipole_decomp_extinct}.
Fig.~\ref{fig:multipole_decomp_extinct}a-c show the case of a full cylinder in vacuum, illuminated from the top by a plane wave of linear polarization along $X$. 
The internal field vectors and the field intensity in and around the nano-disc are shown in Fig.~\ref{fig:multipole_decomp_extinct}a-b for the MD resonance at $\lambda_0=554$\,nm, which can be easily identified in the extinction spectra shown in Fig.~\ref{fig:multipole_decomp_extinct}c. 
The electric field vectors inside the nano-disc form a vortex in the $XZ$ plane, which induces an effective magnetic dipole moment along $Y$.
We now cut a hole in the silicon, through the vortex, as illustrated in figure~\ref{fig:multipole_decomp_extinct}d (hole at the right side of the disc, parallel to $Y$, with a diameter of $D_{\text{hole}}=50\,$nm). 
We see that the vortex at the magnetic resonance wavelength is now strongly perturbed (see in particular Fig.~\ref{fig:multipole_decomp_extinct}e). 
This results in a significant reduction of the magnetic dipole contribution to the extinction, as can be seen in the spectra of figure~\ref{fig:multipole_decomp_extinct}f. 
Simultaneously, an electric near-field hotspot is induced around the hole. 
The shift of the resonances to shorter wavelengths can be attributed to the reduced amount of material in comparison with the full disc.

The \pygdm-script to reproduce the results shown in figure~\ref{fig:multipole_decomp_extinct} can be found \href{https://wiechapeter.gitlab.io/pyGDM2-doc/examples/example09b_multipole_decomposition_hole.html}{in the online documentation at this link}.

\subsection{Magnetic field inside the nanostructure}

\begin{figure*}[t]
	\centering
	\includegraphics[width=\textwidth]{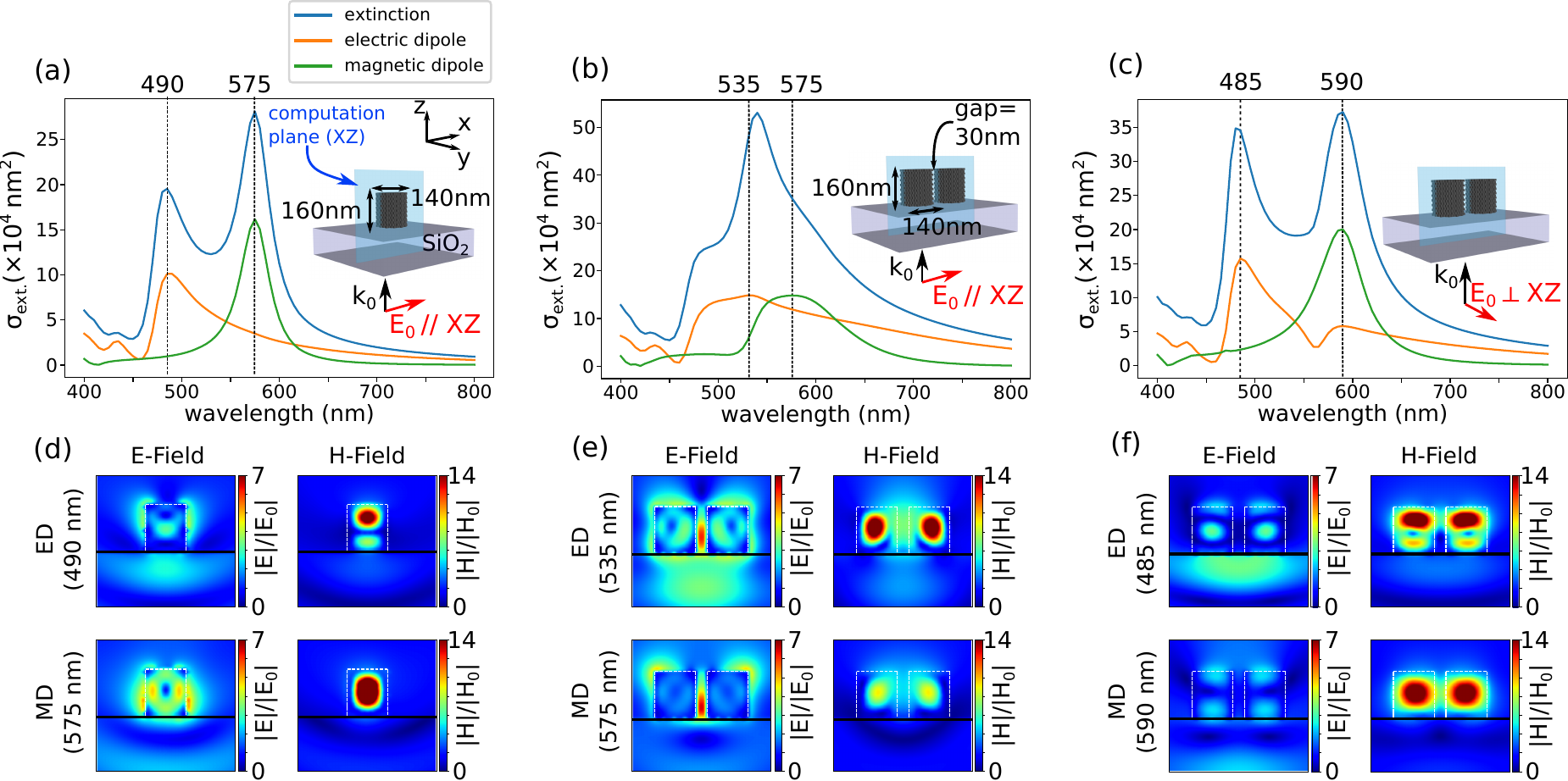}
	\caption{
		Simulated extinction spectra and near-field amplitude maps for a single silicon cylinder and a cylinder dimer lying on a silica substrate ($n_1=1.45$) surrounded by air ($n_2=1$). 
		Each cylinder, discretized by $N=1490$ unit cells on a hexagonal compact grid, has a diameter $D=140$\,nm, a height $H=160$\,nm and, for the dimer, a gap $G=30$\,nm. 
		The structures are illuminated by a plane wave from below (through the substrate).
		(a) total extinction cross section (blue curve), as well as the electric (orange) and magnetic (green) dipole contributions are calculated. Linear polarization along  $X$ (see inset of (a)). 
		(d) Normalized amplitude maps of the electric and magnetic fields inside and around the structures (in the $XZ$ plane) at the electric and magnetic resonances. 
		The positions of the resonances are highlighted by the vertical dashed lines in (a). 
		(b), (e) and (c), (f): Same as (a),(d) but for a dimer illuminated by a parallel (b), (e) or a perpendicular linear polarization (c), (f). 
		The maps show areas of $450\times425\,\text{nm}^2$. 
		The outline of the respective structure is indicated by a dashed white line, the air-silica interface with a black solid line.
	}\label{fig:NF_MAP_E_H}
\end{figure*}

Calculating only the electric response can be done by inversion of $\mathbf{M}^{\text{EE}}$ in equation~(\ref{eq:MEE_MHE_for_inversion}).
However, if the magnetic field inside a nanostructure needs to be known, we require the inverse of matrix $\mathbbm{M}$ in equation (\ref{eq:F-MF0}).
Standard matrix inversion techniques scale with the third power. In consequence, doubling the size of the matrix means the inversion takes 8 times longer (while the matrix occupies four times the memory).
Fortunately, a block-matrix can be efficiently inverted block-wise using the Hans Boltz method \cite{cormenIntroductionAlgorithms2001}. 
In our case, assuming non-magnetic media ($\chi_{\text{m}}=0$), we obtain the inverse of $\mathbbm{M}$ in equation (\ref{eq:F-MF0}) as
\begin{equation}\label{eq:K_blocks_nomag}
	\pmb{\mathbbm{K}}=\left(\begin{matrix}
		\mathbf{K}^{\text{EE}} & 0 \\
		-\mathbf{M}^{\text{HE}}\cdot \mathbf{K}^{\text{EE}} & \mathbf{I}_{3\text{N}}
	\end{matrix}\right)
\end{equation}
with $\mathbf{K}^{\text{EE}}$ the \textit{electric-electric} generalized field propagator \cite{martinGeneralizedFieldPropagator1995}. 
Hence, the $6N\times 6N$ problem reduces to the inversion of a $3N\times 3N$ matrix and an additional product of two matrices of the same size. 
It is therefore of comparable computational cost as the resolution of the electric-electric problem alone \cite{wiechaPyGDMPythonToolkit2018}.
However, calculating the internal magnetic field in addition to the electric field requires around twice as much memory, in order to store the additional $3N\times 3N$ block at the lower left of Eq. (\ref{eq:K_blocks_nomag}).
Therefore the H-field calculation is by default deactivated in \pygdm\ but can be enabled by passing \textit{\texttt{calc\_H=True}} as parameter to the \texttt{core.scatter} function.

As an example we reproduce in Fig.~\ref{fig:NF_MAP_E_H} the internal magnetic field in a silicon disc and in silicon disc-dimers as reported earlier  by Wang et al. \cite{wangBroadbandOpticalScattering2014}.
The silicon cylinders are placed on a silica substrate ($n_1=1.45$) and are embedded in air ($n_2=1$).
The silicon cylinders have a diameter $D=140$\,nm and a height $H=160$\,nm. 
In case of the disc-dimer both cylinders are separated by a gap of $G=30$\,nm.
We discretize the cylinders on a hexagonal compact lattice (with $N=1490$ mesh cells).
We use plane wave illumination at normal incidence from inside the substrate, polarized along the O$x$-axis (Fig.~\ref{fig:NF_MAP_E_H}a, \ref{fig:NF_MAP_E_H}d and \ref{fig:NF_MAP_E_H}b, \ref{fig:NF_MAP_E_H}e) or along the Oy-axis (Fig. \ref{fig:NF_MAP_E_H}c and \ref{fig:NF_MAP_E_H}f).

Figure~\ref{fig:NF_MAP_E_H}a-c show the total extinction curves (blue spectra), and additionally the contributions due to the effective electric and magnetic dipole moments (orange and green curves, respectively).
This allows us to determine where the structures have electric dipole (ED) and magnetic dipole (MD) resonances.

While the $X$-polarized dimer has a totally different optical response, the dimer illuminated by $Y$-polarized light yields similar optical spectra as the isolated cylinder.
The reason is that under $Y$-polarization, the electric field is perpendicular to the axis of the dimer, which reduces coupling. 
At the respective electric and magnetic dipole resonances we computed the normalized amplitudes of the electric and magnetic fields, shown in the bottom subplots of figure~\ref{fig:NF_MAP_E_H}.
The spectra and field amplitude maps are in good agreement with results obtained by Finite Difference Time Domain (FDTD) simulations \cite{bakkerMagneticElectricHotspots2015, albellaLowLossElectricMagnetic2013, wangBroadbandOpticalScattering2014}.

The \pygdm-script to reproduce the results shown in figure~\ref{fig:NF_MAP_E_H} can be found \href{https://wiechapeter.gitlab.io/pyGDM2-doc/examples/example12_internal_H_field.html}{in the online documentation under this link}.

\begin{figure}[t!]
	\centering
	\includegraphics[width=0.85\columnwidth]{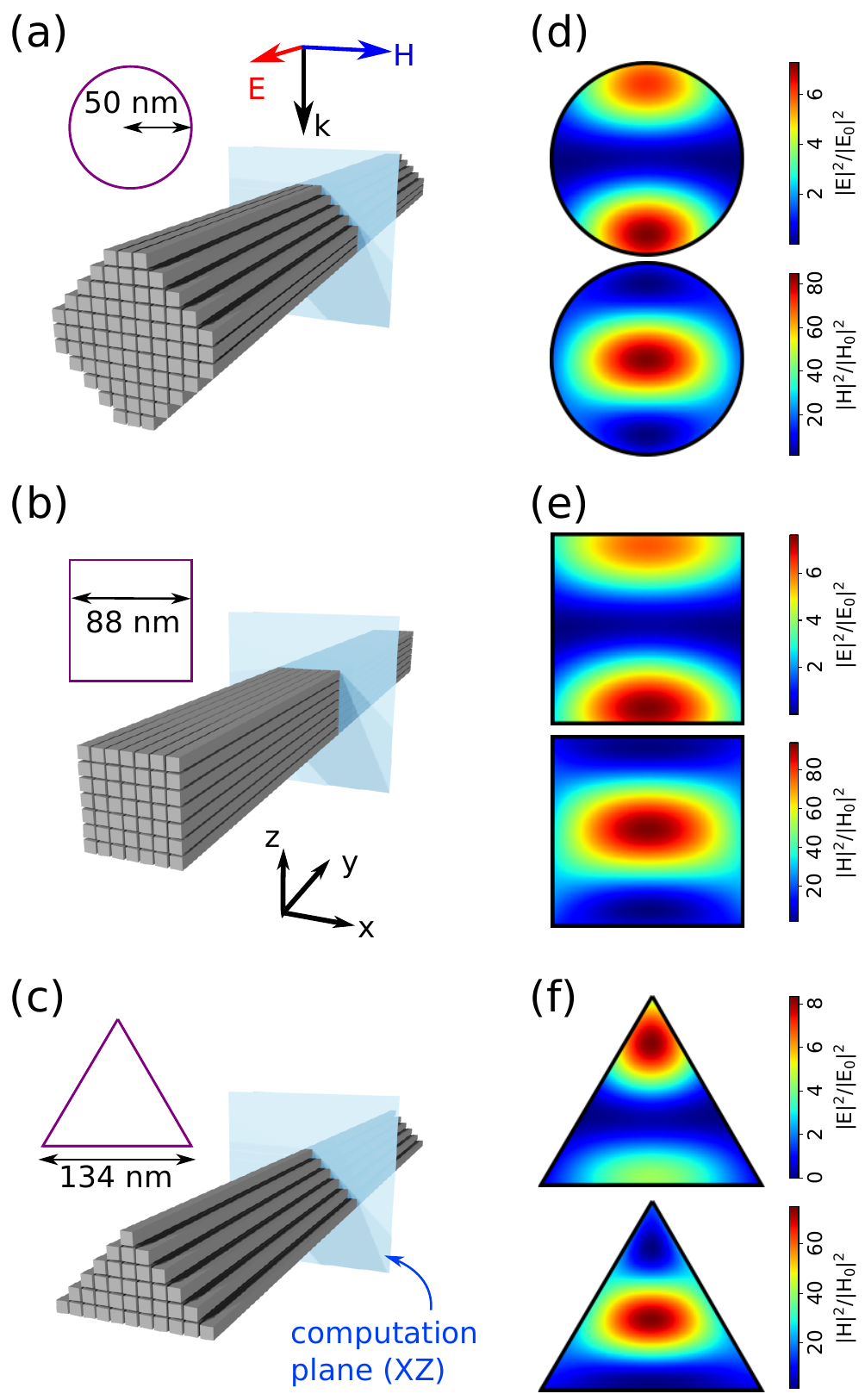}
	\caption{
		2D GDM simulations of the field intensity distribution inside infinitely long silicon nanowires (SiNWs) of different shapes: circular (a), square (b) and triangular (c). The sizes are chosen in order to excite the magnetic  dipole Mie resonance at $\lambda_0 = 550\,$nm (normal incidence plane wave illumination with linear polarization along the NW axis).
		The corresponding electric and magnetic field intensity maps inside the SiNWs are shown in (d-f) in the upper, respectively lower subplots.
		The SiNWs are in vacuum. Note that for illustrative purposes the sketches (a-c) depict a significantly coarser discretization as used for the actual simulations.
	}\label{fig:2D_sim_SiNWs}
\end{figure}

\subsection{2D simulations}

\pygdm\ was entirely rewritten in pure python. In this process, the definitions of the Green's dyads used to describe the environment was implemented in a new class. 
This allows to flexibly use different sets of field susceptibilities. So far this comprises the original set of tensors for 3D simulations in an environment with up to 3 layers in the quasistatic near-field approximation.
New in \pygdm\ is now a class providing field susceptibilities for 2D simulations, hence with one axis being infinitely long. 
In this case the nano-structure is discretized on a 2D square mesh using ``line-dipoles'' and light-emission is described by cylindrical waves. The according vacuum Green's dyads can be derived by an integration of the 3D Green's dyad along the infinite axis (for details see \ref{sec:2DGreensDyads}).
Like the 3D dyads, we provide a quasi-static approximation for a layered environment with up to three layers based on the  method of mirror dipoles. Consequently the same limitations hold as in the case of the quasistatic 3D tensors.

In \pyGDM\ the usage of 2D tensors is very simple and is basically nothing more than replacing the 3D \textit{dyads} class by the equivalent class for 2D. 
An example of the simulation of silicon nanowires of different shapes is depicted in figure~\ref{fig:2D_sim_SiNWs}. 
The simulations show silicon nanowires (SiNWs) of circular (\ref{fig:2D_sim_SiNWs}a), square (\ref{fig:2D_sim_SiNWs}b) and triangular (\ref{fig:2D_sim_SiNWs}c) cross section. 
The SiNWs are illuminated from above by a plane wave of wavelength $\lambda_0=550\,$nm, linearly polarized along the NW axis (transverse magnetic, TM). 
Fig.~\ref{fig:2D_sim_SiNWs}d-f show the electric (top panels) and magnetic (bottom panels) field intensity distributions inside the nanowire. The nanowire dimensions are chosen such that the magnetic dipole mode is excited.
We note that the results agree qualitatively and quantitatively with FDTD simulations \cite{wiechaStronglyDirectionalScattering2017}.
We emphasize that the full 3D optical response of infinitely long nanowires is completely described considering 2D simulations, which is of strong interest for investigating \textit{e.g.} propagating modes \cite{colasdesfrancsIntegratedPlasmonicWaveguides2009, barthesPurcellFactorPointlike2011}.

The \pygdm-script to reproduce the results shown in figure~\ref{fig:2D_sim_SiNWs} can be found \href{https://wiechapeter.gitlab.io/pyGDM2-doc/examples/example2D_SiNW_shapes.html}{in the online documentation under this link}.

\subsection{EELS / CL}

\begin{figure}[t!]
	\includegraphics[width=.95\columnwidth]{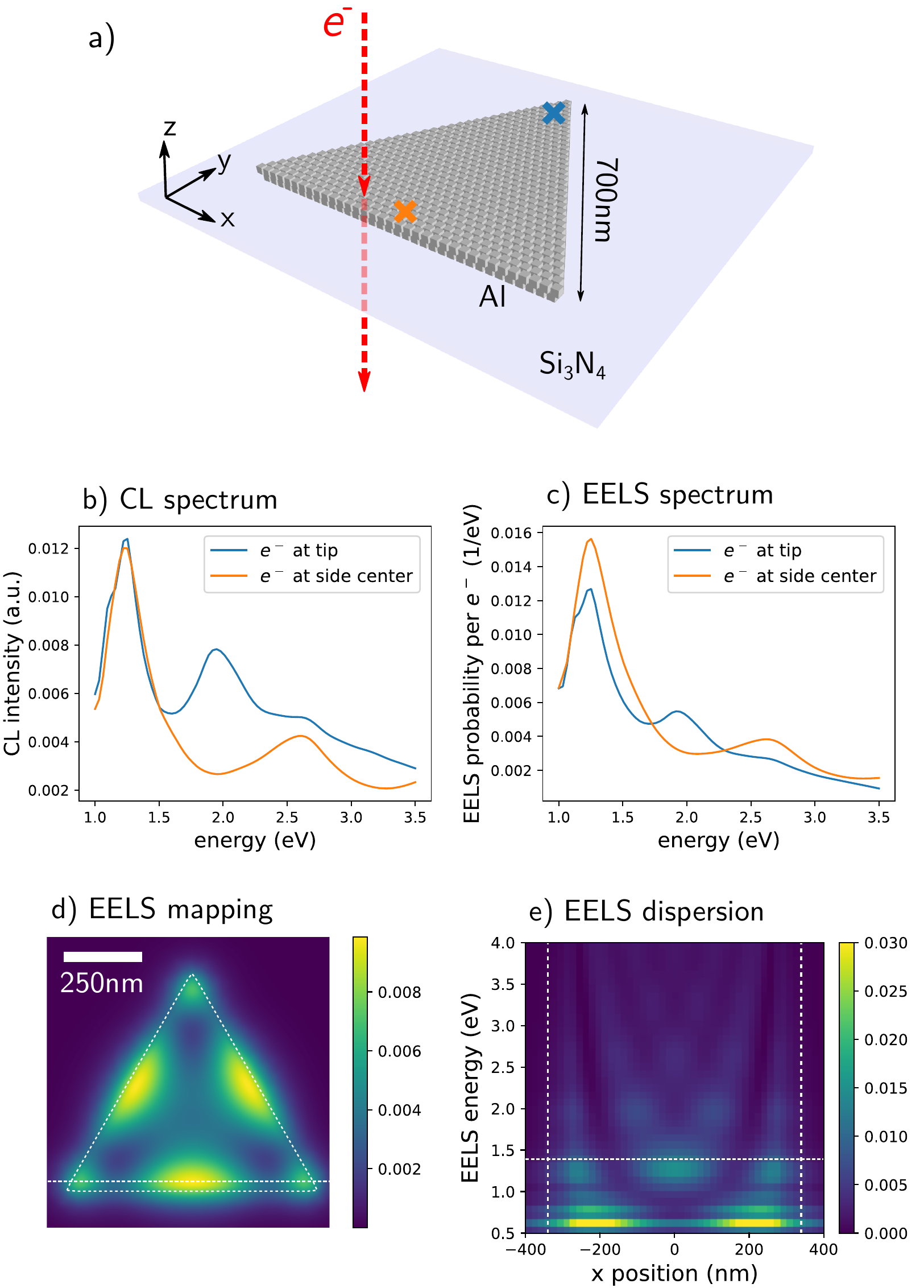}
	\centering
	\caption{
		Example of electron energy loss spectroscopy (EELS) and cathodoluminescence (CL) simulations, reproducing results from Ref.~\cite{camposPlasmonicBreathingEdge2017}.
		A regular aluminum prism of $700\,$nm side length lies on an Si$_3$N$_4$ membrane ($n_1=2$). 
		A normally incident, focused beam of $100$\,keV electrons is raster-scanned along the sample. 
		From the interaction with the plasmonic structure we calculate (b) the CL spectrum at the tip (blue) respectively side center (orange), (c) the EELS spectrum  (again at the tip and side center), (d) the spatially resolved EELS probability for an energy loss of $1.45$\,eV (indicated by a horizontal white dotted line in (e)), as well as (e) the spatial dispersion of the EELS signal for various energies along the bottom side (indicated by a horizontal white dotted line in (d)).
		The colorbar scales in (d-e) represent the energy loss probability per electron and per eV.
	}\label{fig:eels_cl}
\end{figure}

Another new feature is the support of fast-electron illumination and the implementation of corresponding evaluation functions throught the new \texttt{electron} submodule. 
These new functionalities are based on prior theoretical work \cite{arbouetElectronEnergyLosses2014}. 
They allow to simulate electron energy loss spectroscopy (EELS) and cathodoluminescence (CL) experiments.
A recapitulation is given in \ref{sec:fast_electrons_formalism}, the formalism to calculate the energy loss probability is derived in \ref{sec:fast_electrons_formalism}.1.

Figure~\ref{fig:eels_cl} shows an example in which we reproduce results from Ref.~\cite{camposPlasmonicBreathingEdge2017}.
A flat (thickness of $40$\,nm), regular prism made of aluminum with a side length of $700\,$nm lies on a Si$_3$N$_4$ membrane ($n_1=2$), typically used in electron microscopy. It is surrounded by vacuum ($n_2=1$).
A normally incident beam of fast electrons (of $100$\,keV energy) passes through the sample, as illustrated in figure~\ref{fig:eels_cl}a by a red dashed line. 
In a first simulation we focus the electron beam on two positions: either on the tip (blue marker) or on the center of one side (orange marker). 
At those positions we calculate the cathodoluminescence (Fig.~\ref{fig:eels_cl}b) and EELS spectra (Fig.~\ref{fig:eels_cl}c).
In a second step we fix the energy loss to $1.45$\,eV and perform a raster-scan simulation, in which the electron beam scans across the entire structure, as shown in figure~\ref{fig:eels_cl}d. 
We note that when the beam is moving over the structure, it is important to keep a constant distance between the electron beam and the meshcells' centers, in order to obtain a smooth mapping. To this end we implemented a tool \texttt{tools.adapt\_map\_to\_structure\_mesh}, which re-maps positions close to a meshcell exactly onto the respective closest discretization dipole.

Finally, we combine both types of simulations. With the fast electron beam we perform a line-scan along one side of the Al triangle. 
At each position we calculate an EELS spectrum, which allows to visualize the spatial dispersion of the plasmonic edge modes in the prism (Fig.~\ref{fig:eels_cl}e).

\pygdm-scripts to reproduce the results shown in figure~\ref{fig:eels_cl} can be found in the online documentation under \href{https://wiechapeter.gitlab.io/pyGDM2-doc/examples/exampleFastElec_fast_electrons_ex1.html}{link to example 1}, \href{https://wiechapeter.gitlab.io/pyGDM2-doc/examples/exampleFastElec_fast_electrons_ex2.html}{link to example 2} and \href{https://wiechapeter.gitlab.io/pyGDM2-doc/examples/exampleFastElec_fast_electrons_ex3.html}{link to example 3}.

\subsection{Optical chirality}

\begin{figure*}[h!]
	\includegraphics[width=0.9\textwidth]{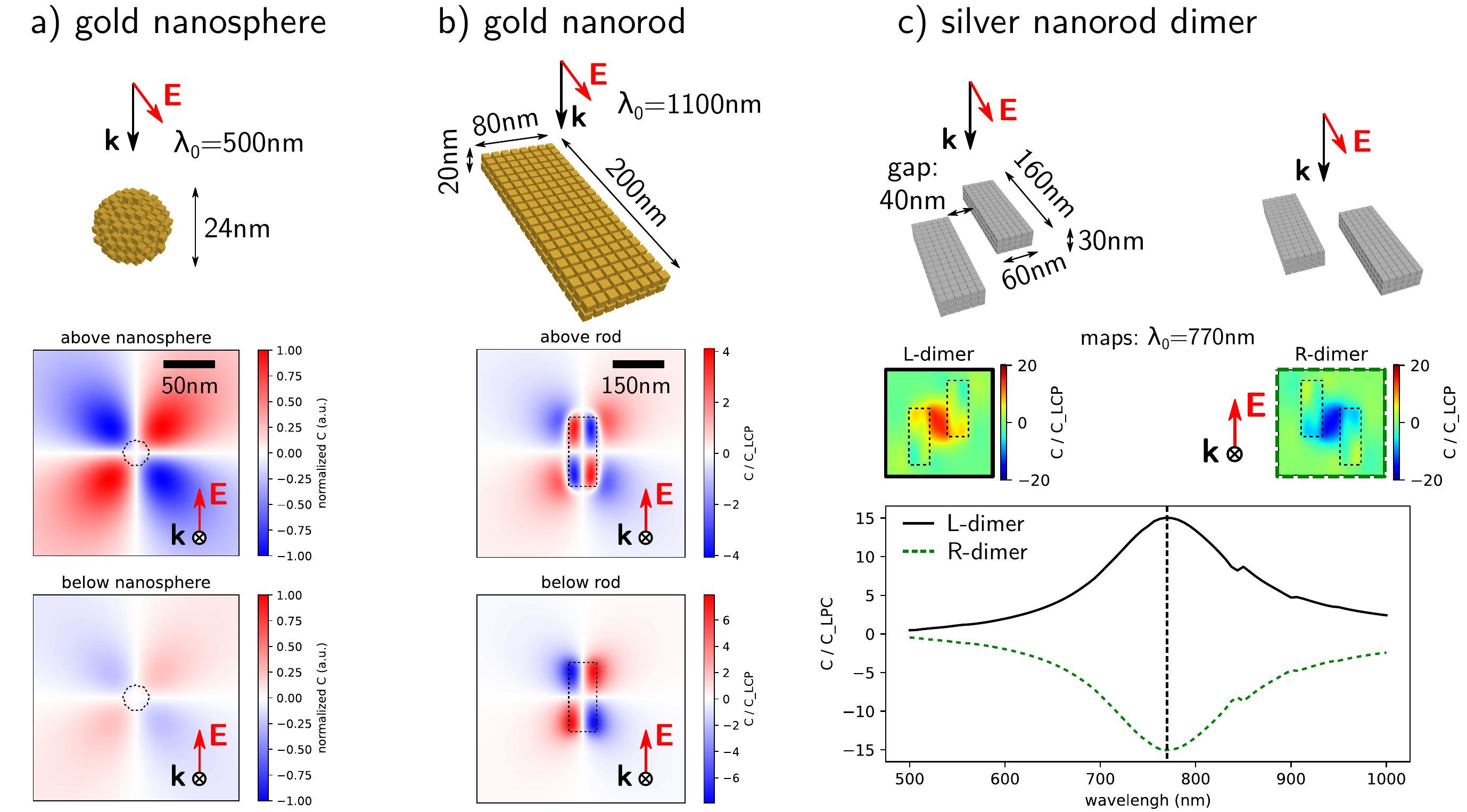}
	\centering
	\caption{
		Examples of calculations of the chirality of the optical near-field. 
		(a-b) Reproduction of results from \cite{schaferlingFormationChiralFields2012}.
		(a) Plane wave illuminated gold nano-sphere as a model for a point dipole response, resulting in a symmetric pattern of the optical chirality close to the structure (here calculated at $\pm 25$\,nm above and below the nanosphere). The illumination wavelength (here $\lambda_0=500\,$nm) does not have a significant impact on the chirality pattern.
		(b) Gold nanorod illuminated by a plane wave at the wavelength of the dipolar plasmon resonance ($\lambda_0=1100\,$nm), linearly polarized along the rod's long axis.
		The chirality is calculated at $\pm 25$\,nm above and below the surface of the nanorod.
		(c) Reproduction of results from \cite{meinzerProbingChiralNature2013}.
		Chirality spectra calculated at a distance of $15\,$nm above the top surface in the center between the two silver nanorods.
		The colormaps show spatial mappings of the chirality in the L-dimer and R-dimer case at the wavelength of maximum chirality ($\lambda_0=770$\,nm, indicated by a vertical dashed line in the bottom plot).
		For simplicity all structures in (a-c) are placed in vacuum, yet the qualitative trends are very well reproduced, in comparison with the original publications \cite{schaferlingFormationChiralFields2012, meinzerProbingChiralNature2013}.
	}\label{fig:chirality}
\end{figure*}

Another new function in \pyGDM\ is the calculation of optical near-field chirality. 
The optical chirality is a scalar quantity which can be interpreted as a measure of the asymmetry in the coupling of left and right handed chiral emitters with the available photonic modes \cite{meinzerProbingChiralNature2013, gorodetskiGeneratingFarFieldOrbital2013}.
It is defined as \cite{tangOpticalChiralityIts2010}
\begin{equation}
	\frac{C}{C_{\text{LCP}}} = - \text{Im}\Big( \mathbf{E}^* \cdot \mathbf{H} \Big) \, ,
\end{equation}
where the superscript asterisk stands for complex conjugation.
\pyGDM\ follows the convention to normalize the chirality $C$ to the value of a left circularly polarized plane wave ($C_{\text{LCP}}$). 
Consequently, left circularly polarized (LCP), respectively right circularly polarized (RCP) plane waves have a chirality of $C_{\text{LCP}} = +1$, respectively $C_{\text{RCP}} = -1$. 
Nanostructures inducing an optical chirality larger than that of a circularly polarized plane wave ($|C|>1$) are often referred to as \textit{super-chiral}.

To demonstrate the new function, we reproduce in figure~\ref{fig:chirality} selected examples of chiral near-fields from literature.
Figure~\ref{fig:chirality}a reproduces the chirality around a point dipole excited by a linearly polarized plane wave. This is illustrated by a small gold nanosphere in vacuum. The calculated optical chirality is in very good agreement  with literature \cite{schaferlingFormationChiralFields2012}.
Figure~\ref{fig:chirality}b illustrates the chirality above and below a gold nanorod, illuminated by a linearly polarized plane wave at the fundamental localized plasmon resonance, again reproducing results from literature with high fidelity \cite{schaferlingFormationChiralFields2012}.
In a third example shown in Fig.~\ref{fig:chirality}c, we calculate the chirality at the center of a silver nanorod dimer, which is either in a left-handed (``L-dimer'') or right-handed (``R-dimer'') configuration. 
The bottom plot shows the spectrum of the respective chiralities (black line: L-dimer, green dashed line: R-dimer). In the insets above the spectra, we show the spatial mapping of $C$ calculated $15$\,nm above the nanorod's top surface. 
The L-handed (left plot) and R-handed (right plot) dimers are illuminated at the wavelength $\lambda_0 = 770\,$nm, leading to a peak of the chirality. 
The results agree well with literature \cite{meinzerProbingChiralNature2013}.

\pygdm-scripts to reproduce the results shown in figure~\ref{fig:chirality} can be found in the online documentation under \href{https://wiechapeter.gitlab.io/pyGDM2-doc/examples/example10a_optical_chirality_1.html}{link to example 1}, \href{https://wiechapeter.gitlab.io/pyGDM2-doc/examples/example10b_optical_chirality_2.html}{link to example 2} and \href{https://wiechapeter.gitlab.io/pyGDM2-doc/examples/example10c_optical_chirality_3.html}{link to example 3}.

\subsection{Field gradients and optical forces}

\begin{figure}[t!]
	\includegraphics[width=0.9\columnwidth]{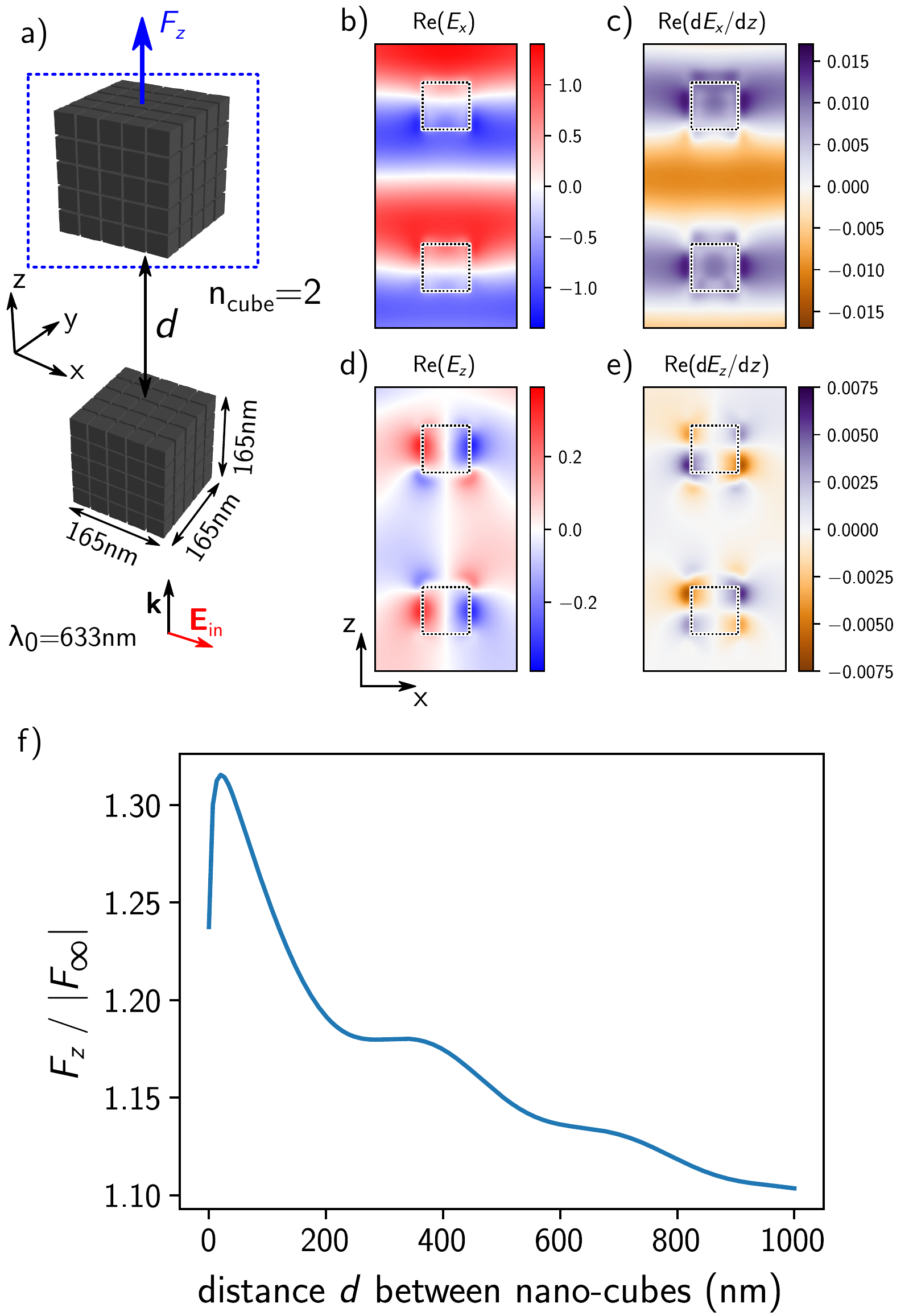}
	\centering
	\caption{
		Calculation of field gradients to deduce optical forces.
		(a) illustration of the problem. Two dielectric nano-cubes ($n_{\text{cube}}=2$) in vacuum are aligned on the $z$-axis with an inter-cube distance $d$. The cubes are illuminated from below with a plane wave of wavelength $\lambda_0=633\,$nm, with linear polarization along $X$. We seek to calculate the optical force $F_z$  acting in $z$-direction on the upper cube.
		(b-e) real part of the optical fields $E_x$ (b) and $E_z$ (d) as well as their spatial derivatives in $z$-direction $\partial E_x / \partial z$ (c) and $\partial E_z / \partial z$ (e). Shown areas are $500\times 1000$\,nm$^2$ in the $y=0$ plane, cutting through the center of the dielectric cubes. The nano-cube outlines are indicated by white/black dotted lines.
		All fieldmaps are in units of the incident field amplitude $E_0$.
		(f) optical force on the upper nano-cube as function of the distance $d$ between the two cubes, normalized to the force at large distance.
	}\label{fig:gradients_forces}
\end{figure}

\begin{figure}[t!]
	\includegraphics[width=0.9\columnwidth]{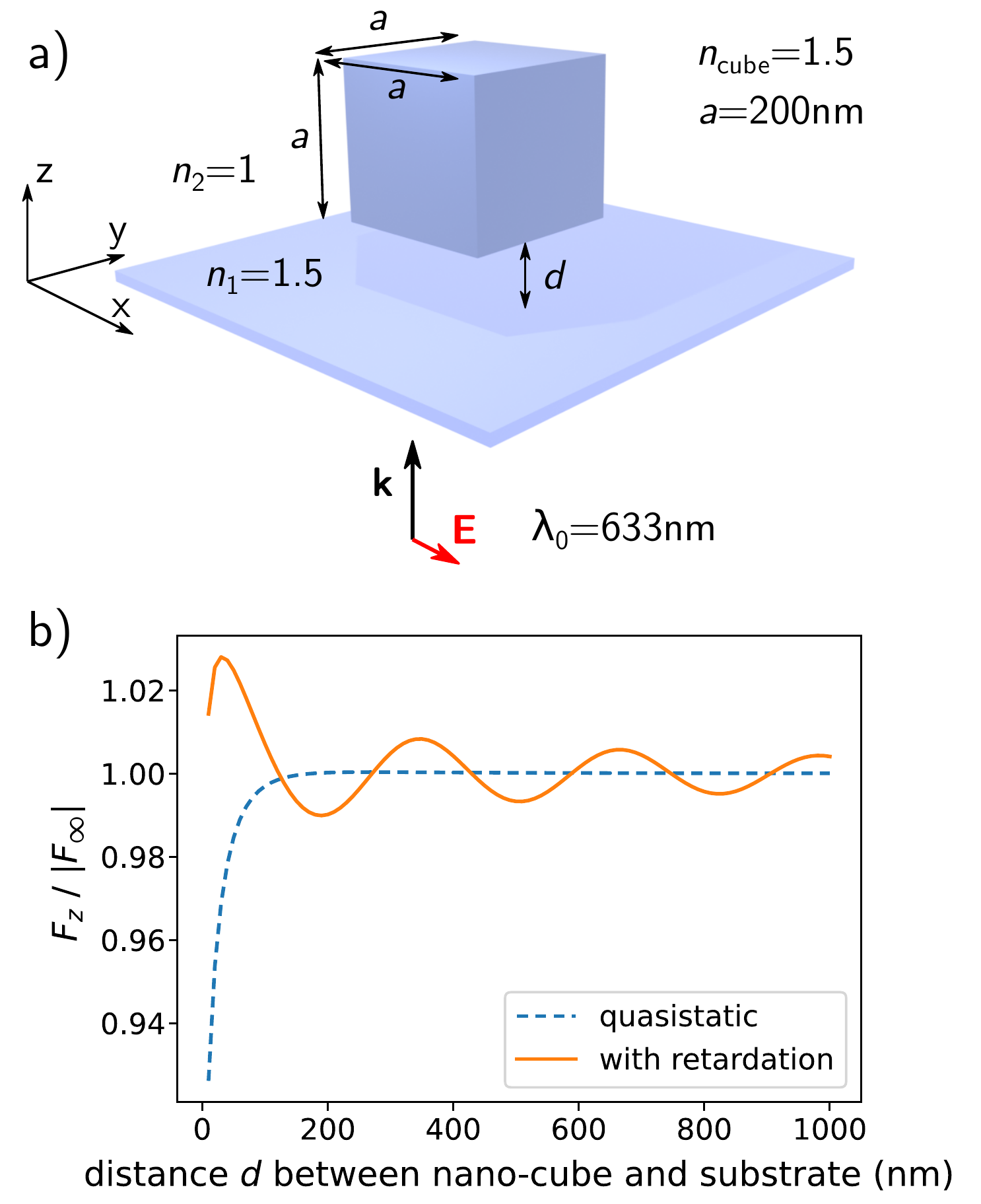}
	\centering
	\caption{
		Calculation of the optical force of a dielectric cube above a substrate.
		(a) illustration of the configuration. A dielectric nano-cube (side length $a=200\,$nm, $n_{\text{cube}}=1.5$) is placed in vacuum ($n_2=1$) at a small distance $d$ above a dielectric substrate of same index ($n_1=1.5$). 
		The system is illuminated from below with a plane wave of linear polarization along $X$ and vacuum wavelength $\lambda_0=633\,$nm. We calculate the optical force $F_z$ on the cube.
		(b) optical force $F_z$ on the nano-cube as function of the distance $d$ to the dielectric interface, normalized to the force at large distance to the surface. 
		The optical force is calculated either using the quasistatic mirror-charge approximation for the interface (blue dashed line) or with a Green's tensor for the surface including retardation (orange solid line).
		The results are in good agreement with similar simulations for a nano-sphere \cite{chaumetCoupledDipoleMethod2000}.
	}\label{fig:forces_substrate}
\end{figure}

A further new feature in \pyGDM\ is a function to calculate the gradients of near-fields in- and outside the nanostructure. 
These gradients are required for instance to calculate the charge distribution inside plasmonic nanostructures \cite{martyChargeDistributionInduced2010} or to compute optical forces \cite{girardTheoreticalAnalysisLightinductive1994, chaumetCoupledDipoleMethod2000}.
In \pygdm, we implement a simple numerical center differentiation, which approximates the derivative with second order accuracy \cite{pressNumericalRecipes3rd2007}:
\begin{equation}
	\frac{\partial f(x)}{\partial x} \approx \frac{f(x+\Delta) - f(x-\Delta)}{2 \Delta}
\end{equation}
By default, the discretization stepsize is used as value for $\Delta$, so inside a structure the fields are taken at each meshpoint. 
Outside the structure the field can be evaluated at arbitrary positions around the evaluation location. Here the user may optionally choose a value smaller than the stepsize for $\Delta$.
Note that the gradients inside the structure can be obtained also using a self-consistent formulation, which would however require a coupled system of size $12 N \times 12 N$, involving the second derivative of the field susceptibilities ($3\times 3\times 3\times 3$ tensors) \cite{chaumetCoupledDipoleMethod2000}.

Being able to compute the field gradients at the position of a dipole, the optical force acting on this dipole of moment $\mathbf{p}(\omega)$ can be obtained through \cite{chaumetCoupledDipoleMethod2000}
\begin{equation}
	F_i = \frac{1}{2}\, \text{Re} \left( \sum\limits_{j=1}^{3} p_j(\omega) \frac{\partial E^*_j(\omega)}{\partial e_i}  \right) \, ,
\end{equation}
where the superscript asterisk $^*$ indicates complex conjugation, $i,j \in \{1,2,3\}$ stand for the Cartesian directions ($x,y$ and $z$), and $e_i$ corresponds to the Cartesian direction of the force vector, along which we also differentiate the electric field at the position of the dipole.
The force acting on a larger nanostructure can be obtained by summation of the forces on each individual mesh cell.
Note that the force on a nanoparticle can also be obtained through the Maxwell stress tensor by integration of the electromagnetic fields on a surface enclosing the particle \cite{chaumetCoupledDipoleMethod2000}. However, in the GDM the gradients method can take advantage of the pre-calculated internal fields and is therefore usually faster.

In figure~\ref{fig:gradients_forces}a we illustrate a system of two dielectric nano-cubes of refractive index $n=2$, $165\,$nm side length and variable distance $d$ between each other. 
The cubes are placed in vacuum and illuminated by a plane wave from below, polarized along $x$.
Due to symmetry the $E_y$ component is always zero. In figure~\ref{fig:gradients_forces}b and \ref{fig:gradients_forces}d we show the real part of the non-zero field components $E_x$, respectively $E_z$. Fig.~\ref{fig:gradients_forces}c and \ref{fig:gradients_forces}e show the according spatial derivative along $z$, required to calculate the $z$ component of the optical force.
Figure~\ref{fig:gradients_forces}f shows the force $F_z$ on the upper of the two dielectric cubes.

A second example is illustrated in figure~\ref{fig:forces_substrate}a. 
A dielectric cube of index $n_{\text{cube}}=1.5$ and side length $a=200\,$nm is kept at short distance $d$ above a dielectric substrate of same refractive index ($n_1=1.5$) and illuminated from below with a plane wave ($\lambda_0=633\,$nm, polarization along $X$).
In Fig.~\ref{fig:forces_substrate}b the force along $z$ is shown for increasing distance $d$, either calculated with the mirror-charge approximation (blue dashed line) or using the retarded propagator for the dielectric interface via the \textsf{pyGDM2\_retard} package. 
The results are in excellent agreement with a similar configuration from literature, using a sphere instead of a cube \cite{chaumetCoupledDipoleMethod2000}.
The force between the two cubes (Fig.~\ref{fig:gradients_forces}) behaves actually very similarly to the optical force acting on the nano-cube close to a substrate (Fig.~\ref{fig:forces_substrate}), with the difference of an additional decreasing component in the former case. 
This can be explained by the small size of the second nano-cube in comparison with the infinite extension of the surface.

We note, that the gradients inside plasmonic nanostructures can also be used to calculate the charge distribution \cite{martyChargeDistributionInduced2010}.
Due to the very short skin depth in metals, the field gradients can be very strong which poses difficulties together with the usually relatively coarse volume discretization in the GDM. While this often does not affect the accuracy of results that average the internal fields (like the optical forces, far-fields or also near-fields \textit{outside} the structure), the internal fields and their gradients can appear noisy which may pose problems for computations of quantities \textit{inside} the nanostructure on a small scale, such as the charge density, in which cases fine meshings need to be used.

The \pygdm-script to reproduce the results shown in figure~\ref{fig:gradients_forces} can be found \href{https://wiechapeter.gitlab.io/pyGDM2-doc/examples/example11_optical_forces.html}{in the online documentation under this link}. 
For a script to reproduce figure~\ref{fig:forces_substrate} see \href{https://wiechapeter.gitlab.io/pyGDM2-doc/examples/exampleRetard_optical_forces.html}{this link}.

\subsection{Scattering into a substrate}

\begin{figure}[t!]
	\includegraphics[width=\columnwidth]{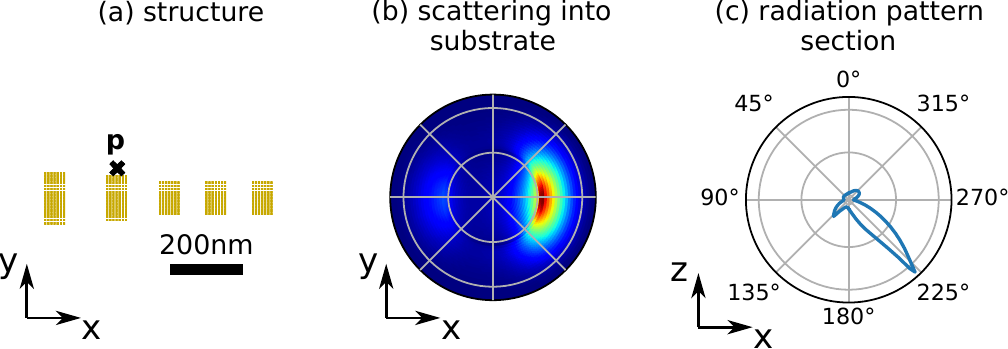}
	\centering
	\caption{
		Radiation pattern of a Yagi Uda antenna in vacuum ($n_2=1$) lying on a substrate of refractive index $n_1=1.5$. 
		(a) Top view of the structure. 
		The structure consists in five nanorods of radius $R=30$~nm, the lengths are respectively $L_1=170$~nm for the reflector rod (left most), $L_2=145$~nm for the feed rod (closest to the electric dipole) and $L_3=115$~nm for the guide rods (three others).
		The dipole is oriented along the Oy axis and emits light at $\lambda_0=820$~nm.
		(b) Top view of the radiation pattern intensity distribution in the substrate. 
		(c) Angular radiation pattern in the plane XZ, the maximum of emission is centered around the critical angle $\theta_{\text{c}}=222\degree$.
	}\label{fig:yagi_uda_eo}
\end{figure}

By calculating the scattered intensity in the far-field region on a spherical surface around the nanostructure, the radiation pattern of a nanostructure can be deduced.
In the new version of \pyGDM\ the asymptotic far-field propagators are fully implemented for a single interface (see \ref{sec:weyl_representation}), hence far-field patterns of scattering into a substrate can now be calculated.

In Figure \ref{fig:yagi_uda_eo}, we present a simulation of the radiation pattern of a gold Yagi-Uda antenna in vacuum ($n_2=1$) laying on a silica substrate ($n_1=1.5$), similar to the configuration described in the reference \cite{curtoUnidirectionalEmissionQuantum2010}.
It consists of 5 nanorods of radius $R=30$\,nm. 
From left to right, we have the reflector rod of length $L_1=170$\,nm,
the feed rod of length $L_2=145$\,nm and the last three, of length $L_3=115$\,nm, are parasitic elements serving as directors (see Fig.~\ref{fig:yagi_uda_eo}a).
An electric dipole $\mathbf{p}$, of emission wavelength $\lambda_0=820$\,nm, is positioned in the vicinity of the feed.
The rods are discretized on cubic mesh with a discretization step of $10\,$nm.

We obtain results in agreement with literature \cite{curtoUnidirectionalEmissionQuantum2010, taminiauEnhancedDirectionalExcitation2008, wiechaDesignPlasmonicDirectional2019}, as we find that the emission of the emitter-antenna system points unidirectionally towards the positive $X$ axis (see Fig.~\ref{fig:yagi_uda_eo}b). 
Additionally, the change in refractive index between the vacuum and the substrate bends the direction of radiation to the critical angle at $\theta_{\text{c}}=\text{arcsin}(n_1/n_2)=222\degree$ (Fig.~\ref{fig:yagi_uda_eo}c).
The nano Yagi-Uda antenna makes it possible to emit predominantly in the direction of the positive $X$ axis.
The dipole emitter couples to the feed rod, which strongly enhances the signal due to the Purcell effect. 
Then, as a result of the specifically tailored inter-rods distances, the other elements impose the directivity of emission due to constructive interference only towards the positive $X$ direction.

The \pygdm-script to reproduce the results shown in figure~\ref{fig:yagi_uda_eo} can be found \href{https://wiechapeter.gitlab.io/pyGDM2-doc/examples/example13_plasmonic_YagiUda_directional_antenna.html}{in the online documentation under this link}.

Please note that \pyGDM\ includes a module for evolutionary optimization of photonic nanostructures. Using this functionality we have demonstrated inverse design of directional nano-antennas \cite{wiechaDesignPlasmonicDirectional2019}.
The analysis of the near-field amplitude and phase clearly reveals the role of the antenna elements even for the complex optimized nanostructure that we obtained.


\subsection{Quantum emitter decay rate inside nanostructures}

\begin{figure}[t!]
	\includegraphics[width=\columnwidth]{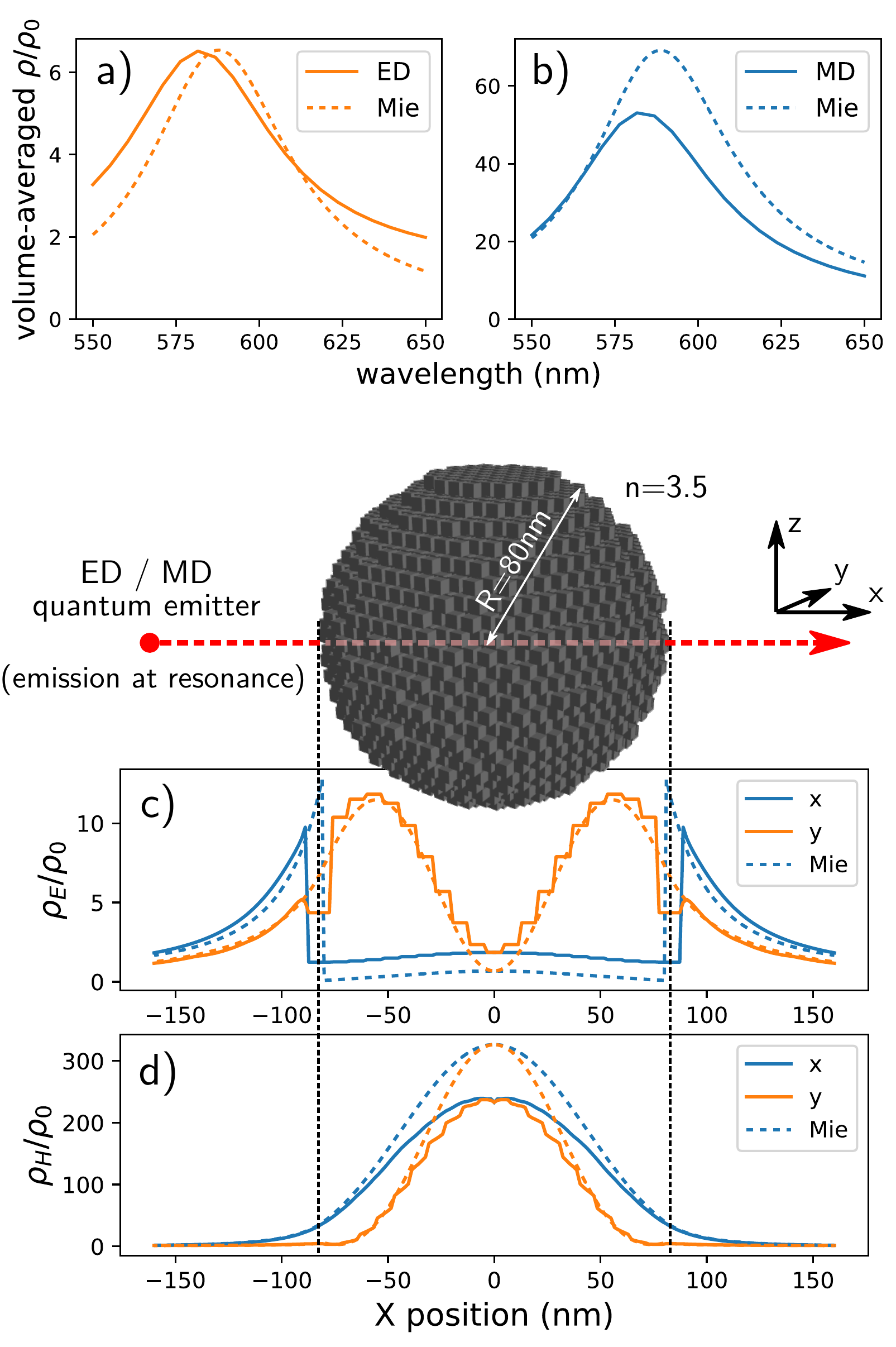}
	\centering
	\caption{
		Decay rate inside a dielectric sphere (refractive index $n=3.5$) of radius $R=80\,$nm, placed in vacuum.
		Spectrally resolved decay rate $\rho$, averaged over the entire volume of the sphere for (a) electric dipole (ED) transitions and (b) magnetic dipole (MD) transitions. Solid lines correspond to the GDM simulation, dashed lines are values obtained by Mie theory.
		(c-d) show the decay rate modification along a line parallel to the $X$ axis through the center of the sphere. Blue lines indicate an $x$-oriented, orange lines a $y$-oriented dipole transition, values from Mie theory are again shown by dashed lines.
		(c) electric, and (d) magnetic dipole transition, calculated at the respective maximum decay rate enhancement.
	}\label{fig:decay_Mie}
\end{figure}

Like the rest of the code, in the new version of \pyGDM\ the \texttt{core.decay\_rate} function has been entirely rewritten in pure python. 
A first noteworthy improvement is a significant optimization of the algorithm, in which now the required field susceptibilties are pre-calculated, which hugely accelerates the numerical evaluation of the volume integral.
Furthermore, like all other \pygdm\ functions, the decay-rate calculation is now capable to handle multi-material structures and materials with non-isotropic permittivity.
Finally, it is now possible to calculate both, the electric and magnetic decay rates also \textit{inside} nanostructures.
\pygdm\ uses a formalism based on field-susceptibilities for the derivation of the photonic local density of states (LDOS), as detailed in references~\cite{wiechaPyGDMPythonToolkit2018, wiechaDecayRateMagnetic2018, majorelQuantumTheoryNearfield2020, carminatiElectromagneticDensityStates2015}.

We demonstrate the possibility to calculate magnetic decay rates inside nanostructures by a comparison to Mie theory \cite{kimClassicalDecayRates1988, schmidtDielectricAntennasSuitable2012}.
We calculate the normalized electric and magnetic LDOS $\rho/\rho_0$ averaged over the particle volume of a dielectric sphere of refractive index $n=3.5$ and radius $R=80\,$nm in air. This models the decay rate of a dielectric particle doped with electric or magnetic emitters.

We note that the sphere is discretized by around 7000 meshpoints, which would have resulted in many days of simulation time with the former code for numerical integration. It becomes feasible with the optimized implementation, which runs fully parallelized and for the described problem takes in the order of 15 minutes per wavelength running on 8 CPU cores (3rd generation AMD Ryzen).
The corresponding spectra are shown in figure~\ref{fig:decay_Mie}a, respectively~\ref{fig:decay_Mie}b.
Despite a slight resonance shift the agreement is very good for the electric LDOS, and of the correct order of magnitude for the magnetic case.
To investigate the origin of the discrepancy in the magnetic case, we calculate the E- and H-LDOS along a line parallel to $OX$ through the center of the nanosphere (figures~\ref{fig:decay_Mie}c-d).
The \pyGDM-simulated electric LDOS reproduces well Mie theory (Fig.~\ref{fig:decay_Mie}c), but more than 20\% error is found for the magnetic contribution (Fig.~\ref{fig:decay_Mie}d). 
On the one hand we attribute this discrepancy to the non-perfect spherical shape of the particle that is meshed using cells with a stepsize of $8\,$nm, because the resonance position strongly depends on the shape. 
On the other hand, the magnetic LDOS is deduced by repropagating the electric Green's tensor \cite{wiechaDecayRateMagnetic2018}, which can explain the larger error in the interior of the structure. 
In conclusion, the LDOS can be calculated with a good qualitative accuracy.
However, in demanding situations like when high refractive indices are used or for round geometries like spheres, which are a challenging geometry for the GDM's volume discretization, the associated limitations of \pyGDM\ should be kept in mind.

The \pygdm-script to reproduce the results shown in figure~\ref{fig:decay_Mie} can be found \href{https://wiechapeter.gitlab.io/pyGDM2-doc/examples/example08b_decay_rate_2_Mie.html}{in the online documentation under this link}.

\subsection{Plasmonic properties of doped dielectric nanostructures}

\begin{figure}[t!]
	\includegraphics[width=\columnwidth]{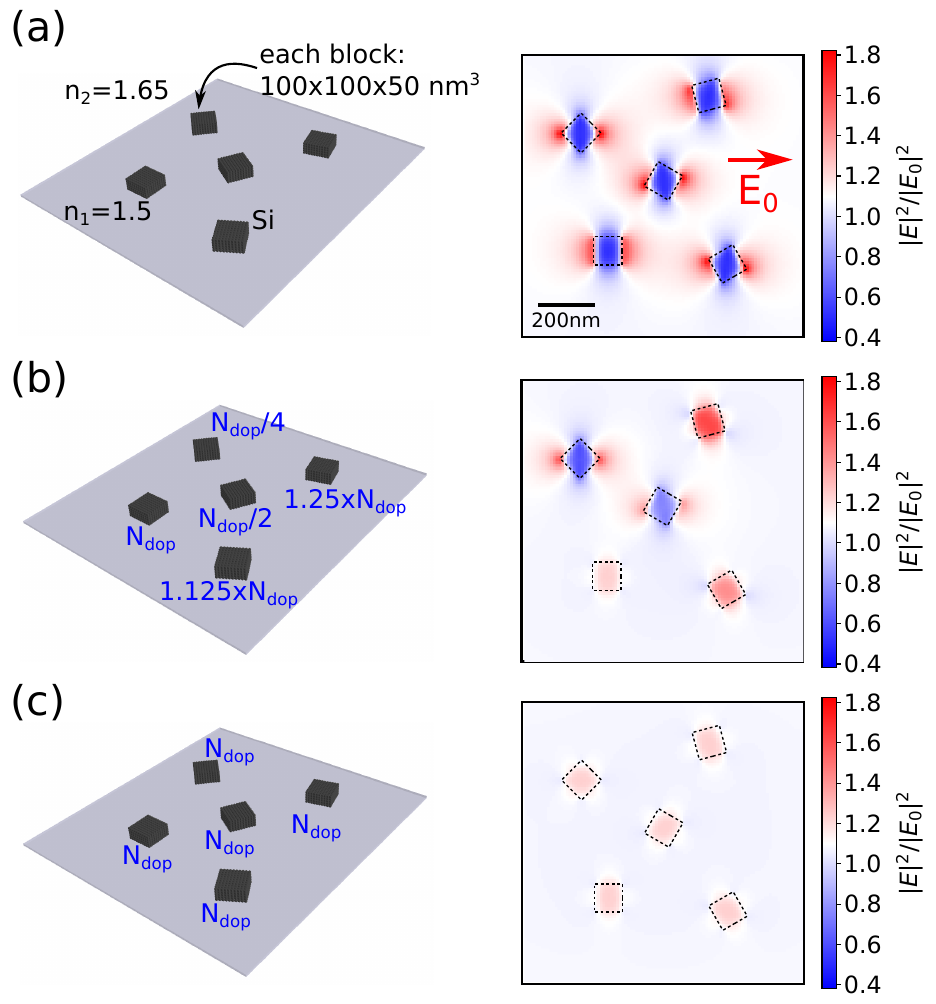}
	\centering
	\caption{
		Evolution of the near-field pattern when progressively doping five silicon pads until a carrier density of $N_{\text{dop}}=5\times 10^{20}$\,cm$^{-3}$ is reached (window size: $1000\times1000$\,nm$^2$).
		(a) Geometry consisting of five undoped silicon blocks of size $100\times100\times50$\,nm$^3$ in an environment $n_2=1.65$ and lying on a substrate $n_1=1.5$, illuminated by a descending plane wave.
		The electric near-field intensity map computed $10$\,nm above the top surface of the pads at $\lambda_0=2754$\,nm is shown on the right.
		(b) same simulation as (a), but with various different doping concentrations. Each pad's dopant concentration relative to $N_{\text{dop}}$ is indicated by a blue label.
		(c) same as (a), but each silicon pad is doped with the concentration $N_{\text{dop}}$, leading to minimum near-field contrast.
	}\label{fig:NF_doped_pads}
\end{figure}

\begin{figure*}[t!]
	\includegraphics[width=.9\linewidth]{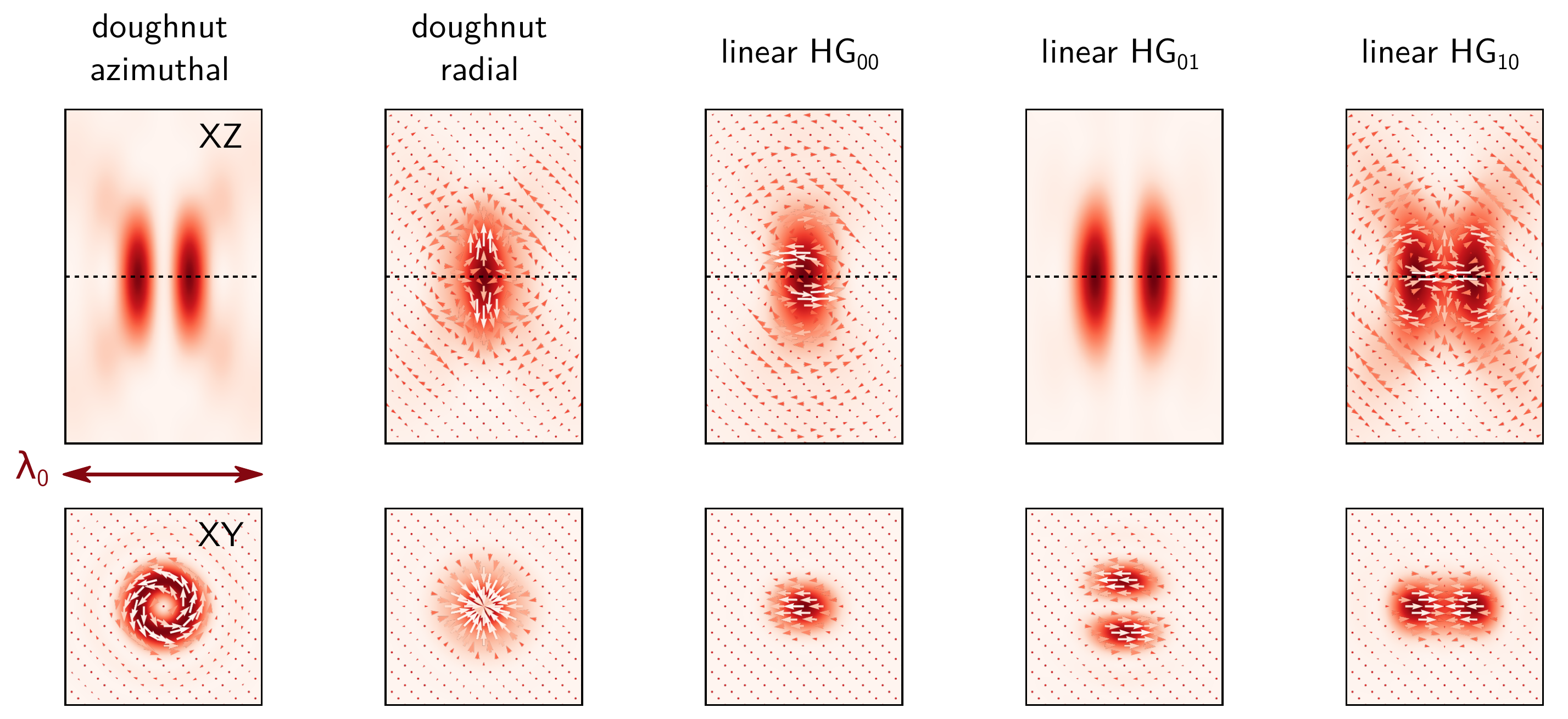}
	\centering
	\caption{
		Electric field intensity plots of the new added focused vectorbeams. 
		From left to right: Doughnut modes (azimuthal and radial polarizations) and Hermite-Gauss modes of the first orders, all calculated for focus by a numerical aperture $\text{NA}=1.4$ and incidence from positive towards negative $z$. 
		Top panels show the plane $y=0$, bottom plots show the focal plane at $z=0$, which is indicated by a dotted black line in the top panels. 
		The real part of the electric field vectors is indicated by small arrows.
	}\label{fig:vectorbeams_fieldmaps}
\end{figure*}

Doped semi-conductors nanostructure has open the way towards tunable plasmonic \cite{lutherLocalizedSurfacePlasmon2011}.
Doping consists of introducing impurities into a pure material in order to modify its electronic and optical properties.
The impurities provide free carriers which form a free electron cloud, that enables a plasmonic response of the material.
A new class of such doped dielectric materials has been added to \pyGDM , which adds a supplemental Drude term to the permittivity of the undoped material, based on a Drude-Lorentz model \cite{majorelTheoryPlasmonicProperties2019}
\begin{equation}\label{eq:constante_dielec_Drude}
	\epsilon(\omega) = \epsilon_{\text{int}}(\omega)-\frac{\omega_{\text{p}}^2}{\omega^2+i\gamma\omega}\,,
\end{equation}
where $\epsilon_{\text{int}}(\omega)$ is the dielectric constant of the intrinsic dielectric, $\gamma$ is the damping rate due to the collisions, and the plasma frequency $\omega_{\text{p}}$ is related to the concentration of free carriers $N_{\text{dop}}$ according to
\begin{equation}\label{eq:wp_metaux}
	\omega_{\text{p}}=\sqrt{\frac{4\pi N_{\text{dop}}e^2}{m^*}}
\, ,
\end{equation}
where $e$ and $m^*$ are respectively the elementary charge and the effective mass of the carriers.

We added three material-classes \texttt{materials.hyperdoped\_xxx} to \pyGDM .
One adds the doping-induced plasmonic response to a tabulated dielectric permittivity which is loaded from an external user-defined text file. 
The other two are specific cases, namely silicon as well as a dielectric material of constant refractive index.

In figure~\ref{fig:NF_doped_pads}a, we show a simulation of the electric near-field intensity maps $10$\,nm above the top surface of an array of five undoped silicon pads. 
The structures are embedded in an environment of refractive index $n_2=1.65$, laying on a substrate with $n_1=1.5$, all illuminated by a descending plane wave of wavelength $\lambda_0=2754$\,nm.
In figure~\ref{fig:NF_doped_pads}b, we observe a strong change in the near-field pattern as we gradually increase the doping of the pads.
The optical contrast decreases as the doping approaches $N_{\text{dop}}=5\times 10^{20}$\,cm$^{-3}$.
When the doping is further increased, the near-field contrast increases again, yet with an opposite sign.
The fact that the nanostructures with a doping of $N_{\text{dop}}=5\times 10^{20}$\,cm$^{-3}$ seem to disappear (Fig.~\ref{fig:NF_doped_pads}c) is due to an index matching of the real part of the refractive index of the object with that of the environment at a specific doping concentration \cite{teulleVisibilityPlasmonicParticles2013}.

The \pygdm-script to reproduce the results shown in figure~\ref{fig:NF_doped_pads} can be found \href{https://wiechapeter.gitlab.io/pyGDM2-doc/examples/example14_doped_dielectrics.html}{in the online documentation under this link}.

\subsection{New illuminations: Vectorbeams}

\begin{figure}[t!]
	\includegraphics[width=.95\columnwidth]{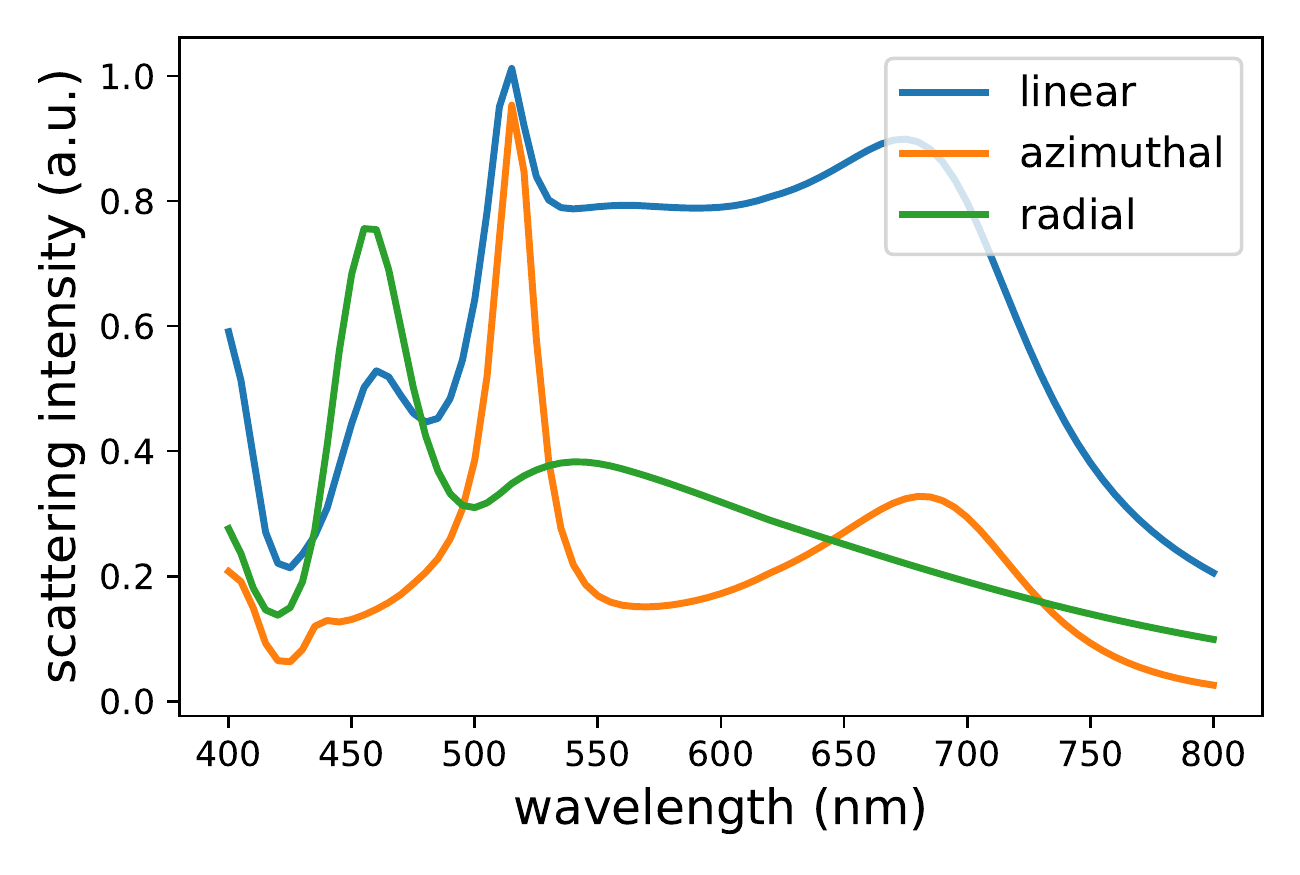}
	\centering
	\caption{
		Scattering from a silicon nanosphere of diameter $D=170$\,nm in oil ($n_{\text{oil}}=1.48$), illuminated by different polarized vectorbeams through a numerical aperture of $\text{NA}=1.2$.
		The blue line corresponds to a linear polarized TEM$_{00}$ beam, the orange line to azimuthal polarization and the green line to radial polarization.
	}\label{fig:vectorbeams_silicon_sphere}
\end{figure}

In the \pygdm\ update we implemented a few additional focused vectorbeams, following chapter 3.6 of the textbook of Novotny and Hecht \cite{novotnyPrinciplesNanooptics2006}. 
We now provide generator functions for doughnut beams with azimuthal and radial polarization as well as linearly polarized Hermite-Gauss beams for the orders HG$_{00}$, HG$_{01}$ and HG$_{10}$. 
Focusing through a numerical aperture of $\text{NA} = 1.4$ is shown in figure~\ref{fig:vectorbeams_fieldmaps}.
We note that \pyGDM\ implements a full tight focus model of these beams, including the correct $E_z$ component. 
So far the vector-beams are limited to homogeneous environments, but we are developing an implementation of taking into account transmission and reflection at an interface, for which an experimental version is already available, which currently undergoes extensive testing.

To demonstrate the focused vectorbeam illuminations, we show in figure~\ref{fig:vectorbeams_silicon_sphere} scattering spectra of a silicon nanosphere of diameter $D=170$\,nm in a homogeneous oil environment ($n_{\text{oil}}=1.48$).
The sphere is illuminated by either a Gaussian beam of HG$_{00}$ mode and linear polarization (blue line), or by a doughnut mode of either azimuthal (orange line) or radial (green line) polarization.
The \pygdm\ simulations reproduce well recently reported experimental results \cite{mannaSelectiveExcitationEnhancement2020}.

A \pygdm-script to reproduce the results shown in figures~\ref{fig:vectorbeams_fieldmaps} and~\ref{fig:vectorbeams_silicon_sphere} can be found \href{https://wiechapeter.gitlab.io/pyGDM2-doc/examples/example15_vectorbeams.html}{in the online documentation under this link}.

\subsection{Tools for the analysis of optical interaction effects}

\begin{figure}[t!]
	\includegraphics[width=0.9\columnwidth]{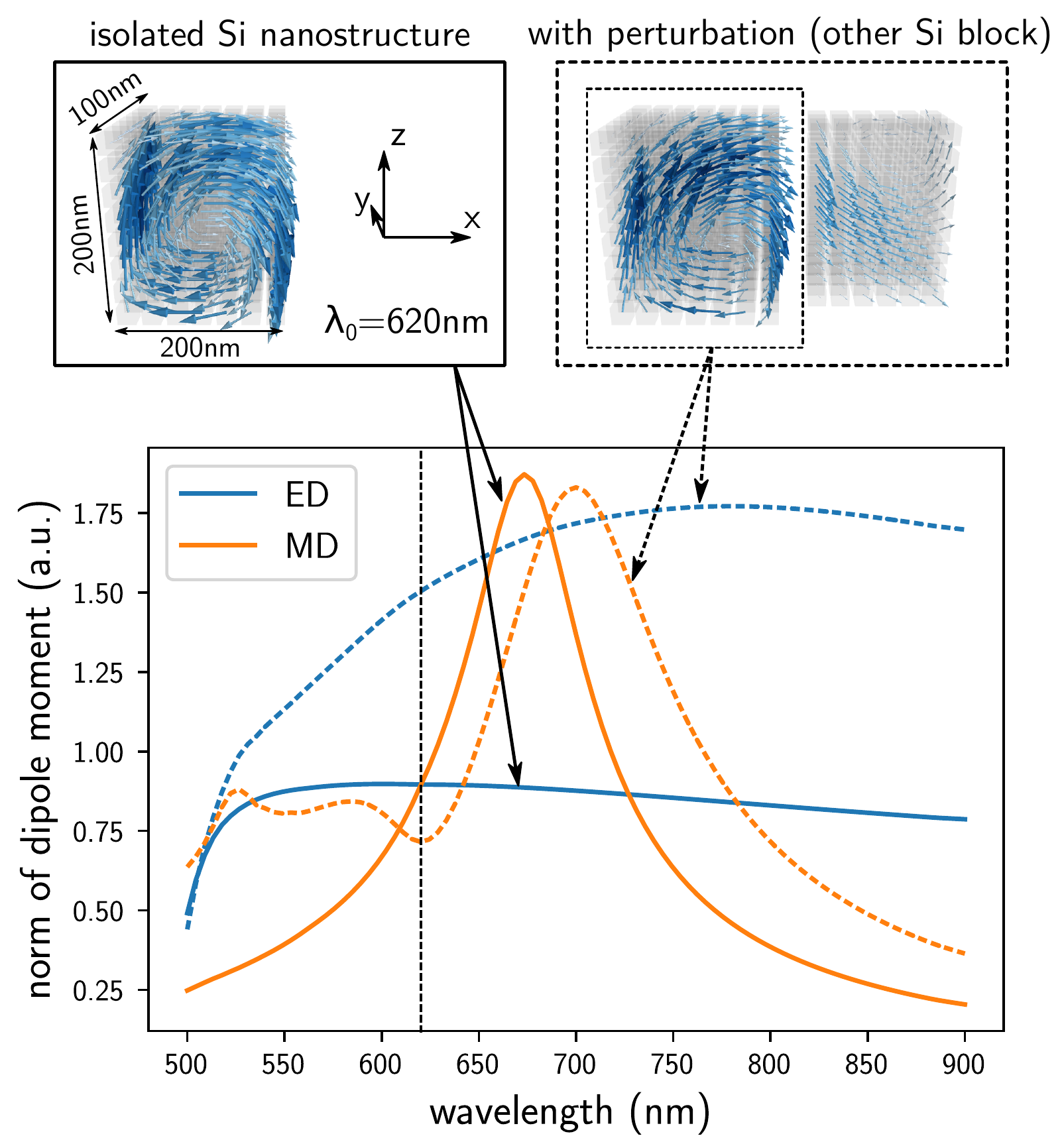}
	\centering
	\caption{
		Electric (ED) and magnetic dipole moment (MD) analysis of a silicon nano-block. 
		The block is either isolated (top left sketch), or perturbed by a second, smaller silicon nanoblock (top right illustration).
		The simulations are performed in vacuum with $X$ polarized plane wave illumination from the top.
		To assess the impact of the perturbation on the electric and magnetic dipole moments (MD) inside the larger Si block, the tool ``\texttt{tools.split\_simulation}'' is used to split off a sub-part of the structure after simulating the coupled system of two blocks (dashed box in top right sketch) in order to analyze the effect of the perturbation.
		The bottom plot compares the dipole moment magnitudes of the isolated case (solid lines) with the perturbed silicon nanoblock (dashed lines). 
		This analysis reveals a shift of the magnetic dipole resonance, some new emerging magnetic contributions (between $500\,$nm and $600\,$nm) as well as a significant increase in the induced electric dipole moment as result of the perturbation by a second silicon nanostructure.
		A vertical dashed line indicates the wavelength $\lambda_0=620\,$nm of the field distributions shown in the top figures.
	}\label{fig:ED_MD_splitsim}
\end{figure}

\begin{figure*}[t!]
	\includegraphics[width=\linewidth]{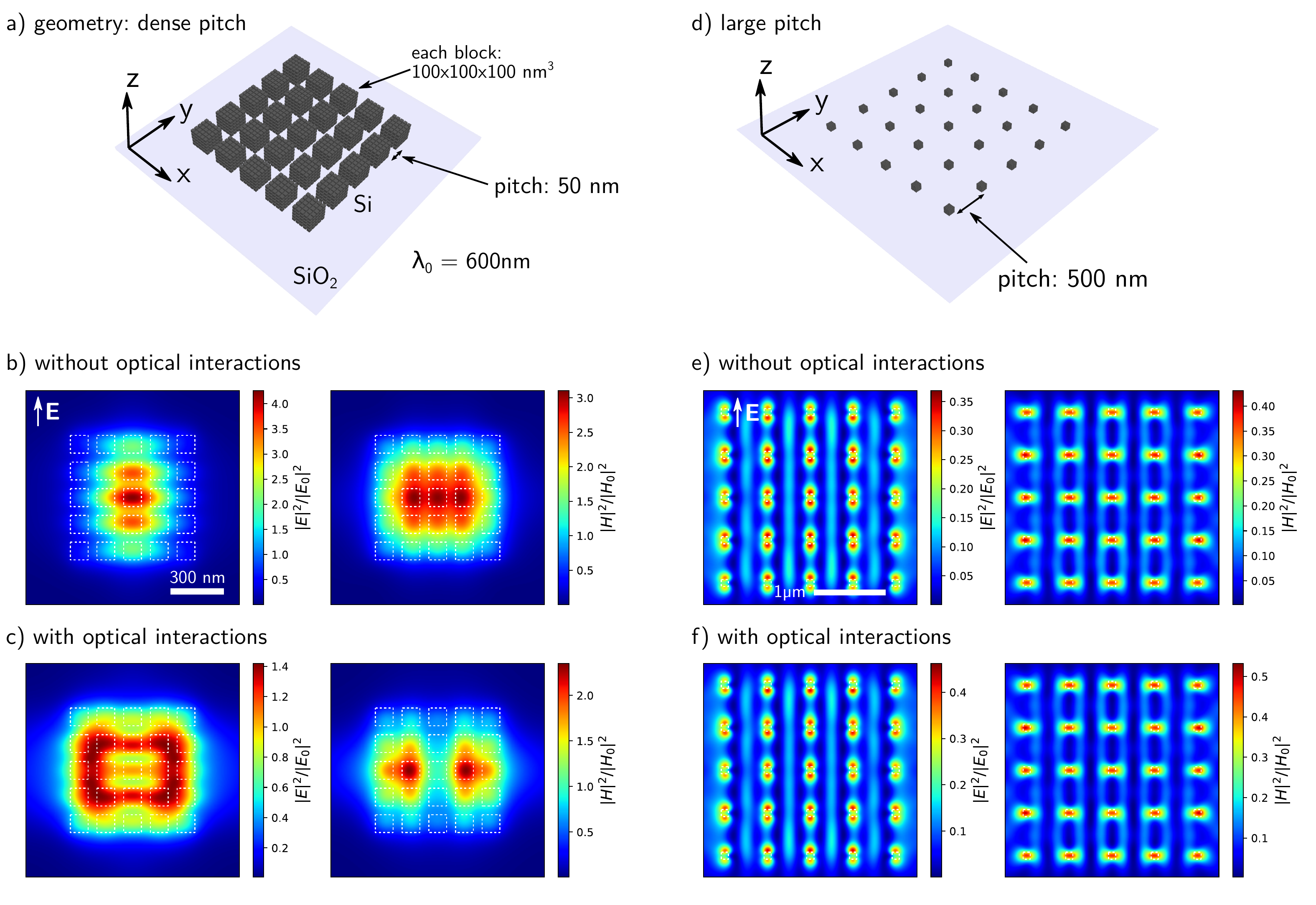}
	\centering
	\caption{
		Impact of optical interactions on an array of silicon nano-blocks.
		(a) shows the considered geometry, consisting of a square array of 25 silicon nano-cubes of $100\,$nm side length, lying on a glass substrate ($n_1=1.45$). 
		The spacing between the cubes is $50\,$nm.
		The array is illuminated from below by a $\lambda_0=600\,$nm plane wave, linearly polarized along $y$.
		(b-c) show electric (left) and magnetic (right) near-field intensity maps, calculated $50\,$nm above the top-surface of the blocks, on $1.2 \times 1.2\,$\textmu m$^2$ large areas.
		(b) simulation without optical interactions at the excitation, using the new ``\texttt{tools.combine\_simulations}'' function.
		(c) full simulation with all optical interactions.
		(d-f) same as (a-c) but with a larger pitch of $500\,$nm. Maps in (e-f) are $3\times 3$\,\textmu m$^2$ large.
	}\label{fig:optical_interactions_array}
\end{figure*}

Several tools have been added to \pygdm\ in the past 3 years. 
Along with tools to rotate and shift structures (\texttt{structures.rotate} and \texttt{structures.shift}), we added a tool to combine simulations (\texttt{tools.combine\_simulations}, assuming identical configurations for each simulation), as well as to split the geometry of a simulation (\texttt{tools.split\_simulation}).
These tools allow in particular to manipulate simulation objects with already calculated fields, splitting or combining geometries whilst preserving the formerly calculated optical response. This allows to analyze the impact of optical interactions and near-field coupling in an ensemble of several nanostructures.

We demonstrate a possible use scenario of the new function \texttt{tools.split\_simulation} in figure~\ref{fig:ED_MD_splitsim}. 
On the left we show the real part of the electric field inside an isolated silicon cube under $X$ polarized plane wave illumination at $\lambda_0=620\,$nm. 
On the right we show a second simulation with the same silicon object but including a second, smaller silicon block close to the original one. 
This additional nanostructure perturbs the optical response of the larger block. Using \texttt{tools.split\_simulation} we can remove the small block from the simulation once the internal fields are calculated, and calculate further observables from the perturbed internal fields, that have been modified by the optical interactions.
The bottom plot in figure~\ref{fig:ED_MD_splitsim} shows the norm of the electric (blue) and magnetic (orange) dipole moments of the unperturbed (solid lines) as well as of the perturbed block (dashed lines).
We find that the small perturbation has mainly a quantitative effect on the electric dipole resonance, whereas the magnetic dipole resonance is also strongly modified in qualitative sense. The resonance is redshifted and a few side-modes appear in the short wavelength region as a result of the presence of the small neighbor silicon block.

In a reciprocal approach to splitting simulations, it is possible to combine several simulations using \texttt{tools.combine\_simulations}. Those simulations can contain the already calculated optical response of their respective nanostructures, which will be all combined in a new simulation object. 
The fields in such a combined simulation represent then an approximation without optical interactions between the constituents.
This is illustrated in figure~\ref{fig:optical_interactions_array} by the example of an array of $5\times 5$ silicon nano-cubes. 
As depicted in Fig.~\ref{fig:optical_interactions_array}a, the cubes of $100\,$nm side length and with a mutual spacing of $50$\,nm are deposited on a glass substrate, surrounded by air. 
A plane wave of $\lambda_0=600\,$nm and linear polarization along $Y$ illuminates the structure from below.
First, each cube is described by a separate \pygdm\ simulation object. 
After calculating the individual cubes' optical responses, these 25 simulation objects are combined via \texttt{tools.combine\_simulations}. 
The resulting simulation contains the optical response in the Born approximation, hence without multiple scattering or near-field coupling between the cubes.
This combined simulation can subsequently be used in \pygdm\ to derive further quantities. 
Figure \ref{fig:optical_interactions_array}b shows intensity maps of the electric (left) and magnetic (right) near-field, calculated $50$\,nm above the top surface of the (non-interacting) blocks.
By running \texttt{scatter()} on the combined simulation object, we can then compute the self-consistent response of the full structure, hence including all optical interactions between the cubes. 
The resulting intensity maps are shown in figure~\ref{fig:optical_interactions_array}c. 
The comparison with \ref{fig:optical_interactions_array}b demonstrates that near-field interactions can play a crucial role in the optical excitation of dense assemblies of nano-structures.

Figure~\ref{fig:optical_interactions_array}d-f shows the exact same simulations as \ref{fig:optical_interactions_array}a-c, with the difference of a significantly larger pitch between the silicon nano-cubes ($500$\,nm distance instead of $50\,$nm). 
The comparison of simulations without and with optical interactions in figure~\ref{fig:optical_interactions_array}e and~\ref{fig:optical_interactions_array}f indicates a very weak optical interaction between the distant cubes, since the near-field maps are almost identical in this case.
In such a de-coupled scenario, the \texttt{tools.combine\_simulations} function can also be used to cost-efficiently calculate large arrays of non-interacting particles to evaluate interference effects, for instance in gratings.

The \pygdm-script to reproduce the results shown in figure~\ref{fig:ED_MD_splitsim} can be found \href{https://wiechapeter.gitlab.io/pyGDM2-doc/tutorials/09_split_simulations.html}{in the online documentation under this link}. For figure~\ref{fig:optical_interactions_array}, see \href{https://wiechapeter.gitlab.io/pyGDM2-doc/tutorials/10_combine_simulations.html}{this link}.

%% file: 06_bugfixes.tex
\section{All changes and bugfixes since version 1.0}

The new features and improvements are summarized in table~\ref{tab:new_features}, respectively table~\ref{tab:improvements}. Important bugfixes concerning the physics are listed in table~\ref{tab:bugfixes}.

\begin{table}[b!]
	\renewcommand\arraystretch{1.2}
	\centering
	\begin{small}
		\begin{tabular}{l}
			\textbf{New Functionalities} \\
			\midrule
			$\bullet$ 2D simulations \\
			$\bullet$ Substrates with retardation effects \\
			$\bullet$ Fast electrons illumination: EELS and CL \\
			$\bullet$ Internal H-field \\
			$\bullet$ Fields gradients, optical forces \\
			$\bullet$ Decay-rate for multi-material structures \\
			$\bullet$ Magnetic decay-rate inside structures \\
			$\bullet$ Multipole decomposition (ED / MD) \\
			$\bullet$ Combine and split simulations \\
			$\bullet$ pyGDM-UI (Graphical user interface, \textit{still experimental}) \\
		\end{tabular}    
	\end{small}
	\caption{Summary of new functionalities implemented in pyGDM.}\label{tab:new_features}
\end{table}

\begin{table}[b!]
	\renewcommand\arraystretch{1.2}
	\centering
	\begin{small}
		\begin{tabular}{l}
			\textbf{Improvements} \\
			\midrule
			$\bullet$ Automatic mesh-type recognition \\
			$\bullet$ Multi-materials nanostructures \\
			$\bullet$ Proper distinction inside / outside object in NF calculations \\
			$\bullet$ Structures supported in any layer of the environment \\
			$\bullet$ Dispersive environments \(n_{\text{env}}(\omega) \) \\
			$\bullet$ New hardcoded materials: SiO$_2$, TiO$_2$, doped dielectrics \\
			$\bullet$ Optimized LDOS (decay rate) calculation \\
			$\bullet$ Structure consistency test at initialization \\
			$\bullet$ Support for GPU solver to accelerate simulations \\
			$\bullet$ Avoid unnecessary calculations in \texttt{nearfield} / \texttt{farfield} \\
		\end{tabular}
	\end{small}
	\caption{Summary of improvements compared to the initial version of pyGDM.}\label{tab:improvements}
\end{table}

\begin{table}[b!]
	\renewcommand\arraystretch{1.8}
	\centering
	\begin{small}
		\begin{tabular}{l c c}
			\textbf{extinct} & \multicolumn{2}{c}{Prefactor for environments \(\epsilon \neq 1\)}\\
			\midrule
			\textbf{farfield} & \multicolumn{2}{c}{\makecell{Asymptotic propagator \\ for environments \(\epsilon \neq 1\)}}\\
			\textbf{heat} & \multicolumn{2}{c}{Correct equations implementation}\\
			& \multicolumn{2}{c}{Tensorial alpha support}\\
			\midrule
			\textbf{material} & \multicolumn{2}{c}{Correct $\epsilon$ for Ag}\\
			\midrule
			\textbf{visu} & \multicolumn{2}{c}{\makecell{Correct field-intensity\\ calculation}}\\
			& \multicolumn{2}{c}{\makecell{Non-cubic mesh\\ in \texttt{visu.visu\_contour}}}\\
			\midrule
			\textbf{fields} & \multicolumn{2}{c}{\makecell{Correct H-field phase\\ in plane wave and gaussian}}\\
			& \multicolumn{2}{c}{\makecell{arbitrary incident\\ angles for plane wave}}\\
			& \multicolumn{2}{c}{\makecell{correct 3-layer environments\\ in plane wave}}\\
			& \multicolumn{2}{c}{\makecell{arbitrary polarization states}}\\
		\end{tabular}    
	\end{small}
	\caption{Summary of major bug fixes to different pyGDM functions.}\label{tab:bugfixes}
\end{table}

%% file: Appendix_Nearfield_Farfield_retardation.tex
\section{Expressions of the field susceptibilities used in \pygdm}

\subsection{Cartesian representation of the free space field susceptibilities}

The solutions of the wave equations~(\ref{eq:wave-equations}) are \cite{patouxPolarizabilitiesComplexIndividual2020,girardFieldsNanostructures2005,wiechaDecayRateMagnetic2018}
\begin{subequations}\label{eq:props_1}
  \begin{align}
  \label{G0EE}
    &\mathbf{G}_{0}^{\text{EE}}(\mathbf{r}, \mathbf{r}', \omega)= \left[ \dfrac{\nabla\nabla}{\epsilon_{\text{env}}}+\mu_{\text{env}}k_{0}^{2}\mathbf{I} \right]\dfrac{e^{ik\vert\mathbf{r}-\mathbf{r}'\vert}}{\vert\mathbf{r}-\mathbf{r}'\vert}\\[0.75ex]
    \label{G0EH}
    &\mathbf{G}_{0}^{\text{EH}}(\mathbf{r}, \mathbf{r}', \omega)=ik_{0}\nabla\times\dfrac{e^{ik\vert\mathbf{r}-\mathbf{r}'\vert}}{\vert\mathbf{r}-\mathbf{r}'\vert}\\[0.75ex]
    \label{G0HH}
    &\mathbf{G}_{0}^{\text{HH}}(\mathbf{r}, \mathbf{r}', \omega)= \left[ \dfrac{\nabla\nabla}{\mu_{\text{env}}}+\epsilon_{\text{env}}k_{0}^{2}\mathbf{I} \right]\dfrac{e^{ik\vert\mathbf{r}-\mathbf{r}'\vert}}{\vert\mathbf{r}-\mathbf{r}'\vert}\\[0.75ex]
    \label{G0HE}
    &\mathbf{G}_{0}^{\text{HE}}(\mathbf{r}, \mathbf{r}', \omega)=-ik_{0}\nabla\times\dfrac{e^{ik\vert\mathbf{r}-\mathbf{r}'\vert}}{\vert\mathbf{r}-\mathbf{r}'\vert}
  \end{align}
\end{subequations}
with $\mathbf{r}=(x,y,z)$, $\mathbf{r}'=(x',y',z')$, $k_0$ is the wave vector in the vacuum and $k=\sqrt{\mu_{\text{env}}\epsilon_{\text{env}}}k_0$ the wave vector in the environment.
As mentionned in the main document, all materials (gases or solids) encountered in nature have no intrinsic response to the magnetic field at optical frequencies.
That means $\mu_{\text{env}}=1$, which leads to a wave vector in the environment $k=\sqrt{\epsilon_{\text{env}}}k_0=n_{\text{env}}k_0$. 
In Eq.~(\ref{eq:props_1}), if we apply the different nabla operators to the scalar Green's function, we obtain the field susceptibilities as a sum of tensors
\begin{subequations}\label{eq:props_2}
  \begin{multline}
  \label{G0EE_2}
    \mathbf{G}_{0}^{\text{EE}}(\mathbf{r}, \mathbf{r}', \omega)= \\
    \left[ -k^{2}\mathbf{T}_{1}^{\text{EE}}(\mathbf{R})-ik\mathbf{T}_{2}^{\text{EE}}(\mathbf{R})+\mathbf{T}_{3}^{\text{EE}}(\mathbf{R})\right]\dfrac{e^{ikR}}{\epsilon_{\text{env}}}
    \end{multline}
    \begin{equation}
    \label{G0HH_2}
    \mathbf{G}_{0}^{\text{HH}}(\mathbf{r}, \mathbf{r}', \omega)=\epsilon_{\text{env}}\mathbf{G}_{0}^{\text{EE}}(\mathbf{r}, \mathbf{r}', \omega)
    \end{equation}
    \begin{multline}
    \label{G0HE_2}
    \mathbf{G}_{0}^{\text{HE}}(\mathbf{r}, \mathbf{r}', \omega)=\\
    \left[ n_{\text{env}}k_0^{2}\mathbf{T}_{1}^{\text{HE}}(\mathbf{R})+ik_0\mathbf{T}_{2}^{\text{HE}}(\mathbf{R})\right]e^{ikR}
  	\end{multline}
    \begin{equation}
        \label{G0EH_2}
    \mathbf{G}_{0}^{\text{EH}}(\mathbf{r}, \mathbf{r}', \omega)=-\mathbf{G}_{0}^{\text{HE}}(\mathbf{r}, \mathbf{r}', \omega)
    \end{equation}
\end{subequations}
with $\mathbf{R}=\mathbf{r}-\mathbf{r}'=(\Delta_x,\Delta_y,\Delta_z)$ the relative distance between the source at the position $\mathbf{r}'$ and the evaluation point $\mathbf{r}$.
The different dyadic tensors give the electromagnetic field contributions in the far field ($\mathbf{T}_{1}^{\text{EE}}(\mathbf{R})$, $\mathbf{T}_{1}^{\text{HE}}(\mathbf{R})\propto 1 / R$), the intermediate ($\mathbf{T}_{2}^{\text{EE}}(\mathbf{R})$, $\mathbf{T}_{2}^{\text{HE}}(\mathbf{R})\propto 1 / R^2$) and near field ($\mathbf{T}_{3}^{\text{EE}}(\mathbf{R})\propto 1 / R^3$) regions.
They are defined as \cite{girardFieldsNanostructures2005, novotnyPrinciplesNanooptics2006}
\begin{subequations}\label{tensors_def}
	\begin{align}\label{T1EE}
		&\mathbf{T}_1^{\text{EE}}(\mathbf{R})=\frac{\mathbf{R}\mathbf{R}-\mathbf{I}R^2}{R^3},\\
	\label{T2EE}
		&\mathbf{T}_2^{\text{EE}}(\mathbf{R})=\frac{3\mathbf{R}\mathbf{R}-\mathbf{I}R^2}{R^4},\\
	\label{T3EE}
		&\mathbf{T}_3^{\text{EE}}(\mathbf{R})=\frac{3\mathbf{R}\mathbf{R}-\mathbf{I}R^2}{R^5},
	\end{align}
\end{subequations}
and
\begin{subequations}\label{tensors_def_2}
	\begin{align}\label{T1HE}
		&\mathbf{T}_1^{\text{HE}}(\mathbf{R})=\frac{1}{R^2}\left(
		\begin{matrix}
		0 & -\Delta_z & \Delta_y\\
		\Delta_z & 0 & -\Delta_x\\
		-\Delta_y & \Delta_x & 0
		\end{matrix}\right),\\
	\label{T2HE}
		&\mathbf{T}_2^{\text{HE}}(\mathbf{R})=\frac{1}{R^3}\left(
		\begin{matrix}
		0 & -\Delta_z & \Delta_y\\
		\Delta_z & 0 & -\Delta_x\\
		-\Delta_y & \Delta_x & 0
		\end{matrix}\right).
	\end{align}
\end{subequations}

\subsection{Weyl representation of the field susceptibilities of a flat surface}\label{sec:weyl_representation}

We consider in the following a single, flat interface at $z=0$, separating the dipole environment (of refractive index $n_2$) and a substrate (of refractive index $n_1$).
For the description of radiation from oscillating dipoles across planar interfaces, the Cartesian forms of the free space susceptibilities are exact but not well suited for this boundary conditions problem.
It is easier to use electromagnetic fields with wavefronts parallel to the interface (plane wave for a flat interface; spherical wave for a spherical particle interface).
Due to the broken symmetry along the $z$-axis, a plane wave expansion of the surface wave vector $\mathbf{k}_{\parallel}=(\mathbf{k}_{\text{x}},\mathbf{k}_{\text{y}})$ is performed.
Thus the susceptibilities associated with the surface are, as presented in the references \cite{agarwalQuantumElectrodynamicsPresence1975, novotnyPrinciplesNanooptics2006}, developed in 2D-integrals of the shape 
\begin{multline}
  \label{eq:G_2_layer}
    \mathbf{G}_{\text{2-layer}}^{\alpha\beta^{+/-}}(\mathbf{r}, \mathbf{r}', \omega)=\\ 
    \dfrac{i}{2\pi}\int \mathrm{d}\mathbf{k}_{\parallel}\,F^{+/-}(\mathbf{r}, \mathbf{r}', \omega)\mathbf{N}_{\alpha\beta}^{+/-}(\mathbf{k}_{\parallel}, \omega).
    \end{multline}
For all the terms, the ``+'' superscript means that we consider the case above the interface (in the medium containing the structure) and ``-'' under it (in the substrate).
The $\mathbf{N}_{\alpha\beta}^{+/-}(\mathbf{k}_{\parallel}, \omega)$ terms are ($3\times 3$) matrices, ${k_{\text{z}}^{-}}=\sqrt{\epsilon_1 k_{0}^{2}-k_{\parallel}^{2}}$ and ${k_{\text{z}}^{+}}=\sqrt{\epsilon_2 k_{0}^{2}-k_{\parallel}^{2}}$, with $\epsilon_1=n_1^2$.
For the electric-electric coupling term, the definition of the matrices $\mathbf{N}_{\alpha\beta}^{+/-}(\mathbf{k}_{\parallel}, \omega)$ are mentionned in the references~\cite{girardFieldsNanostructures2005} ($\mathbf{N}_{\alpha\beta}^{+}$, see Eq.~(51) in~\cite{girardFieldsNanostructures2005}) and \cite{colasdesfrancsTheoryNearfieldOptical2002} ($\mathbf{N}_{\alpha\beta}^{-}$, Eq.~(A5) in~\cite{colasdesfrancsTheoryNearfieldOptical2002}).
The spatial term $F^{+}(\mathbf{r}, \mathbf{r}', \omega)=e^{i\left[\mathbf{k}_{\parallel}\cdot(\mathbf{l}-\mathbf{l}')+k_{\text{z}}^{+}(z+z') \right]}$ (respectively  
$F^{-}(\mathbf{r}, \mathbf{r}', \omega)=e^{i\left[\mathbf{k}_{\parallel}\cdot(\mathbf{l}-\mathbf{l}')-(k_{\text{z}}^{-}z-k_{\text{z}}^{+}z') \right]}$) depend on the position of the dipole $\mathbf{r}'$ above the interface, and the evaluation position $\mathbf{r}$ above (respectively below) the surface.

The equation (\ref{eq:G_2_layer}) gives the electric field everywhere in space, nevertheless that requires numerical integrations in the complex plane to deduce the nine elements of the matrix. 
In the reference \cite{paulusAccurateEfficientComputation2000}, all the details on how to perform the calculation efficiently are described.
The expression (\ref{eq:G_2_layer}) is an integral of the shape:
\begin{equation}
G_{\text{s}}(\mathbf{r}, \mathbf{r}', \omega)=\dfrac{i}{2\pi}\int_{0}^{+\infty}\!\!\mathrm{d}k_{\parallel}\int_{0}^{2\pi} \!\! g(k_{\parallel},\alpha)e^{irf(k_{\parallel},\alpha)} \, \mathrm{d}\alpha 
\end{equation}
where the vector $\mathbf{k}_{\parallel}$ is expressed in cylindrical coordinates $(k_{\parallel}, \alpha)$.
This integration can be avoided in the near-field and far-field regions. 
In the near-field region (small values of r) we use the electrostatic limit $(c\rightarrow\infty)$ and in the far-field region (large values of r) we use the retarded asymptotic limit defined in the references \cite{bornPrinciplesOpticsElectromagnetic1999}:
\begin{equation}\label{eq:lim_asymp}
G_{\text{s}}(\mathbf{r}, \mathbf{r}', \omega)\sim_{\infty}\dfrac{1}{\vert\tan\theta\vert}g(\tilde{k}_{\parallel},\tilde{\alpha})\dfrac{e^{ink_{0}r}}{r}.
\end{equation}
In the previous equation, r and $\theta$ are the spherical components of the evaluation position $\mathbf{r}=(r,\theta,\varphi)$, $n$ is the refractive index of the medium considered and $(\tilde{k}_{\parallel},\tilde{\alpha})$ defined the saddle point of the function $f(k_{\parallel},\alpha)$
\begin{equation}
\dfrac{\partial f}{\partial k_{\parallel}}(\tilde{k}_{\parallel},\tilde{\alpha})=0\quad\text{and}\quad\dfrac{\partial f}{\partial \alpha}(\tilde{k}_{\parallel},\tilde{\alpha})=0.
\end{equation}
The calculus is fully explained in Ref.~\cite{colasdesfrancsOptiqueSublongueurOnde2002}, here we present only the electrostatic and asymptotic expressions of the surface Green's dyadic tensor, which are implemented in \pyGDM .
As previously, due to the non-magnetic response of the environment, we get interest only to the ``EE'' and ``HE'' surface dyadic tensors.

\subsubsection{Non-retarded form and electrostatic limit}

In the electrostatic limit we suppose $c\rightarrow\infty$, hence $k_0\rightarrow0$, which simplifies the problem drastically. 
For the electric-electric tensors that leads to the following expressions
\begin{equation}\label{G_2layer_EE_elec+}
\mathbf{G}_{\text{2-layer}}^{\text{EE}^{\text{+,stat}}}(\mathbf{r}, \mathbf{r}', \omega)=\Delta_{\text{surf}}^{+} \dfrac{\mathbf{T}_{3}(\mathbf{R})}{\epsilon_{\text{env}}}
\cdot\left( \begin{matrix}
-1 & 0 & 0\\
0 & -1 & 0\\
0 & 0 & 1
\end{matrix}\right),
\end{equation}
\begin{equation}\label{G_2layer_EE_elec-}
\mathbf{G}_{\text{2-layer}}^{\text{EE}^{\text{-,stat}}}(\mathbf{r}, \mathbf{r}', \omega)=\Delta_{\text{surf}}^{-} \mathbf{T}_{3}(\mathbf{R}),
\end{equation}
with $\Delta_{\text{surf}}^{+}=(\epsilon_1-\epsilon_2)/(\epsilon_1+\epsilon_2)$ and $\Delta_{\text{surf}}^{-}=2/(\epsilon_1+\epsilon_2)$.

The magnetic-electric case vanishes in the electrostatic limit
\begin{equation}\label{G_2layer_HE_elec}
\begin{aligned}
\mathbf{G}_{\text{2-layer}}^{\text{HE}^{\text{+,stat}}}(\mathbf{r}, \mathbf{r}', \omega)=\mathbf{0}\\
\mathbf{G}_{\text{2-layer}}^{\text{HE}^{\text{-,stat}}}(\mathbf{r}, \mathbf{r}', \omega)=\mathbf{0}.
\end{aligned}
\end{equation}
These results can be obtained equally with the image dipole theory \cite{jacksonClassicalElectrodynamics1999}.

\subsubsection{Retarded asymptotic form}

The retarded expressions of the electric-electric Green's dyad, above and below the surface, are obtained by using the equation (\ref{eq:lim_asymp}).
These tensors make it possible to easily obtain the electric field emitted by an electric dipole, in the vicinity of an interface, far from it. 
The tensors depend mainly of the spherical components $(r,\theta,\varphi)$ of the evaluation position, above the surface it is expressed as
\begin{subequations}\label{eq:Ss_+inf}
	\begin{align}
	G_{\text{2-layer,xx}}^{\text{EE}^{+\infty}} &=\mathcal{F}_0( r_{\text{p}}\cos^{2}\theta\cos^{2}\varphi-r_{\text{s}}\sin^{2}\varphi) \label{eq:EE_2layer_above_xx}\\
	G_{\text{2-layer,xy}}^{\text{EE}^{+\infty}} &=\mathcal{F}_0 (r_{\text{p}}\cos^{2}\theta+r_{\text{s}})\cos\varphi\sin\varphi       \label{eq:EE_2layer_above_xy}\\
	G_{\text{2-layer,xz}}^{\text{EE}^{+\infty}} &=\mathcal{F}_0 r_{\text{p}}\sin\theta\cos\theta\cos\varphi \label{eq:EE_2layer_above_xz}\\
	G_{\text{2-layer,yx}}^{\text{EE}^{+\infty}} &= G_{\text{2-layer,xy}}^{\text{EE}^{+\infty}} \label{eq:EE_2layer_above_yx}\\
	G_{\text{2-layer,yy}}^{\text{EE}^{+\infty}} &=\mathcal{F}_0( r_{\text{p}}\cos^{2}\theta\sin^{2}\varphi-r_{\text{s}}\cos^{2}\varphi) \label{eq:EE_2layer_above_yy}\\
	G_{\text{2-layer,yz}}^{\text{EE}^{+\infty}} &=\mathcal{F}_0 r_{\text{p}}\sin\theta\cos\theta\sin\varphi      \label{eq:EE_2layer_above_yz}\\
	G_{\text{2-layer,zx}}^{\text{EE}^{-\infty}} &=-G_{\text{2-layer,xz}}^{\text{EE}^{+\infty}} \label{eq:EE_2layer_above_zx}\\
	G_{\text{2-layer,zy}}^{\text{EE}^{+\infty}} &=-G_{\text{2-layer,yz}}^{\text{EE}^{+\infty}} \label{eq:EE_2layer_above_zy}\\
	G_{\text{2-layer,zz}}^{\text{EE}^{+\infty}} &=-\mathcal{F}_0 r_{\text{p}}\sin^{2}\theta \label{eq:EE_2layer_above_zz}
	\end{align}
\end{subequations}
with 
\begin{equation}
\begin{aligned}
\mathcal{F}_0=-k_{0}^{2}\dfrac{e^{in_2 k_{0}r}}{r}e^{-in_2 k_{0}\sin\theta(x'\cos\varphi+y'\sin\varphi)}\times
\\
e^{in_2 k_{0}\cos\theta z'}
\end{aligned}
\end{equation}
%
%
%
%
%
$r_{\text{p}}$ and $r_{\text{s}}$ are the reflection coefficients, depending on the refractive index of the media from either side of the interface, given by \cite{novotnyPrinciplesNanooptics2006, colasdesfrancsOptiqueSublongueurOnde2002}
\begin{equation}
\begin{aligned}
&r_{\text{p}}=\dfrac{\epsilon_1 n_2\cos\theta-\epsilon_2(\epsilon_1-\epsilon_2\sin^{2}\theta)^{1/2}}{\epsilon_1 n_2\cos\theta+\epsilon_2(\epsilon_1-\epsilon_2\sin^{2}\theta)^{1/2}},\\
&r_{\text{s}}=\dfrac{n_2\cos\theta-(\epsilon_1-\epsilon_2\sin^{2}\theta)^{1/2}}{n_2\cos\theta+(\epsilon_1-\epsilon_2\sin^{2}\theta)^{1/2}}.
\end{aligned}
\end{equation}
A completely similar form is obtained for the electric-electric surface susceptibility below the interface
\begin{subequations}\label{eq:Ss_-inf}
	\begin{align}
	G_{\text{2-layer,xx}}^{\text{EE}^{-\infty}} &=\mathcal{F}_1 (\dfrac{\Phi_{\text{p}}}{\epsilon_2}\cos^{2}\varphi+\Phi_{\text{s}}\sin^{2}\varphi)\cos\theta \label{eq:EE_2layer_under_xx}\\
	G_{\text{2-layer,xy}}^{\text{EE}^{-\infty}} &=\mathcal{F}_1 (\dfrac{\Phi_{\text{p}}}{\epsilon_2}-\Phi_{\text{s}})\cos\theta\sin\varphi\cos\varphi       \label{eq:EE_2layer_under_xy}\\
	G_{\text{2-layer,xz}}^{\text{EE}^{-\infty}} &=\mathcal{F}_1 \tau_{\text{p}}\dfrac{\epsilon_1}{\epsilon_2}\sin\theta\cos\theta\cos\varphi \label{eq:EE_2layer_under_xz}\\
	G_{\text{2-layer,yx}}^{\text{EE}^{-\infty}} &= G_{\text{2-layer,xy}}^{\text{EE}^{-\infty}} \label{eq:EE_2layer_under_yx}\\
	G_{\text{2-layer,yy}}^{\text{EE}^{-\infty}} &=\mathcal{F}_1 (\dfrac{\Phi_{\text{p}}}{\epsilon_2}\sin^{2}\varphi+\Phi_{\text{s}}\cos^{2}\varphi)\cos\theta \label{eq:EE_2layer_under_yy}\\
	G_{\text{2-layer,yz}}^{\text{EE}^{-\infty}} &=\mathcal{F}_1 \tau_{\text{p}}\dfrac{\epsilon_1}{\epsilon_2}\sin\theta\cos\theta\sin\varphi      \label{eq:EE_2layer_under_yz}\\
	G_{\text{2-layer,zx}}^{\text{EE}^{-\infty}} &=-\mathcal{F}_1 \dfrac{\Phi_{\text{p}}}{\epsilon_2}\sin\theta\cos\varphi \label{eq:EE_2layer_under_zx}\\
	G_{\text{2-layer,zy}}^{\text{EE}^{-\infty}} &=-\mathcal{F}_1 \dfrac{\Phi_{\text{p}}}{\epsilon_2}\sin\theta\sin\varphi \label{eq:EE_2layer_under_zy}\\
	G_{\text{2-layer,zz}}^{\text{EE}^{-\infty}} &=-\mathcal{F}_1 \tau_{\text{p}}\dfrac{\epsilon_1}{\epsilon_2}\sin^{2}\theta \label{eq:EE_2layer_under_zz}
	\end{align}
\end{subequations}
%
%
%
%
%
with
\begin{equation}
\begin{aligned}
\mathcal{F}_1 = k_{0}^{2}\dfrac{e^{in_1k_{0}r}}{r}e^{-in_1k_{0}\sin\theta(x'\cos\varphi+y'\sin\varphi)}\times\\
e^{ik_{0}(\epsilon_2-\epsilon_1\sin^{2}\theta)^{1/2} z'}
\end{aligned}
\end{equation}
and
\begin{equation}
\begin{aligned}
&\Phi_{\text{s}}=\dfrac{n_1\tau_{\text{s}}}{(\epsilon_2-\epsilon_1\sin^{2}\theta)^{1/2}},\ \tau_{\text{s}}=1-\Delta_{\text{s}},\\[12pt]
&\Phi_{\text{p}}=n_1\tau_{\text{p}}(\epsilon_2-\epsilon_1\sin^{2}\theta)^{1/2},\ \tau_{\text{p}}=\Delta_{\text{p}}+1
\end{aligned}
\end{equation}
where
\begin{equation}
\begin{aligned}
&\Delta_{\text{p}}=\dfrac{-n_1\epsilon_2\cos\theta-\epsilon_1(\epsilon_2-\epsilon_1\sin^{2}\theta)^{1/2}}{-n_1\epsilon_2\cos\theta+\epsilon_1(\epsilon_2-\epsilon_1\sin^{2}\theta)^{1/2}}\\[12pt]
&\Delta_{\text{s}}=\dfrac{-n_1\cos\theta-(\epsilon_2-\epsilon_1\sin^{2}\theta)^{1/2}}{-n_1\cos\theta+(\epsilon_2-\epsilon_1\sin^{2}\theta)^{1/2}}.
\end{aligned}
\end{equation}
In the far region, or propagating region, the wave vector, the electric field and magnetic field respect the right-hand rule.
Therefore, we easily calculate the magnetic field in this region by using the well-known equation
\begin{equation}
\mathbf{H}=\mathbf{k}\times\mathbf{E}.
\end{equation}
%

%% file: Appendix_2D_tensors.tex
\section{2D field susceptibilities}\label{sec:2DGreensDyads}

A further new feature in \pyGDM\ is the possibility to simulate two-dimensional nanostructures like nanowires, which are infinitely long along one dimension. In this case, the geometry is disretized using ``line-dipoles'', and light-emission is described by cylindrical waves.

We assume a nanostructure geometry which is ``infinitely long'', hence is invariant along one axis. 
We will use the \(y\) axis in the following. 
In this case, all dependencies along this axis are identical to the time-harmonicity of the incident field and are therefore explicit. 
With \(\mathbf{r}^{2D} = x\hat{e}_x + z\hat{e}_z\) the two-dimensional electric polarization at a ``3D'' location \(\mathbf{r}\) can be written:

\begin{equation}
	\mathbf{P}^{2D} (\mathbf{r}) = \mathbf{P}^{2D} (\mathbf{r}^{2D}) \exp(\text{i} k_{y,0} y)
\end{equation}
where \(k_{y,0}\) is the wavevector component along the infinite axis.
In the \(2D\) field calculations, we discretize the nanostructure into lines of identical dipoles, infinitely extending along the \(y\)-axis. 
Hence an integration along these dipole lines needs to be performed. 
This integration can be included in the 2D Green's Dyads. 
Using the Dyad \( \mathbf{G}_0^{\text{EE}}\) from equation~\eqref{eq:props_1}, gives:

\begin{equation} \label{eq:G02DEE_integration}
\begin{split}
\mathbf{G}_0^{2D,\text{EE}} ( & \mathbf{r}^{2D}, \mathbf{r}'^{,2D} ) = \\
= &\int\limits_{-\infty}^{\infty} \mathbf{G}_0^{\text{EE}} (\mathbf{r}, \mathbf{r}') e^{\text{i} k_{y,0} y'} \text{d}y' \\
= &\frac{1}{\epsilon_{\text{env}}} \Big( k^2 \mathbf{I} + \nabla\nabla \Big)\int\limits_{-\infty}^{\infty} G_0(\mathbf{r}, \mathbf{r}' ) e^{\text{i} k_{y,0} y'} \text{d}y'
\end{split}
\end{equation}
where \(\mathbf{r}^{2D}\) and \(\mathbf{r}'^{,2D}\) are the 2D positions of the observer and the emitting dipole-line, respectively.

The \(y'\)-integral over the vacuum Green's function can be expressed by the following identity via the zero-order Hankel function of the first kind \(H_0^{(1)}\) \cite{martinElectromagneticScatteringPolarizable1998, gradshteynTableIntegralsSeries2007}:
\begin{multline}
	\int\limits_{-\infty}^{\infty} \frac{e^{\text{i}k \sqrt{R^2 + (y-y')^2}}}{\sqrt{R^2 + (y-y')^2}}   e^{\text{i} k_{y,0} y'}\text{d}y' = \\
	\text{i} \pi H_0^{(1)} \Big( k_r\, R \Big) e^{\text{i} k_{y,0} y}
\end{multline}
%
where we define the in-plane distance as \(R = |\mathbf{R}^{2D}| = |\mathbf{r}^{2D}-\mathbf{r}'^{,2D}|\), and
\begin{equation}
k_{r} = \sqrt{k^2 - k_{y,0}^2}\, .
\end{equation}

Now, by including the explicit \(y\)-dependence in the differential operator for the \(2D\) problem
\begin{equation}
\begin{aligned}
\nabla & = \partial \hat{e}_x/ \partial x + \partial \hat{e}_z/ \partial z + \text{i}k_{y,0} \hat{e}_y \\
& = \nabla_{\rho} + \text{i}k_{y,0} \hat{e}_y
\end{aligned}
\end{equation}
we obtain
\begin{multline} \label{eq:G_2D_EE_matrix}
\mathbf{G}_0^{2D,\text{EE}} (\mathbf{r}^{2D}, \mathbf{r}'^{,2D} ) 
= \\
\frac{\text{i}\pi}{\epsilon_{\text{env}}}
\begin{pmatrix}
k^2 + \frac{\partial^2}{\partial x^2} &  
			\text{i}k_{y,0} \frac{\partial}{\partial x} &
						\frac{\partial^2}{\partial x \partial z } \\
\text{i}k_{y,0} \frac{\partial}{\partial x} &
			k^2 - k_{y,0}^2 &
						\text{i}k_{y,0} \frac{\partial}{\partial z} \\
\frac{\partial^2}{\partial z \partial x } &
			\text{i}k_{y,0} \frac{\partial}{\partial z} &
						k^2 + \frac{\partial^2}{\partial z^2} \\
\end{pmatrix} 
\\
\times H_0^{(1)}\Big(k_r R\Big)e^{\text{i}k_{y,0}y} \, .
\end{multline}
With the derivatives of the \(n\)-th order Hankel functions of the first kind
\begin{equation}
\frac{\text{d}}{\text{d}a} H_n^{(1)} \left( a \right) = \frac{n H_n^{(1)}(a)}{a} - H_{n+1}^{(1)}(a)\, ,
\end{equation}
and the recurrence relation
\begin{equation}
H_n^{(1)} \left( a \right) = \frac{a}{2n} \left( H_{(n-1)}^{(1)} \left( a \right) + H_{(n+1)}^{(1)} \left( a \right) \right)\, ,
\end{equation}
we arrive finally at the following set of components for the 2D Green's dyad, where we assume that the observer is located in the plane \(y=0\):

\begin{multline}
G_{0,xx}^{2D,\text{EE}} = \text{i}\pi\, \Big(k_0^2 - k_{r}^2 / (2\epsilon_{\text{env}})\Big)\, H_0^{(1)}(k_{r}\, R)\\
+ \text{i}\pi\, \frac{k_{r}^2  \left( 2 (\Delta x / R)^2 - 1 \right) }{ 2\epsilon_{\text{env}} }\, H_2^{(1)}(k_{r}\, R)
\end{multline}

\begin{equation}
G_{0,xy}^{2D,\text{EE}} = \pi\, \frac{k_{r}\, k_{y,0}\, \Delta x / R}{\epsilon_{\text{env}}}\, H_1^{(1)}(k_{r}\, R)
\end{equation}

\begin{equation}
G_{0,xz}^{2D,\text{EE}} = \text{i}\pi\, \frac{k_{r}^2\, \Delta x \Delta z / R^2}{\epsilon_{\text{env}}}\, H_2^{(1)}(k_{r}\, R)
\end{equation}

\begin{equation}
G_{0,yy}^{2D,\text{EE}} = \text{i}\pi\, \Big(k_0^2 - k_{y,0}^2 / \epsilon_{\text{env}}\Big)\, H_0^{(1)}(k_{r}\, R)
\end{equation}

\begin{equation}
G_{0,yz}^{2D,\text{EE}} = \pi\, \frac{k_{r}\, k_{y,0} \Delta z/R}{\epsilon_{\text{env}}}\,  H_1^{(1)}(k_{r}\, R)
\end{equation}

\begin{multline}
G_{0,zz}^{2D,\text{EE}} = \text{i}\pi\, \Big(k_0^2 - k_{r}^2 / (2\epsilon_{\text{env}})\Big)\, H_0^{(1)}(k_{r}\, R) \\
+ \text{i}\pi\, \frac{k_{r}^2 \left( 2(\Delta z/R)^2 - 1 \right)}{2\epsilon_{\text{env}}}\, H_2^{(1)}(k_{r}\, R)
\end{multline}

\begin{equation}
\begin{split}
G_{0,yx}^{2D,\text{EE}} = G_{0,xy}^{2D,\text{EE}}\, \\
G_{0,zx}^{2D,\text{EE}} = G_{0,xz}^{2D,\text{EE}}\, \\
G_{0,zy}^{2D,\text{EE}} = G_{0,yz}^{2D,\text{EE}}\, 
\end{split}
\end{equation}
with \(\Delta x = x-x'\) and \(\Delta z = z-z'\).

Analogously, for the mixed electric-magnetic Green's tensor we obtain:
\begin{equation} \label{eq:G02dHE_integration}
\begin{split}
\mathbf{G}_0^{2D,\text{HE}} ( & \mathbf{r}^{2D}, \mathbf{r}'^{,2D} ) = \\
= &\int\limits_{-\infty}^{\infty} \mathbf{G}_0^{\text{HE}} (\mathbf{r}, \mathbf{r}') e^{\text{i} k_{y,0} y'} \text{d}y' \\
= & - \text{i} k_0 \nabla \times \int\limits_{-\infty}^{\infty} G_0(\mathbf{r}, \mathbf{r}') e^{\text{i} k_{y,0} y'} \text{d}y' \\
= & k_0 \pi 
\begin{pmatrix}
0                            & \frac{-\partial}{\partial z} & \frac{\partial}{\partial y} \\
 \frac{\partial}{\partial z} & 0                      & \frac{-\partial}{\partial x} \\
\frac{-\partial}{\partial y} &  \frac{\partial}{\partial x} & 0\\
\end{pmatrix}
H_0^{(1)}\Big(k_r R\Big)e^{\text{i}k_{y,0}y}
\end{split}
\end{equation}
which explicitly gives
\begin{subequations}\label{eq:2D_GHE_components}
	\begin{align}
	G_{0,xy}^{2D,\text{HE}} &= \pi  \frac{k_0 k_r \Delta z}{R}\,  H_1^{(1)} \left( k_r\, R \right)       \label{eq:2dHExy}\\
	G_{0,xz}^{2D,\text{HE}} &= \text{i} \pi k_0 k_{y,0}\, H_0^{(1)} \left( k_r\, R \right)       \label{eq:2dHExz}\\
	G_{0,yz}^{2D,\text{HE}} &= \pi  \frac{k_0 k_r \Delta x}{R} \, H_1^{(1)} \left( k_r\, R \right)      \label{eq:2dHEyz}\\
	G_{0,yx}^{2D,\text{HE}} &= -G_{0,xy}^{2D,\text{HE}}      \label{eq:2dHEyx}\\
	G_{0,zx}^{2D,\text{HE}} &= -G_{0,xz}^{2D,\text{HE}}      \label{eq:2dHEzx}\\
	G_{0,zy}^{2D,\text{HE}} &= -G_{0,yz}^{2D,\text{HE}}      \label{eq:2dHEzy}\\
	G_{0,xx}^{2D,\text{HE}} & =	G_{0,yy}^{2D,\text{HE}} = G_{0,zz}^{2D,\text{HE}} = 0\label{eq:2dHE_diag} \, .
	\end{align}
\end{subequations}

Without loss of generality, 2D simulations in \pyGDM\ are restricted to the \(y=0\) plane. 
For the evaluation of the fields at positions $y\neq 0$, the tensors simply need to be multiplied by the \(y\)-dependent phase factor of the incident field.

\subsection*{2D self-terms}\label{sec:2D_selfterms}
The electric-electric self-term components write \cite{martinElectromagneticScatteringPolarizable1998}:
\begin{multline}
\text{norm}_{x} = \text{norm}_{z} = \frac{- 4 \pi} {2\epsilon_{\text{env}} S_{\text{cell}}} + \\
4 \text{i} \pi^2 \left( 2 k_0^2 - \frac{k_{r} ^2}{\epsilon_{\text{env}}} \right)
\dfrac{H_1^{(1)}\left(k_{r} d/\sqrt{\pi}\right) + 2\text{i}/(\pi k_{r}) }{4 S_{\text{cell}} k_{r} }
\end{multline}
and
\begin{multline}
\text{norm}_{y} = 4 \text{i} \pi^2 \left( k_0^2 - \frac{k_{y,0}^2}{\epsilon_{\text{env}}} \right)\ \\
\dfrac{H_1^{(1)}\left(k_{r} d/\sqrt{\pi}\right) + 2\text{i}/(\pi k_{r}) }{2 S_{\text{cell}} k_{r} }
\end{multline}

where \(S_{\text{cell}} = d^2\) is the surface covered by a square mesh-cell with side length \(d\). 
Similar to the 3D Green's tensors, the magnetic-electric self-terms equal zero.
Note that in the current version of \pygdm , 2D simulations are only possible on a square mesh.

%% file: Appendix_fast_electrons.tex
\section{Fast electron illumination: Computation of electron energy losses and cathodoluminescence}\label{sec:fast_electrons_formalism}

The pyGDM toolbox now provides built-in routines to simulate the results of Electron Energy Loss Spectroscopy (EELS) or Cathodoluminescence (CL) experiments.

\begin{figure}[ht]
	\centering\includegraphics[width=5.5cm]{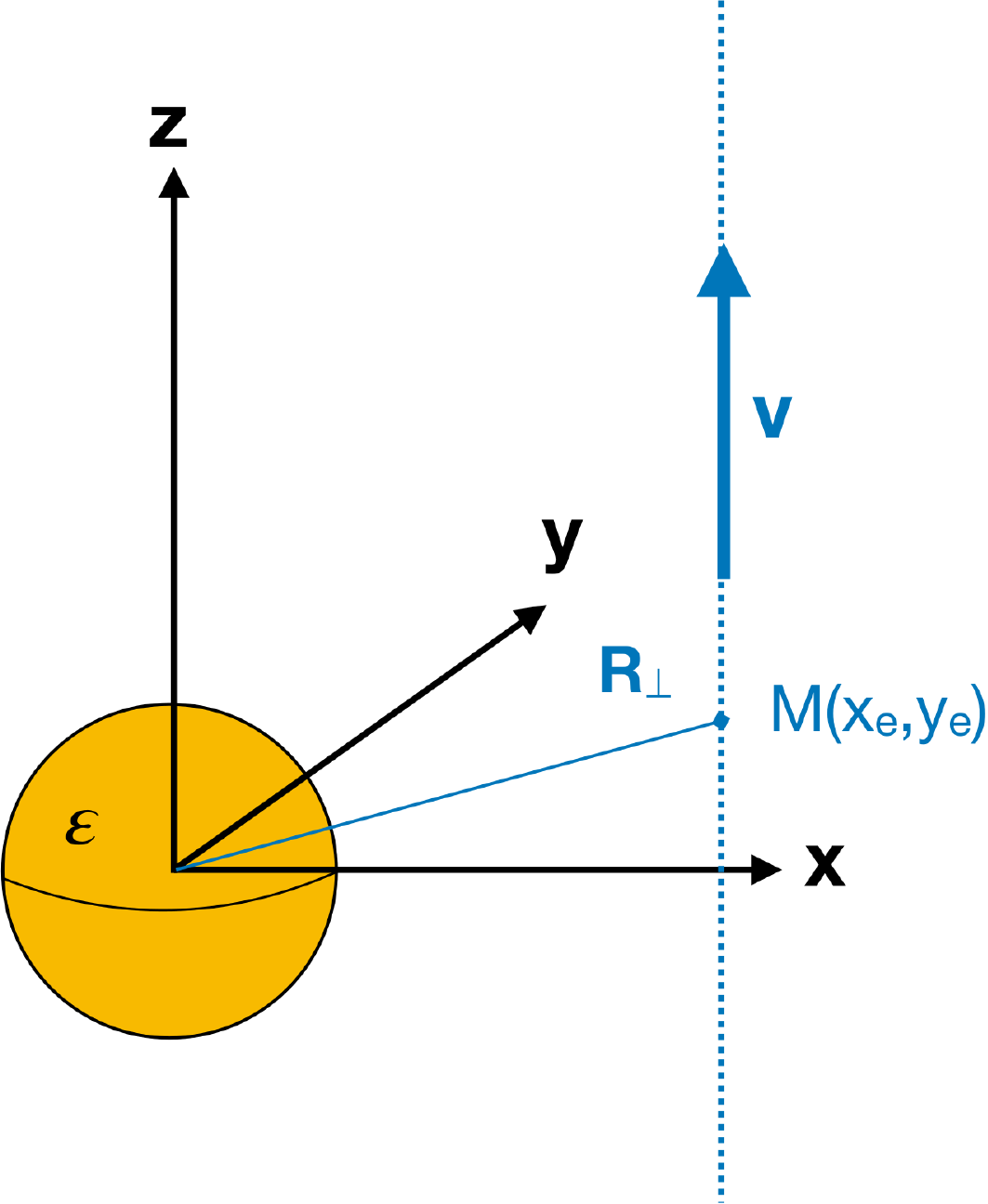}
	\caption{A fast electron is incident on a nanostructure and crosses the (OXY) plane at ${\bf R_{\perp}} = (x_e, y_e )$.
		We compute the electron energy losses experienced by the swift particle in the framework of the Green Dyadic Method.
	}
	\label{fig:EELS_sim_sketch}
\end{figure}

We consider in the following the situation represented in Figure~\ref{fig:EELS_sim_sketch}. 
A fast electron with charge ($-e$) is traveling along the (OZ) axis towards positive z i.e ${\bf v} = + v {\bf e_z}$.
The electron passes in the vicinity of a nanostructure.
It is incident normally on the sample plane at coordinates ${\bf R_{\perp}} = (x_e, y_e )$.

Classically, the energy losses experienced by a fast electron are interpreted as follows :
(i) A moving charge is the source of an electromagnetic field which polarizes the neighbouring nano-objects. 
(ii) In return, the polarized nano-objects radiate an electric field which acts on the electron movement.
The energy losses are a due to the work of the Lorentz force on the moving charge.
The loss probability per incident electron per unit angular frequency is given by the following formula \cite{garciadeabajoOpticalExcitationsElectron2010}:
\begin{equation}
		\Gamma(\omega) = \frac{e}{\hbar \pi \omega} \int \Re \Big ( e^{-\imath \omega t} {\bf v}.{\bf E_{ind}}({\bf r}(t), \omega) \Big   ) dt
\end{equation}

$\Gamma(\omega)$ can be directly computed by (i) repropagating the electric field at each point of the electron trajectory and (ii) summing up the different contributions to the energy losses along the trajectory.
In a previous version of our simulation tools, this was done using adaptative discretization of the trajectory of the electron \cite{arbouetElectronEnergyLosses2014}.
In the following, we derive a concise expression for the energy losses in the framework of the Green Dyadic Method.
This expression is now used in the EELS function of the pyGDM toolbox enabling rapid computations of EELS spectra and maps.

\subsection{Electron Energy Losses using the Green Dyadic Method}
The following alternative strategy can be adopted to obtain $\Gamma(\omega)$.
The induced electric field can be expressed as a function of the vacuum propagator $S_0$ and the polarization distribution inside the nanoparticle $\bf{p}(\bf{r'},\omega)$~\cite{girardFieldsNanostructures2005}:

\begin{eqnarray*}
	\Gamma(\omega) & =  & \frac{e}{\hbar \pi \omega} \int \Re \left ( e^{- \imath \omega t} {\bf v}.{\bf E_{ind}({\bf r}(t), \omega}) \right  ) dt \\
	& = & \frac{e}{\hbar \pi \omega} \int_{-\infty}^{+\infty} \Re \Bigg \{ e^{-\imath \omega t} {\bf v}. \\
	&&\left ( \int_{V'} 
	S_0 ({\bf r}(t), {\bf r'}, \omega).{\bf p}({\bf r'},\omega) {\bf dr'} 
	\right )
	\Bigg \} dt
\end{eqnarray*}
In the latter, $V'$ is the volume of the nanoparticle.
We can then exchange the integration on time and $\bf{r'}$. 
Taking into account the fact that $z = vt$ we have:

\begin{multline*}
	\Gamma(\omega) =  \frac{e v}{\hbar \pi \omega} \Re \Bigg \{ \int_{V'}  {\bf e_z}. \\
	 \left ( \int_t    e^{- \imath \omega z(t)/v}
	S_0 (({\bf R},z(t)), {\bf r'}, \omega) dt \right ).
	 {\bf p}({\bf r'},\omega) {\bf dr'} 
	\Bigg \}
\end{multline*}

The electric-electric vacuum Green's tensor is given by (Eq.~(\ref{eq:props_1})):
\begin{multline}
	S_0 ({\bf r}, {\bf r'}, \omega) = \mathbf{G}_{0}^{\text{EE}}(\mathbf{r}, \mathbf{r}', \omega)= 
	\\
	\left \{ k_0^2  \texttt{I} + \frac{1}{\varepsilon} \nabla \nabla \right \} \frac{e^{\imath k | {\bf r} - {\bf r'}|}}{| {\bf r} - {\bf r'}|}
\end{multline}

which leads to:

\begin{multline*}
	\Gamma(\omega)  =   \frac{e}{\hbar \pi \omega} \Re \Bigg \{ \int_{V'}  {\bf e_z}. \\
	\left ( \int_z    e^{- \imath \omega z/v}
	\left \{ k_0^2 \texttt{I}  + \frac{1}{\varepsilon} \nabla \nabla \right \} \frac{e^{\imath k | {\bf r} - {\bf r'}|}}{| {\bf r} - {\bf r'}|}|
	d z\right ).\\
	{\bf p}({\bf r'}, \omega) {\bf dr'}  
	\Bigg \}
\end{multline*}

or more explicitly: 

\begin{multline*}
	\Gamma(\omega)  =   \frac{e}{\hbar \pi \omega} \Re \Bigg \{ \int_{V'}   \Bigg ( \int_z    e^{- \imath \omega z/v}
	{\bf e_z}. \\
	\begin{bmatrix}
		k_0^2 +  \frac{1}{\varepsilon} \frac{\partial^2}{\partial x^2}  & \frac{1}{\varepsilon} \frac{\partial^2}{\partial x \partial y }  &    \frac{1}{\varepsilon}  \frac{\partial^2}{\partial x \partial z } \\
		\frac{1}{\varepsilon}  \frac{\partial^2}{\partial y \partial x } &  k_0^2 +   \frac{1}{\varepsilon} \frac{\partial^2}{\partial y^2}  &  \frac{1}{\varepsilon}  \frac{\partial^2}{\partial y \partial z }  \\
		\frac{1}{\varepsilon}  \frac{\partial^2}{\partial z \partial x }   &    \frac{1}{\varepsilon}  \frac{\partial^2}{\partial z \partial y }   &   k_0^2 +  \frac{1}{\varepsilon} \frac{\partial^2}{\partial z^2} 
	\end{bmatrix}
\\
	\frac{e^{\imath k | {\bf r} - {\bf r'}|}}{| {\bf r} - {\bf r'}|}
	d z.
	\begin{bmatrix}
		p_x({\bf r'}, \omega) \\
		p_y({\bf r'}, \omega)\\
		p_z({\bf r'}, \omega)
	\end{bmatrix}
	\Bigg )
	{\bf dr'}  
	\Bigg\}
\end{multline*}

The term between parenthesis can be expressed as:
\begin{eqnarray*}
	A  &=&  A_1 + A_2 + A_3 + A_4 \\
	&=& \frac{1}{\varepsilon}  p_x \int_{-\infty}^{+\infty} e^{- \imath \omega z/v} \frac{\partial^2}{\partial z \partial x}\left(\frac{e^{\imath k R}}{R}\right) dz \\
	&&+\   \frac{1}{\varepsilon}  p_y \int_{-\infty}^{+\infty} e^{- \imath \omega z/v} \frac{\partial^2}{\partial z \partial y}\left(\frac{e^{\imath k R}}{R}\right) dz \\
	&&+\ k_0^2  p_z \int_{-\infty}^{+\infty} e^{- \imath \omega z/v} \frac{e^{\imath k R}}{R} dz \\
	&&+\ \frac{1}{\varepsilon}  p_z \int_{-\infty}^{+\infty} e^{- \imath \omega z/v} \frac{\partial^2}{\partial z^2}\left(\frac{e^{\imath k R}}{R}\right) dz
\end{eqnarray*}

Each of these 4 terms $A_i$ can be calculated by integration by parts and connected to the following integral:
\begin{equation*}
	I = \int_{-\infty}^{+\infty} e^{- \imath \omega z/v} \frac{e^{\imath k R}}{R} dz 
\end{equation*}
which gives
\begin{equation}
		I = 2 e^{- \imath \omega z'/v} K_0 \left(\sqrt{ \left (\frac{\omega}{v} \right )^2 - k^2} \; |{\bf R} - {\bf R'}| \right)
\end{equation}
with $R = |{\bf r} - {\bf r'}|$ and $K_n$ the modified Bessel function of the second kind and order $n$.
We get the following expressions for the $A_i$ terms:
\begin{align}
	A_1 &=  
	\begin{multlined}[t][0.7\columnwidth]- 2 \, \imath \, e^{- \imath \omega z'/v}  \frac{\omega^2}{\varepsilon \gamma v^2}  \,\frac{x - x'}{ |{\bf R} - {\bf R'}|} \\
	K_1 (k_{\rho} \; |{\bf R} - {\bf R'}| ) \, p_x 
	\end{multlined}\\
	A_2 &= 
	\begin{multlined}[t][0.7\columnwidth] - 2 \, \imath \, e^{- \imath \omega z'/v}  \frac{\omega^2}{\varepsilon \gamma v^2}  \,\frac{y - y'}{ |{\bf R} - {\bf R'}|} \\
	K_1 (k_{\rho} \; |{\bf R} - {\bf R'}| ) \, p_y 
	\end{multlined}\\
	A_3 &= 2 k_0^2  e^{- \imath \omega z'/v} K_0 (k_{\rho} \; |{\bf R} - {\bf R'}| ) \,  p_z  \\
	A_4 &= -2  \left(\frac{\omega^2}{ \varepsilon v^2} \right) 
	e^{- \imath \omega z'/v} K_0 (k_{\rho} \; |{\bf R} - {\bf R'}| )  \, p_z
\end{align}
with $k_z = \omega / v$, $k_{\rho} = \sqrt{k_z^2 - k^2}$, $k = k_0 \sqrt{\varepsilon}$, $\gamma = 1/\sqrt{1-\varepsilon v^2/ c^2}$ and $k_{\rho} = \omega/v \gamma$ .

After some algebra we obtain finally the following expression for the energy losses:

%
\begin{multline*}
	\Gamma(\omega)  =   \frac{2 e}{\hbar \pi \omega} \Re \Bigg \{ \int_{V'}   
	e^{- \imath \omega z'/v} 
	\Bigg[
	\\
	- \Big(\frac{\omega^2}{ \varepsilon \gamma^2 v^2}\Big) K_0 (k_{\rho} \; |{\bf R} - {\bf R'}| ) p_z  \\
	 - \imath \frac{\omega^2}{\varepsilon \gamma v^2} 
	\, \frac{p_x (x - x') + p_y (y - y') }{ |{\bf R} - {\bf R'}|} 
	\\
	K_1 (k_{\rho} \; |{\bf R} - {\bf R'}| )
	\Bigg]
	{\bf dr'}  
	\Bigg \}  \nonumber
\end{multline*}
In GDM, the used Fourier transform is defined with a $2 \pi$ factor difference from the convention used for instance by J. Garcia de Abajo (see \cite{arbouetElectronEnergyLosses2014} for more information).
The latter defines the Fourier transform as:
\begin{eqnarray*}
	E(r,t) &=& \int \frac{d \omega}{2 \pi}E(r,\omega) e^{-\imath \omega t} \\
	E(r,\omega) &=& \int dt E(r,t) e^{\imath \omega t}
\end{eqnarray*}
whereas in the GDM it is defined as:
\begin{eqnarray*}
	E(r,t) &=& \int d \omega E(r,\omega) e^{-\imath \omega t} \\
	E(r,\omega) &=& \frac{1}{2\pi} \int dt E(r,t) e^{\imath \omega t}
\end{eqnarray*}
The formula for the loss probability per unit frequency that is implemented in pyGDM is therefore:
%
\begin{multline}
	\Gamma(\omega) = \frac{4 e}{\hbar  \omega} \Re \Bigg \{ \int_{V'}   
	e^{- \imath \omega z'/v} 
	\Bigg[
	\\
	- \left(\frac{\omega^2}{ \varepsilon \gamma^2 v^2}\right) K_0 (k_{\rho} \; |{\bf R} - {\bf R'}| ) p_z \\
	 -  \imath \frac{\omega^2}{\varepsilon \gamma v^2} 
	\, \frac{p_x (x - x') + p_y (y - y') }{ |{\bf R} - {\bf R'}|} 
	\\
	K_1 (k_{\rho} \; |{\bf R} - {\bf R'}| )
	\Bigg]
	{\bf dr'} 
	\Bigg \}
	\label{eq:EELS-proba}
\end{multline}
It connects the electron energy losses to the three components of the polarization induced inside the nanostructure by the moving charge. 

Please note that equation~(\ref{eq:EELS-proba}) diverges for meshpoints on the beam trajectory because of the divergence of the modified Bessel function of the second kind (see also~\cite{arbouetElectronEnergyLosses2014}). 
To calculate equation~(\ref{eq:EELS-proba}) for an electron beam passing through the structure, we therefore simply remove all meshpoints from the numerical integration, that are closer to the beam than one discretization step. 
Furthermore, in order to avoid numerical noise due to the discrete mesh when scanning the beam above the structure, we added a new function \texttt{tools.adapt\_map\_to\_structure\_mesh} to pyGDM, which adjusts all $(X, Y)$  raster-scan positions onto the closest mesh coordinate. We recommend using this tool for EELS and CL raster-scan simulations. The example scripts contain an according description.

We also want to mention that electron propagation through a lossy material yields additional losses, the probability of which is described by the so-called ``bulk loss probability'' $\Gamma_{\text{bulk}}(\omega)$, that is proportional to the thickness of the traversed material and to the imaginary part of $1/\epsilon(\omega)$. 
This contribution is not taken into account in the pyGDM simulations, but it could be straightforwardly computed, if required. More details can be found in Refs.~\cite{garciadeabajoOpticalExcitationsElectron2010, bernasconiWhereDoesEnergy2017}.

In the next section we give a simpler expression of this result for a dipolar particle.

\subsection{Case of a very small particle}

We demonstrate the calculation now for the case of a small nanoparticle that can be approximated by a polarizability $\alpha$ located at the origin. Then we can write the components of the polarization induced inside this nanoparticle by an electron beam crossing the (OXY) plane at location $(d,0)$ as:
\begin{equation*}
	{\bf p} = \alpha \, {\bf E_{el}} \, .
\end{equation*}
The electric field ${\bf E_{el}}$ created at the location of the nano-object by a swift electron travelling along (OZ) towards positive z  is:
\begin{eqnarray*}
	p_x &=&  \frac{e \omega}{\pi v^{2} \gamma \varepsilon} \alpha K_1 \left(\frac{\omega d}{\gamma v} \right) \\
	p_y &=& 0 \\
	p_z &=&  \imath \frac{e \omega}{\pi v^{2} \gamma^2 \varepsilon} \alpha K_0 \left(\frac{\omega d}{\gamma v}\right)
\end{eqnarray*}
Inserting these expressions in formula (\ref{eq:EELS-proba}) yields:
%
%
\begin{eqnarray*}
	\Gamma(\omega)   &=&    \frac{4 e}{\hbar  \omega} \Re \Bigg \{ 
	\Bigg[
	- \left(\frac{\omega^2}{ \varepsilon \gamma^2 v^2}\right)  K_0 (k_{\rho} \;d) p_z \\
	& & \hspace{2cm}- 
	\imath \frac{\omega^2}{\varepsilon \gamma v^2} 
	\, (p_x 
	K_1 (k_{\rho} \; d))
	\Bigg]
	\Bigg \}  \\ \nonumber
	&=& 
	\frac{4 e}{\hbar  \omega} \Re \Bigg \{ 
	\Bigg[- \imath \frac{e \omega}{\pi v^{2} \gamma^2 \varepsilon} \alpha
	\left(\frac{\omega^2}{ \varepsilon \gamma^2 v^2}\right)  K_0^2 (k_{\rho} \;d)  \\
	&&  \hspace{2cm} - \imath \frac{e \omega}{\pi v^{2} \gamma \varepsilon} \alpha
	\frac{\omega^2}{\varepsilon \gamma v^2} \, K_1^2 (k_{\rho} \; d)
	\Bigg]
	\Bigg \}  
\end{eqnarray*}

\begin{multline*}
	\Gamma(\omega)   =  
	\frac{1}{\pi \hbar} \left(\frac{2 e \omega}{v^2 \gamma \varepsilon} \right)^2  \frac{K_0^2 (k_{\rho} \; d)} {\gamma^2} \Im (\alpha) 
	\\ 
	+ \frac{1}{\pi \hbar} \left(\frac{2 e \omega}{v^2 \gamma \varepsilon} \right)^2  K_1^2 (k_{\rho} \; d)  \Im (\alpha)
\end{multline*}
Which finally leads to
\begin{multline}
		\Gamma(\omega)   =   
		\frac{1}{\pi \hbar} \left(\frac{2 e \omega}{v^2 \gamma \varepsilon} \right)^2   \Im (\alpha)
		\Bigg\{
		K_1^2 (k_{\rho} \; d) + \\
		\frac{K_0^2 (k_{\rho} \; d)} {\gamma^2} 
		\Bigg\}
\end{multline}

We see here that assuming a dipolar response for the nanoparticle in formula (\ref{eq:EELS-proba}) yields identical results as formulas (34) and (35) of reference \cite{garciadeabajoOpticalExcitationsElectron2010}.

The computation of cathodoluminescence spectra and maps follows the same lines as detailed in reference \cite{arbouetElectronEnergyLosses2014}.
In short, pyGDM computes the power radiated by the polarization distribution, induced inside the sample by the moving charge from the integral of the Poynting vector on a sphere, centered on the sample.

%% file: Appendix_technical_details.tex
\section{Technical details}\label{sec:technical_detail}

\subsection{Simulation size}
The maximum number of discretization cells depends on the available RAM. 
The memory requirement scales with the square of the discretization points.
8GB of RAM are enough to treat around 7000 meshpoints.
32GB can host around 15k meshpoints, and 128GB will suffice for around 30k meshpoints.
The computation time scales with the cube of the number of cells, therefore for larger structures the simulation time will become the bottleneck in systems with enough memory. For instance, a model with 25k meshpoints takes several days per wavelength on a recent 36 core intel server CPU. In consequence the number of meshpoints for practical simulations is limited to \(\approx 10000-15000\).

\subsection{Dependencies}

\pygdm\ is written in pure python. 
It uses several third-party tools and libraries for acceleration, processing and visualization.
All dependencies are exclusively open-source and can be installed via ``pip'':
\paragraph{Required third party python packages}

\begin{itemize}
	\setlength\itemsep{.1em}
	\item ``numba'' (just-in-time compiler for acceleration)
	\item ``numpy'' (numerical stack)
	\item ``scipy'' (various efficient scientific algorithm)
\end{itemize}

\paragraph{Optional python packages}

\begin{itemize}
	\setlength\itemsep{.1em}
	\item ``matplotlib'' (for 2D visualizations)
	\item ``mayavi'' (for 3D visualizations)
	\item ``cupy'' (CUDA-based GPU solver)
	\item ``pytables'' (to save simulations results in the efficient hdf5 data format)
	\item ``mpi4py'' (for MPI parallelized spectra calculation)
	\item ``pyGMO'' (required by the evolutionary optimization submodule)
	\item ``PIL'' (required by some image processing tools)
\end{itemize}

\subsection{Compiling, installation}\label{sec:installation}

We provide the latest stable version via the pypi repository, installation is therefore easiest via ``pip'':

\vspace{.5\baselineskip}
\indent\hspace{1cm}\texttt{pip install -U pygdm2}

To install the latest development version, download the source code from the \href{}{gitlab repository} and run in a terminal in the \pygdm\ sources root directory

\vspace{.5\baselineskip}
\indent\hspace{1cm}\texttt{pip install -U .}

\vspace{.5\baselineskip}
\noindent
Attention, the trailing point "." is significant in the command, pointing to the current working directory.
For a local installation, add ``\texttt{{-}{-}user}''.

\vspace{.5\baselineskip}
%
%
%
%

\subsection*{Compiling, installation of the retarded Green's dyads module}\label{sec:installation_retard}

The retarded dyads module \href{https://pypi.org/project/pyGDM2-retard/0.2/}{\textsf{pyGDM2-retard}} depends on fortran based binaries and is available as an external python package in order to avoid a binary dependency in the main \pygdm\ package. For windows we provide pre-compiled packages of the latest stable version. On other operating systems like linux the package needs to be compiled. 
Installation is easiest via ``pip'':

\vspace{.5\baselineskip}
\indent\hspace{1cm}\texttt{pip install -U pyGDM2-retard}
\vspace{.5\baselineskip}

\noindent
If no binaries are available on your platform, the code must be locally compiled. If the gfortran compiler is correctly installed this should be done automatically by \textsf{pip} and installation should be possible with the above command. Details and tips concerning the installation can be found in the \href{https://wiechapeter.gitlab.io/pyGDM2-doc/readme.html}{\pygdm\ online documentation}.

\subsection*{Installation of the graphical user interface pyGDM-UI}\label{sec:installation_pyGDM-UI}

We currently work on a \href{https://wiechapeter.gitlab.io/pyGDM2-doc/pygdmui.html}{graphical user interface for pyGDM} based on Qt, it therefore requires \href{https://pypi.org/project/PyQt5/}{pyqt5}.
Currently, also \href{http://docs.enthought.com/mayavi/mayavi/mlab.html}{mayavi} is required, as it is used for several visualizations.
pyGDM-UI is still under development but we provide a working beta-version. It can be installed \href{https://pypi.org/project/pygdmUI/}{from pypi} via pip 

\vspace{.5\baselineskip}
\indent\hspace{1cm}\texttt{pip install -U pygdmUI}
\vspace{.5\baselineskip}

\noindent

\subsection{Minimum working example script}

\begin{figure}[t!]
	\centering
	\includegraphics[width=\linewidth]{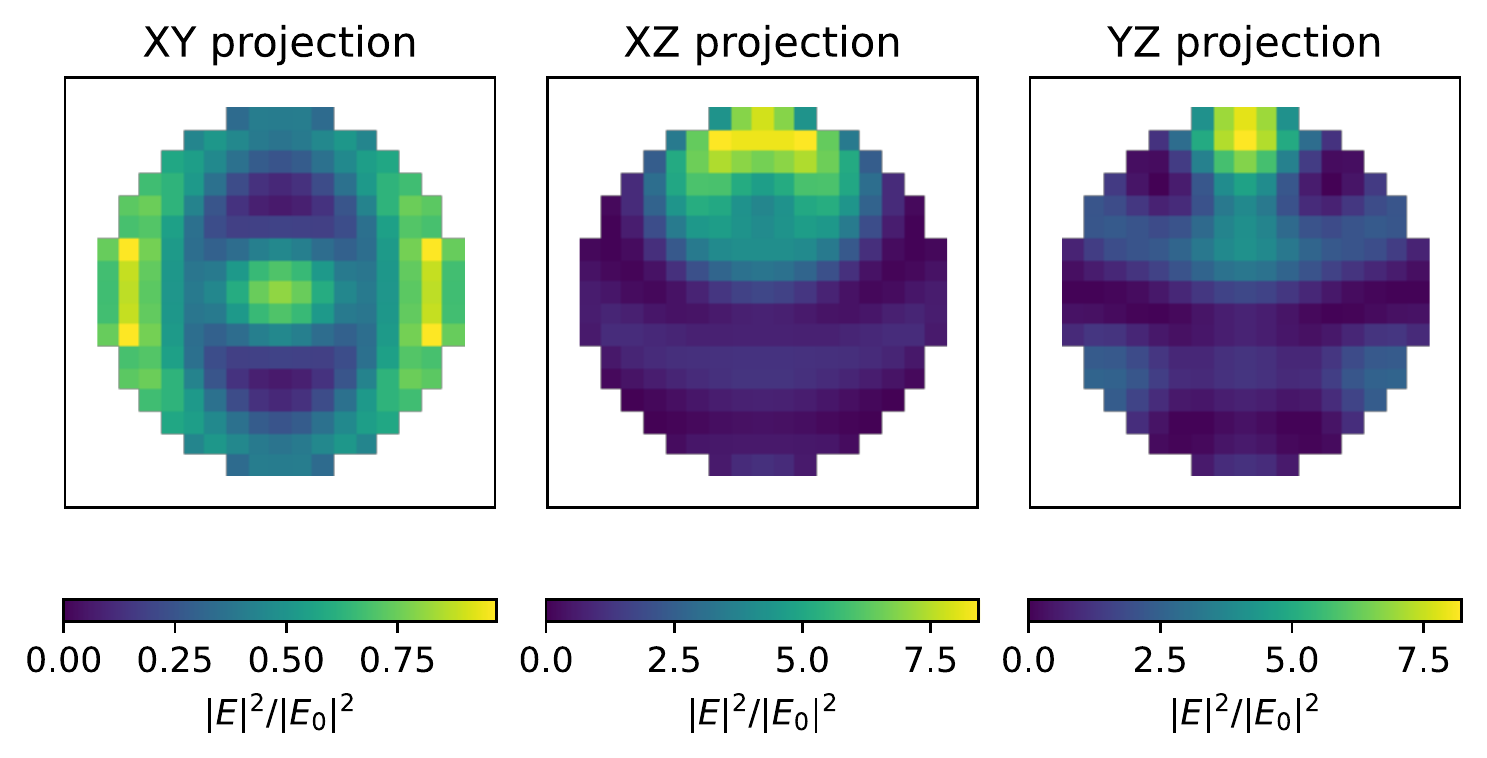}
	\caption{
		Graphical output of the example script given in listing~\ref{lst:simple_example_script}. 
		A dielectric sphere ($n=2$, $r=160$\,nm) in vacuum is illuminated from below by a  plane wave of linear polarization along $X$ and wavelength $\lambda_0=400\,$nm.
		Each plot shows the E-field intensity inside the sphere on a slice through its center on an area of $400\times 400\,$nm$^2$.
	}\label{fig:example_script_output}
\end{figure}
%

\begin{lstlisting}[language=Python, caption={Minimum example script. The plots generated by the script are shown in Fig.~\ref{fig:example_script_output}.}, label={lst:simple_example_script}]
	from pyGDM2 import structures
	from pyGDM2 import materials
	from pyGDM2 import fields
	from pyGDM2 import propagators
	from pyGDM2 import core
	from pyGDM2 import visu
	
	## --- simulation setup ---
	## structure: sphere of 160nm radius,
	## constant dielectric function (n=2),
	step = 20
	geometry = structures.sphere(step, R=8.2, mesh='cube')
	material = materials.dummy(2.0)
	struct = structures.struct(step, geometry, material)
	
	## incident field: plane wave, 400nm, lin. pol. || x
	field_generator = fields.plane_wave
	wavelengths = [400]   # nm
	kwargs = dict(inc_angle=0, inc_plane='xz', theta=0)
	efield = fields.efield(field_generator, 
	wavelengths=wavelengths, kwargs=kwargs)
	
	## environment: vacuum
	n1 = 1.0
	dyads = propagators.DyadsQuasistatic123(n1)
	
	## create simulation object
	sim = core.simulation(struct, efield, dyads)

	## --- run the simulation ---
	core.scatter(sim)

	## --- plot nearfield intensity inside sphere ---
	## using first (of one) field-config (=index 0)
	## slice through sphere center
	visu.vectorfield_color_by_fieldindex(sim, 0, 
							projection='XY', slice_level=160)
	visu.vectorfield_color_by_fieldindex(sim, 0, 
							projection='XZ', slice_level=0)
	visu.vectorfield_color_by_fieldindex(sim, 0, 
							projection='YZ', slice_level=0)
	
\end{lstlisting}

\subsection{Technical documentation of main classes and functions}

For a detailed documentation of all functions and classes as well as many commented examples and tutorials, please visit  the online documentation \href{https://wiechapeter.gitlab.io/pyGDM2-doc}{https://wiechapeter.gitlab.io/pyGDM2-doc}.
We provide here merely a very brief description of the core classes and functions.


\subsection*{Main classes:}

\functiondescription{o}{core.simulation}{}{
	{\textit{struct:} instance of \object{structures.struct}}
	{\textit{efield:} instance of \object{fields.efield}}
	{\textit{dyads:} class defining simulation environment}
}

\functiondescription{o}{structures.struct}{}{
	{\textit{step:} discretization stepsize (in nm)}
	{\textit{geometry:} list of meshpoint coordinates \((x,y,z)\) (in nm)}
	{\textit{material:} structure material dispersion, instance of \object{materials.CLASS} or list of materials (one for each meshpoint)}
}

\functiondescription{o}{fields.efield}{}{
	{\textit{field\_generator:} field generator function (e.g. from \function{fields} module)}
	{\textit{wavelengths:} list of wavelengths at which to do the simulation (in nm)}
	{\textit{kwargs (optional):} dict (or list of dict) with further kwargs for the field generator}
}

\functiondescription{o}{propagators.DyadsQuasistatic123}{}{
	{\textit{n1:} refractive index for layer 1 (substrate)}
	{\textit{n2:} ref. index for layer 2 (center layer)}
	{\textit{n3:} ref. index for layer 3 (top cladding)}
	{\textit{spacing:} thickness of layer 2}
	{\textit{radiative\_correction:} whether or not to use the radiative correction term}
}

\subsection*{Main core functions:}

\functiondescription{f}{core.scatter / core.scatter\_mpi}{}{
	{\textit{sim:} instance of \object{core.simulation}}
	{\textit{method (optional):} inversion method, default: ``lu''}
	{\textit{calc\_H (optional):} calculate internal magnetic field, default: ``False''}
}

\functiondescription{f}{core.decay\_rate}{}{
	{\textit{sim:} instance of \object{core.simulation}}
	{\textit{wavelength} target wavelength}
	{\textit{r\_probe} list of coordinates to evaluate}
	{\textit{method (optional):} inversion method, default: ``scipyinv''}
}

%% file: 2021_wiecha_pygdm_update.bbl
\begin{thebibliography}{10}
\expandafter\ifx\csname url\endcsname\relax
  \def\url#1{\texttt{#1}}\fi
\expandafter\ifx\csname urlprefix\endcsname\relax\def\urlprefix{URL }\fi
\providecommand{\bibinfo}[2]{#2}
\providecommand{\eprint}[2][]{\url{#2}}

\bibitem{maxwellDynamicalTheoryElectromagnetic1865}
\bibinfo{author}{Maxwell, J.~C.}
\newblock \bibinfo{title}{A {{Dynamical Theory}} of the {{Electromagnetic
  Field}}}.
\newblock \emph{\bibinfo{journal}{Philosophical Transactions of the Royal
  Society of London}} \textbf{\bibinfo{volume}{155}}, \bibinfo{pages}{459--512}
  (\bibinfo{year}{1865}).

\bibitem{oskooiMEEPFlexibleFreesoftware2010}
\bibinfo{author}{Oskooi, A.~F.} \emph{et~al.}
\newblock \bibinfo{title}{{{MEEP}}: {{A}} flexible free-software package for
  electromagnetic simulations by the {{FDTD}} method}.
\newblock \emph{\bibinfo{journal}{Computer Physics Communications}}
  \textbf{\bibinfo{volume}{181}}, \bibinfo{pages}{687--702}
  (\bibinfo{year}{2010}).

\bibitem{draineDiscretedipoleApproximationScattering1994}
\bibinfo{author}{Draine, B.~T.} \& \bibinfo{author}{Flatau, P.~J.}
\newblock \bibinfo{title}{Discrete-dipole approximation for scattering
  calculations}.
\newblock \emph{\bibinfo{journal}{Journal of the Optical Society of America A}}
  \textbf{\bibinfo{volume}{11}}, \bibinfo{pages}{1491} (\bibinfo{year}{1994}).

\bibitem{chaumetCoupleddipoleMethodMagnetic2009}
\bibinfo{author}{Chaumet, P.~C.} \& \bibinfo{author}{Rahmani, A.}
\newblock \bibinfo{title}{Coupled-dipole method for magnetic and
  negative-refraction materials}.
\newblock \emph{\bibinfo{journal}{Journal of Quantitative Spectroscopy and
  Radiative Transfer}} \textbf{\bibinfo{volume}{110}}, \bibinfo{pages}{22--29}
  (\bibinfo{year}{2009}).

\bibitem{hohenesterMNPBEMMatlabToolbox2012}
\bibinfo{author}{Hohenester, U.} \& \bibinfo{author}{Tr{\"u}gler, A.}
\newblock \bibinfo{title}{{{MNPBEM}} \textendash{} {{A Matlab}} toolbox for the
  simulation of plasmonic nanoparticles}.
\newblock \emph{\bibinfo{journal}{Computer Physics Communications}}
  \textbf{\bibinfo{volume}{183}}, \bibinfo{pages}{370--381}
  (\bibinfo{year}{2012}).

\bibitem{garciadeabajoOpticalExcitationsElectron2010}
\bibinfo{author}{{Garc{\'i}a de Abajo}, F.~J.}
\newblock \bibinfo{title}{Optical excitations in electron microscopy}.
\newblock \emph{\bibinfo{journal}{Reviews of Modern Physics}}
  \textbf{\bibinfo{volume}{82}}, \bibinfo{pages}{209--275}
  (\bibinfo{year}{2010}).
\newblock \eprint{0903.1669}.

\bibitem{demesyAllpurposeFiniteElement2010}
\bibinfo{author}{Dem{\'e}sy, G.}, \bibinfo{author}{Zolla, F.},
  \bibinfo{author}{Nicolet, A.} \& \bibinfo{author}{Commandr{\'e}, M.}
\newblock \bibinfo{title}{All-purpose finite element formulation for
  arbitrarily shaped crossed-gratings embedded in a multilayered stack}.
\newblock \emph{\bibinfo{journal}{JOSA A}} \textbf{\bibinfo{volume}{27}},
  \bibinfo{pages}{878--889} (\bibinfo{year}{2010}).

\bibitem{hoffmannComparisonElectromagneticField2009}
\bibinfo{author}{Hoffmann, J.}, \bibinfo{author}{Hafner, C.},
  \bibinfo{author}{Leidenberger, P.}, \bibinfo{author}{Hesselbarth, J.} \&
  \bibinfo{author}{Burger, S.}
\newblock \bibinfo{title}{Comparison of electromagnetic field solvers for the
  {{3D}} analysis of plasmonic nanoantennas}.
\newblock In \emph{\bibinfo{booktitle}{Modeling {{Aspects}} in {{Optical
  Metrology II}}}}, vol. \bibinfo{volume}{7390}, \bibinfo{pages}{73900J}
  (\bibinfo{publisher}{{International Society for Optics and Photonics}},
  \bibinfo{year}{2009}).

\bibitem{barchiesiComputingOpticalNearfield1996}
\bibinfo{author}{Barchiesi, D.}, \bibinfo{author}{Girard, C.},
  \bibinfo{author}{Martin, O. J.~F.}, \bibinfo{author}{Van~Labeke, D.} \&
  \bibinfo{author}{Courjon, D.}
\newblock \bibinfo{title}{Computing the optical near-field distributions around
  complex subwavelength surface structures: {{A}} comparative study of
  different methods}.
\newblock \emph{\bibinfo{journal}{Physical Review E}}
  \textbf{\bibinfo{volume}{54}}, \bibinfo{pages}{4285--4292}
  (\bibinfo{year}{1996}).

\bibitem{parsonsComparisonTechniquesUsed2010}
\bibinfo{author}{Parsons, J.}, \bibinfo{author}{Burrows, C.~P.},
  \bibinfo{author}{Sambles, J.~R.} \& \bibinfo{author}{Barnes, W.~L.}
\newblock \bibinfo{title}{A comparison of techniques used to simulate the
  scattering of electromagnetic radiation by metallic nanostructures}.
\newblock \emph{\bibinfo{journal}{Journal of Modern Optics}}
  \textbf{\bibinfo{volume}{57}}, \bibinfo{pages}{356--365}
  (\bibinfo{year}{2010}).

\bibitem{gallinetNumericalMethodsNanophotonics2015}
\bibinfo{author}{Gallinet, B.}, \bibinfo{author}{Butet, J.} \&
  \bibinfo{author}{Martin, O. J.~F.}
\newblock \bibinfo{title}{Numerical methods for nanophotonics: Standard
  problems and future challenges}.
\newblock \emph{\bibinfo{journal}{Laser \& Photonics Reviews}}
  \textbf{\bibinfo{volume}{9}}, \bibinfo{pages}{577--603}
  (\bibinfo{year}{2015}).

\bibitem{kuznetsovOpticallyResonantDielectric2016}
\bibinfo{author}{Kuznetsov, A.~I.}, \bibinfo{author}{Miroshnichenko, A.~E.},
  \bibinfo{author}{Brongersma, M.~L.}, \bibinfo{author}{Kivshar, Y.~S.} \&
  \bibinfo{author}{Luk'yanchuk, B.}
\newblock \bibinfo{title}{Optically resonant dielectric nanostructures}.
\newblock \emph{\bibinfo{journal}{Science}} \textbf{\bibinfo{volume}{354}}
  (\bibinfo{year}{2016}).

\bibitem{muhlschlegelResonantOpticalAntennas2005}
\bibinfo{author}{M{\"u}hlschlegel, P.}, \bibinfo{author}{Eisler, H.-J.},
  \bibinfo{author}{Martin, O. J.~F.}, \bibinfo{author}{Hecht, B.} \&
  \bibinfo{author}{Pohl, D.~W.}
\newblock \bibinfo{title}{Resonant {{Optical Antennas}}}.
\newblock \emph{\bibinfo{journal}{Science}} \textbf{\bibinfo{volume}{308}},
  \bibinfo{pages}{1607--1609} (\bibinfo{year}{2005}).

\bibitem{bharadwajOpticalAntennas2009}
\bibinfo{author}{Bharadwaj, P.}, \bibinfo{author}{Deutsch, B.} \&
  \bibinfo{author}{Novotny, L.}
\newblock \bibinfo{title}{Optical {{Antennas}}}.
\newblock \emph{\bibinfo{journal}{Advances in Optics and Photonics}}
  \textbf{\bibinfo{volume}{1}}, \bibinfo{pages}{438--483}
  (\bibinfo{year}{2009}).

\bibitem{wiechaPolarizationConversionPlasmonic2017}
\bibinfo{author}{Wiecha, P.~R.} \emph{et~al.}
\newblock \bibinfo{title}{Polarization conversion in plasmonic nanoantennas for
  metasurfaces using structural asymmetry and mode hybridization}.
\newblock \emph{\bibinfo{journal}{Scientific Reports}}
  \textbf{\bibinfo{volume}{7}}, \bibinfo{pages}{40906} (\bibinfo{year}{2017}).

\bibitem{girardDesigningThermoplasmonicProperties2018}
\bibinfo{author}{Girard, C.}, \bibinfo{author}{Wiecha, P.~R.},
  \bibinfo{author}{Cuche, A.} \& \bibinfo{author}{Dujardin, E.}
\newblock \bibinfo{title}{Designing thermoplasmonic properties of metallic
  metasurfaces}.
\newblock \emph{\bibinfo{journal}{Journal of Optics}}
  \textbf{\bibinfo{volume}{20}}, \bibinfo{pages}{075004}
  (\bibinfo{year}{2018}).

\bibitem{girardShapingManipulationLight2008}
\bibinfo{author}{Girard, C.}, \bibinfo{author}{Dujardin, E.},
  \bibinfo{author}{Baffou, G.} \& \bibinfo{author}{Quidant, R.}
\newblock \bibinfo{title}{Shaping and manipulation of light fields with
  bottom-up plasmonic structures}.
\newblock \emph{\bibinfo{journal}{New Journal of Physics}}
  \textbf{\bibinfo{volume}{10}}, \bibinfo{pages}{105016}
  (\bibinfo{year}{2008}).

\bibitem{viarbitskayaTailoringImagingPlasmonic2013}
\bibinfo{author}{Viarbitskaya, S.} \emph{et~al.}
\newblock \bibinfo{title}{Tailoring and imaging the plasmonic local density of
  states in crystalline nanoprisms}.
\newblock \emph{\bibinfo{journal}{Nature Materials}}
  \textbf{\bibinfo{volume}{12}}, \bibinfo{pages}{426--432}
  (\bibinfo{year}{2013}).

\bibitem{viarbitskayaMorphologyinducedRedistributionSurface2013}
\bibinfo{author}{Viarbitskaya, S.} \emph{et~al.}
\newblock \bibinfo{title}{Morphology-induced redistribution of surface plasmon
  modes in two-dimensional crystalline gold platelets}.
\newblock \emph{\bibinfo{journal}{Applied Physics Letters}}
  \textbf{\bibinfo{volume}{103}}, \bibinfo{pages}{131112}
  (\bibinfo{year}{2013}).

\bibitem{ouldaghaNearFieldPropertiesPlasmonic2014}
\bibinfo{author}{Ould~Agha, Y.}, \bibinfo{author}{Demichel, O.},
  \bibinfo{author}{Girard, C.}, \bibinfo{author}{Bouhelier, A.} \&
  \bibinfo{author}{{Colas des Francs}, G.}
\newblock \bibinfo{title}{Near-{{Field Properties}} of {{Plasmonic
  Nanostructures}} with {{High Aspect Ratio}}}.
\newblock \emph{\bibinfo{journal}{Progress In Electromagnetics Research}}
  \textbf{\bibinfo{volume}{146}}, \bibinfo{pages}{77--88}
  (\bibinfo{year}{2014}).

\bibitem{cucheDipolarRegimeHighorder2017}
\bibinfo{author}{Cuche, A.} \emph{et~al.}
\newblock \bibinfo{title}{Beyond dipolar regime in high-order plasmon mode
  bowtie antennas}.
\newblock \emph{\bibinfo{journal}{Optics Communications}}
  \textbf{\bibinfo{volume}{387}}, \bibinfo{pages}{48--54}
  (\bibinfo{year}{2017}).

\bibitem{cucheNearfieldHyperspectralQuantum2017}
\bibinfo{author}{Cuche, A.} \emph{et~al.}
\newblock \bibinfo{title}{Near-field hyperspectral quantum probing of
  multimodal plasmonic resonators}.
\newblock \emph{\bibinfo{journal}{Physical Review B}}
  \textbf{\bibinfo{volume}{95}}, \bibinfo{pages}{121402}
  (\bibinfo{year}{2017}).

\bibitem{wiechaEvolutionaryMultiobjectiveOptimization2017}
\bibinfo{author}{Wiecha, P.~R.} \emph{et~al.}
\newblock \bibinfo{title}{Evolutionary multi-objective optimization of colour
  pixels based on dielectric nanoantennas}.
\newblock \emph{\bibinfo{journal}{Nature Nanotechnology}}
  \textbf{\bibinfo{volume}{12}}, \bibinfo{pages}{163--169}
  (\bibinfo{year}{2017}).

\bibitem{wiechaStronglyDirectionalScattering2017}
\bibinfo{author}{Wiecha, P.~R.} \emph{et~al.}
\newblock \bibinfo{title}{Strongly {{Directional Scattering}} from {{Dielectric
  Nanowires}}}.
\newblock \emph{\bibinfo{journal}{ACS Photonics}} \textbf{\bibinfo{volume}{4}},
  \bibinfo{pages}{2036--2046} (\bibinfo{year}{2017}).

\bibitem{wiechaOriginSecondharmonicGeneration2016}
\bibinfo{author}{Wiecha, P.~R.}, \bibinfo{author}{Arbouet, A.},
  \bibinfo{author}{Girard, C.}, \bibinfo{author}{Baron, T.} \&
  \bibinfo{author}{Paillard, V.}
\newblock \bibinfo{title}{Origin of second-harmonic generation from individual
  silicon nanowires}.
\newblock \emph{\bibinfo{journal}{Physical Review B}}
  \textbf{\bibinfo{volume}{93}}, \bibinfo{pages}{125421}
  (\bibinfo{year}{2016}).

\bibitem{blackTailoringSecondHarmonicGeneration2015}
\bibinfo{author}{Black, L.-J.} \emph{et~al.}
\newblock \bibinfo{title}{Tailoring {{Second}}-{{Harmonic Generation}} in
  {{Single L}}-{{Shaped Plasmonic Nanoantennas}} from the {{Capacitive}} to
  {{Conductive Coupling Regime}}}.
\newblock \emph{\bibinfo{journal}{ACS Photonics}} \textbf{\bibinfo{volume}{2}},
  \bibinfo{pages}{1592--1601} (\bibinfo{year}{2015}).

\bibitem{wiechaLinearNonlinearOptical2016}
\bibinfo{author}{Wiecha, P.~R.}
\newblock \emph{\bibinfo{title}{Linear and Nonlinear Optical Properties of High
  Refractive Index Dielectric Nanostructures}}.
\newblock \bibinfo{type}{{{PhD}} thesis}, \bibinfo{school}{Universit\'e de
  Toulouse, Universit\'e Toulouse III - Paul Sabatier} (\bibinfo{year}{2016}).

\bibitem{lamNumbaLLVMbasedPython2015a}
\bibinfo{author}{Lam, S.~K.}, \bibinfo{author}{Pitrou, A.} \&
  \bibinfo{author}{Seibert, S.}
\newblock \bibinfo{title}{Numba: A {{LLVM}}-based {{Python JIT}} compiler}.
\newblock In \emph{\bibinfo{booktitle}{Proceedings of the {{Second Workshop}}
  on the {{LLVM Compiler Infrastructure}} in {{HPC}}}}, {{LLVM}} '15,
  \bibinfo{pages}{1--6} (\bibinfo{publisher}{{Association for Computing
  Machinery}}, \bibinfo{address}{{New York, NY, USA}}, \bibinfo{year}{2015}).

\bibitem{martinGeneralizedFieldPropagator1995}
\bibinfo{author}{Martin, O. J.~F.}, \bibinfo{author}{Girard, C.} \&
  \bibinfo{author}{Dereux, A.}
\newblock \bibinfo{title}{Generalized {{Field Propagator}} for
  {{Electromagnetic Scattering}} and {{Light Confinement}}}.
\newblock \emph{\bibinfo{journal}{Physical Review Letters}}
  \textbf{\bibinfo{volume}{74}}, \bibinfo{pages}{526--529}
  (\bibinfo{year}{1995}).

\bibitem{wiechaPyGDMPythonToolkit2018}
\bibinfo{author}{Wiecha, P.~R.}
\newblock \bibinfo{title}{{{pyGDM}}\textemdash{{A}} python toolkit for
  full-field electro-dynamical simulations and evolutionary optimization of
  nanostructures}.
\newblock \emph{\bibinfo{journal}{Computer Physics Communications}}
  \textbf{\bibinfo{volume}{233}}, \bibinfo{pages}{167--192}
  (\bibinfo{year}{2018}).

\bibitem{agarwalQuantumElectrodynamicsPresence1975}
\bibinfo{author}{Agarwal, G.~S.}
\newblock \bibinfo{title}{Quantum electrodynamics in the presence of
  dielectrics and conductors. {{I}}. {{Electromagnetic}}-field response
  functions and black-body fluctuations in finite geometries}.
\newblock \emph{\bibinfo{journal}{Physical Review A}}
  \textbf{\bibinfo{volume}{11}}, \bibinfo{pages}{230--242}
  (\bibinfo{year}{1975}).

\bibitem{jacksonClassicalElectrodynamics1999}
\bibinfo{author}{Jackson, J.~D.}
\newblock \emph{\bibinfo{title}{Classical {{Electrodynamics}}}}
  (\bibinfo{publisher}{{Wiley}}, \bibinfo{year}{1999}),
  \bibinfo{edition}{third} edn.

\bibitem{patouxPolarizabilitiesComplexIndividual2020}
\bibinfo{author}{Patoux, A.} \emph{et~al.}
\newblock \bibinfo{title}{Polarizabilities of complex individual dielectric or
  plasmonic nanostructures}.
\newblock \emph{\bibinfo{journal}{Physical Review B}}
  \textbf{\bibinfo{volume}{101}}, \bibinfo{pages}{235418}
  (\bibinfo{year}{2020}).
\newblock \eprint{1912.04124}.

\bibitem{sersicMagnetoelectricPointScattering2011}
\bibinfo{author}{Sersic, I.}, \bibinfo{author}{Tuambilangana, C.},
  \bibinfo{author}{Kampfrath, T.} \& \bibinfo{author}{Koenderink, A.~F.}
\newblock \bibinfo{title}{Magnetoelectric point scattering theory for
  metamaterial scatterers}.
\newblock \emph{\bibinfo{journal}{Physical Review B}}
  \textbf{\bibinfo{volume}{83}}, \bibinfo{pages}{245102}
  (\bibinfo{year}{2011}).

\bibitem{schroterModellingMagneticEffects2003}
\bibinfo{author}{Schr{\"o}ter, U.}
\newblock \bibinfo{title}{Modelling of magnetic effects in near-field optics}.
\newblock \emph{\bibinfo{journal}{The European Physical Journal B}}
  \textbf{\bibinfo{volume}{33}}, \bibinfo{pages}{297--310}
  (\bibinfo{year}{2003}).

\bibitem{girardOpticalMagneticNearfield1997}
\bibinfo{author}{Girard, C.}, \bibinfo{author}{Weeber, J.-C.},
  \bibinfo{author}{Dereux, A.}, \bibinfo{author}{Martin, O. J.~F.} \&
  \bibinfo{author}{Goudonnet, J.-P.}
\newblock \bibinfo{title}{Optical magnetic near-field intensities around
  nanometer-scale surface structures}.
\newblock \emph{\bibinfo{journal}{Physical Review B}}
  \textbf{\bibinfo{volume}{55}}, \bibinfo{pages}{16487--16497}
  (\bibinfo{year}{1997}).

\bibitem{colasdesfrancsOptiqueSublongueurOnde2002}
\bibinfo{author}{{Colas des Francs}, G.}
\newblock \emph{\bibinfo{title}{Optique Sub-Longueur d'onde et Fluorescence
  Mol\'eculaire Perturb\'ee}}.
\newblock \bibinfo{type}{{{PhD}} thesis}, \bibinfo{school}{Universit\'e Paul
  Sabatier Toulouse}, \bibinfo{address}{{CEMES-CNRS}} (\bibinfo{year}{2002}).

\bibitem{girardFieldsNanostructures2005}
\bibinfo{author}{Girard, C.}
\newblock \bibinfo{title}{Near fields in nanostructures}.
\newblock \emph{\bibinfo{journal}{Reports on Progress in Physics}}
  \textbf{\bibinfo{volume}{68}}, \bibinfo{pages}{1883--1933}
  (\bibinfo{year}{2005}).

\bibitem{novotnyPrinciplesNanooptics2006}
\bibinfo{author}{Novotny, L.} \& \bibinfo{author}{Hecht, B.}
\newblock \emph{\bibinfo{title}{Principles of Nano-Optics}}
  (\bibinfo{publisher}{{Cambridge University Press}},
  \bibinfo{address}{{Cambridge ; New York}}, \bibinfo{year}{2006}).

\bibitem{gay-balmazValidityDomainLimitation2000}
\bibinfo{author}{{Gay-Balmaz}, P.} \& \bibinfo{author}{Martin, O. J.~F.}
\newblock \bibinfo{title}{Validity domain and limitation of non-retarded
  {{Green}}'s tensor for electromagnetic scattering at surfaces}.
\newblock \emph{\bibinfo{journal}{Optics Communications}}
  \textbf{\bibinfo{volume}{184}}, \bibinfo{pages}{37--47}
  (\bibinfo{year}{2000}).

\bibitem{novotnyInterferenceLocallyExcited1997}
\bibinfo{author}{Novotny, L.}, \bibinfo{author}{Hecht, B.} \&
  \bibinfo{author}{Pohl, D.~W.}
\newblock \bibinfo{title}{Interference of locally excited surface plasmons}.
\newblock \emph{\bibinfo{journal}{Journal of Applied Physics}}
  \textbf{\bibinfo{volume}{81}}, \bibinfo{pages}{1798--1806}
  (\bibinfo{year}{1997}).

\bibitem{rahmaniFieldPropagatorDressed1997}
\bibinfo{author}{Rahmani, A.}, \bibinfo{author}{Chaumet, P.~C.},
  \bibinfo{author}{{de Fornel}, F.} \& \bibinfo{author}{Girard, C.}
\newblock \bibinfo{title}{Field propagator of a dressed junction:
  {{Fluorescence}} lifetime calculations in a confined geometry}.
\newblock \emph{\bibinfo{journal}{Physical Review A}}
  \textbf{\bibinfo{volume}{56}}, \bibinfo{pages}{3245--3254}
  (\bibinfo{year}{1997}).

\bibitem{paulusAccurateEfficientComputation2000}
\bibinfo{author}{Paulus, M.}, \bibinfo{author}{{Gay-Balmaz}, P.} \&
  \bibinfo{author}{Martin, O. J.~F.}
\newblock \bibinfo{title}{Accurate and efficient computation of the {{Green}}'s
  tensor for stratified media}.
\newblock \emph{\bibinfo{journal}{Physical Review E}}
  \textbf{\bibinfo{volume}{62}}, \bibinfo{pages}{5797--5807}
  (\bibinfo{year}{2000}).

\bibitem{kottmannAccurateSolutionVolume2000}
\bibinfo{author}{Kottmann, J.} \& \bibinfo{author}{Martin, O.}
\newblock \bibinfo{title}{Accurate solution of the volume integral equation for
  high-permittivity scatterers}.
\newblock \emph{\bibinfo{journal}{IEEE Transactions on Antennas and
  Propagation}} \textbf{\bibinfo{volume}{48}}, \bibinfo{pages}{1719--1726}
  (\bibinfo{year}{2000}).

\bibitem{smunevRectangularDipolesDiscrete2015}
\bibinfo{author}{Smunev, D.~A.}, \bibinfo{author}{Chaumet, P.~C.} \&
  \bibinfo{author}{Yurkin, M.~A.}
\newblock \bibinfo{title}{Rectangular dipoles in the discrete dipole
  approximation}.
\newblock \emph{\bibinfo{journal}{Journal of Quantitative Spectroscopy and
  Radiative Transfer}} \textbf{\bibinfo{volume}{156}}, \bibinfo{pages}{67--79}
  (\bibinfo{year}{2015}).

\bibitem{pendryNegativeRefractionMakes2000}
\bibinfo{author}{Pendry, J.~B.}
\newblock \bibinfo{title}{Negative {{Refraction Makes}} a {{Perfect Lens}}}.
\newblock \emph{\bibinfo{journal}{Physical Review Letters}}
  \textbf{\bibinfo{volume}{85}}, \bibinfo{pages}{3966--3969}
  (\bibinfo{year}{2000}).

\bibitem{evlyukhinMultipoleLightScattering2011}
\bibinfo{author}{Evlyukhin, A.~B.}, \bibinfo{author}{Reinhardt, C.} \&
  \bibinfo{author}{Chichkov, B.~N.}
\newblock \bibinfo{title}{Multipole light scattering by nonspherical
  nanoparticles in the discrete dipole approximation}.
\newblock \emph{\bibinfo{journal}{Physical Review B}}
  \textbf{\bibinfo{volume}{84}}, \bibinfo{pages}{235429}
  (\bibinfo{year}{2011}).

\bibitem{cormenIntroductionAlgorithms2001}
\bibinfo{author}{Cormen, T.~H.}, \bibinfo{author}{Cormen, T.~H.},
  \bibinfo{author}{Leiserson, C.~E.}, \bibinfo{author}{Rivest, R.~L.} \&
  \bibinfo{author}{Stein, C.}
\newblock \emph{\bibinfo{title}{Introduction {{To Algorithms}}}}
  (\bibinfo{publisher}{{MIT Press}}, \bibinfo{year}{2001}).

\bibitem{wangBroadbandOpticalScattering2014}
\bibinfo{author}{Wang, C.} \emph{et~al.}
\newblock \bibinfo{title}{Broadband optical scattering in coupled silicon
  nanocylinders}.
\newblock \emph{\bibinfo{journal}{Journal of Applied Physics}}
  \textbf{\bibinfo{volume}{115}}, \bibinfo{pages}{244312}
  (\bibinfo{year}{2014}).

\bibitem{bakkerMagneticElectricHotspots2015}
\bibinfo{author}{Bakker, R.~M.} \emph{et~al.}
\newblock \bibinfo{title}{Magnetic and {{Electric Hotspots}} with {{Silicon
  Nanodimers}}}.
\newblock \emph{\bibinfo{journal}{Nano Letters}} \textbf{\bibinfo{volume}{15}},
  \bibinfo{pages}{2137--2142} (\bibinfo{year}{2015}).

\bibitem{albellaLowLossElectricMagnetic2013}
\bibinfo{author}{Albella, P.} \emph{et~al.}
\newblock \bibinfo{title}{Low-{{Loss Electric}} and {{Magnetic
  Field}}-{{Enhanced Spectroscopy}} with {{Subwavelength Silicon Dimers}}}.
\newblock \emph{\bibinfo{journal}{The Journal of Physical Chemistry C}}
  \textbf{\bibinfo{volume}{117}}, \bibinfo{pages}{13573--13584}
  (\bibinfo{year}{2013}).

\bibitem{colasdesfrancsIntegratedPlasmonicWaveguides2009}
\bibinfo{author}{{Colas des Francs}, G.} \emph{et~al.}
\newblock \bibinfo{title}{Integrated plasmonic waveguides: {{A}} mode solver
  based on density of states formulation}.
\newblock \emph{\bibinfo{journal}{Physical Review B}}
  \textbf{\bibinfo{volume}{80}}, \bibinfo{pages}{115419}
  (\bibinfo{year}{2009}).

\bibitem{barthesPurcellFactorPointlike2011}
\bibinfo{author}{Barthes, J.}, \bibinfo{author}{{Colas des Francs}, G.},
  \bibinfo{author}{Bouhelier, A.}, \bibinfo{author}{Weeber, J.-C.} \&
  \bibinfo{author}{Dereux, A.}
\newblock \bibinfo{title}{Purcell factor for a point-like dipolar emitter
  coupled to a two-dimensional plasmonic waveguide}.
\newblock \emph{\bibinfo{journal}{Physical Review B}}
  \textbf{\bibinfo{volume}{84}}, \bibinfo{pages}{073403}
  (\bibinfo{year}{2011}).

\bibitem{camposPlasmonicBreathingEdge2017}
\bibinfo{author}{Campos, A.} \emph{et~al.}
\newblock \bibinfo{title}{Plasmonic {{Breathing}} and {{Edge Modes}} in
  {{Aluminum Nanotriangles}}}.
\newblock \emph{\bibinfo{journal}{ACS Photonics}} \textbf{\bibinfo{volume}{4}},
  \bibinfo{pages}{1257--1263} (\bibinfo{year}{2017}).

\bibitem{arbouetElectronEnergyLosses2014}
\bibinfo{author}{Arbouet, A.}, \bibinfo{author}{Mlayah, A.},
  \bibinfo{author}{Girard, C.} \& \bibinfo{author}{{Colas des Francs}, G.}
\newblock \bibinfo{title}{Electron energy losses and cathodoluminescence from
  complex plasmonic nanostructures: Spectra, maps and radiation patterns from a
  generalized field propagator}.
\newblock \emph{\bibinfo{journal}{New Journal of Physics}}
  \textbf{\bibinfo{volume}{16}}, \bibinfo{pages}{113012}
  (\bibinfo{year}{2014}).

\bibitem{schaferlingFormationChiralFields2012}
\bibinfo{author}{Sch{\"a}ferling, M.}, \bibinfo{author}{Yin, X.} \&
  \bibinfo{author}{Giessen, H.}
\newblock \bibinfo{title}{Formation of chiral fields in a symmetric
  environment}.
\newblock \emph{\bibinfo{journal}{Optics Express}}
  \textbf{\bibinfo{volume}{20}}, \bibinfo{pages}{26326--26336}
  (\bibinfo{year}{2012}).

\bibitem{meinzerProbingChiralNature2013}
\bibinfo{author}{Meinzer, N.}, \bibinfo{author}{Hendry, E.} \&
  \bibinfo{author}{Barnes, W.~L.}
\newblock \bibinfo{title}{Probing the chiral nature of electromagnetic fields
  surrounding plasmonic nanostructures}.
\newblock \emph{\bibinfo{journal}{Physical Review B}}
  \textbf{\bibinfo{volume}{88}}, \bibinfo{pages}{041407}
  (\bibinfo{year}{2013}).

\bibitem{gorodetskiGeneratingFarFieldOrbital2013}
\bibinfo{author}{Gorodetski, Y.}, \bibinfo{author}{Drezet, A.},
  \bibinfo{author}{Genet, C.} \& \bibinfo{author}{Ebbesen, T.~W.}
\newblock \bibinfo{title}{Generating {{Far}}-{{Field Orbital Angular Momenta}}
  from {{Near}}-{{Field Optical Chirality}}}.
\newblock \emph{\bibinfo{journal}{Physical Review Letters}}
  \textbf{\bibinfo{volume}{110}}, \bibinfo{pages}{203906}
  (\bibinfo{year}{2013}).

\bibitem{tangOpticalChiralityIts2010}
\bibinfo{author}{Tang, Y.} \& \bibinfo{author}{Cohen, A.~E.}
\newblock \bibinfo{title}{Optical {{Chirality}} and {{Its Interaction}} with
  {{Matter}}}.
\newblock \emph{\bibinfo{journal}{Physical Review Letters}}
  \textbf{\bibinfo{volume}{104}}, \bibinfo{pages}{163901}
  (\bibinfo{year}{2010}).

\bibitem{chaumetCoupledDipoleMethod2000}
\bibinfo{author}{Chaumet, P.~C.} \& \bibinfo{author}{{Nieto-Vesperinas}, M.}
\newblock \bibinfo{title}{Coupled dipole method determination of the
  electromagnetic force on a particle over a flat dielectric substrate}.
\newblock \emph{\bibinfo{journal}{Physical Review B}}
  \textbf{\bibinfo{volume}{61}}, \bibinfo{pages}{14119--14127}
  (\bibinfo{year}{2000}).

\bibitem{martyChargeDistributionInduced2010}
\bibinfo{author}{Marty, R.}, \bibinfo{author}{Baffou, G.},
  \bibinfo{author}{Arbouet, A.}, \bibinfo{author}{Girard, C.} \&
  \bibinfo{author}{Quidant, R.}
\newblock \bibinfo{title}{Charge distribution induced inside complex plasmonic
  nanoparticles}.
\newblock \emph{\bibinfo{journal}{Optics Express}}
  \textbf{\bibinfo{volume}{18}}, \bibinfo{pages}{3035--3044}
  (\bibinfo{year}{2010}).

\bibitem{girardTheoreticalAnalysisLightinductive1994}
\bibinfo{author}{Girard, C.}, \bibinfo{author}{Dereux, A.} \&
  \bibinfo{author}{Martin, O. J.~F.}
\newblock \bibinfo{title}{Theoretical analysis of light-inductive forces in
  scanning probe microscopy}.
\newblock \emph{\bibinfo{journal}{Physical Review B}}
  \textbf{\bibinfo{volume}{49}}, \bibinfo{pages}{13872--13881}
  (\bibinfo{year}{1994}).

\bibitem{pressNumericalRecipes3rd2007}
\bibinfo{author}{Press, W.}
\newblock \emph{\bibinfo{title}{Numerical {{Recipes}} 3rd {{Edition}}: {{The
  Art}} of {{Scientific Computing}}}} (\bibinfo{publisher}{{Cambridge
  University Press}}, \bibinfo{year}{2007}).

\bibitem{curtoUnidirectionalEmissionQuantum2010}
\bibinfo{author}{Curto, A.~G.} \emph{et~al.}
\newblock \bibinfo{title}{Unidirectional {{Emission}} of a {{Quantum Dot
  Coupled}} to a {{Nanoantenna}}}.
\newblock \emph{\bibinfo{journal}{Science}} \textbf{\bibinfo{volume}{329}},
  \bibinfo{pages}{930--933} (\bibinfo{year}{2010}).

\bibitem{taminiauEnhancedDirectionalExcitation2008}
\bibinfo{author}{Taminiau, T.~H.}, \bibinfo{author}{Stefani, F.~D.} \&
  \bibinfo{author}{van Hulst, N.~F.}
\newblock \bibinfo{title}{Enhanced directional excitation and emission of
  single emitters by a nano-optical {{Yagi}}-{{Uda}} antenna}.
\newblock \emph{\bibinfo{journal}{Optics Express}}
  \textbf{\bibinfo{volume}{16}}, \bibinfo{pages}{10858--10866}
  (\bibinfo{year}{2008}).

\bibitem{wiechaDesignPlasmonicDirectional2019}
\bibinfo{author}{Wiecha, P.~R.} \emph{et~al.}
\newblock \bibinfo{title}{Design of plasmonic directional antennas via
  evolutionary optimization}.
\newblock \emph{\bibinfo{journal}{Optics Express}}
  \textbf{\bibinfo{volume}{27}}, \bibinfo{pages}{29069--29081}
  (\bibinfo{year}{2019}).

\bibitem{wiechaDecayRateMagnetic2018}
\bibinfo{author}{Wiecha, P.~R.}, \bibinfo{author}{Arbouet, A.},
  \bibinfo{author}{Cuche, A.}, \bibinfo{author}{Paillard, V.} \&
  \bibinfo{author}{Girard, C.}
\newblock \bibinfo{title}{Decay rate of magnetic dipoles near nonmagnetic
  nanostructures}.
\newblock \emph{\bibinfo{journal}{Physical Review B}}
  \textbf{\bibinfo{volume}{97}}, \bibinfo{pages}{085411}
  (\bibinfo{year}{2018}).

\bibitem{majorelQuantumTheoryNearfield2020}
\bibinfo{author}{Majorel, C.}, \bibinfo{author}{Girard, C.},
  \bibinfo{author}{Cuche, A.}, \bibinfo{author}{Arbouet, A.} \&
  \bibinfo{author}{Wiecha, P.~R.}
\newblock \bibinfo{title}{Quantum theory of near-field optical imaging with
  rare-earth atomic clusters}.
\newblock \emph{\bibinfo{journal}{JOSA B}} \textbf{\bibinfo{volume}{37}},
  \bibinfo{pages}{1474--1484} (\bibinfo{year}{2020}).
\newblock \eprint{1912.06023}.

\bibitem{carminatiElectromagneticDensityStates2015}
\bibinfo{author}{Carminati, R.} \emph{et~al.}
\newblock \bibinfo{title}{Electromagnetic density of states in complex
  plasmonic systems}.
\newblock \emph{\bibinfo{journal}{Surface Science Reports}}
  \textbf{\bibinfo{volume}{70}}, \bibinfo{pages}{1--41} (\bibinfo{year}{2015}).

\bibitem{kimClassicalDecayRates1988}
\bibinfo{author}{Kim, Y.~S.}, \bibinfo{author}{Leung, P.~T.} \&
  \bibinfo{author}{George, T.~F.}
\newblock \bibinfo{title}{Classical decay rates for molecules in the presence
  of a spherical surface: {{A}} complete treatment}.
\newblock \emph{\bibinfo{journal}{Surface Science}}
  \textbf{\bibinfo{volume}{195}}, \bibinfo{pages}{1--14}
  (\bibinfo{year}{1988}).

\bibitem{schmidtDielectricAntennasSuitable2012}
\bibinfo{author}{Schmidt, M.~K.} \emph{et~al.}
\newblock \bibinfo{title}{Dielectric antennas - a suitable platform for
  controlling magnetic dipolar emission}.
\newblock \emph{\bibinfo{journal}{Optics Express}}
  \textbf{\bibinfo{volume}{20}}, \bibinfo{pages}{13636} (\bibinfo{year}{2012}).

\bibitem{lutherLocalizedSurfacePlasmon2011}
\bibinfo{author}{Luther, J.~M.}, \bibinfo{author}{Jain, P.~K.},
  \bibinfo{author}{Ewers, T.} \& \bibinfo{author}{Alivisatos, A.~P.}
\newblock \bibinfo{title}{Localized surface plasmon resonances arising from
  free carriers in doped quantum dots}.
\newblock \emph{\bibinfo{journal}{Nature Materials}}
  \textbf{\bibinfo{volume}{10}}, \bibinfo{pages}{361--366}
  (\bibinfo{year}{2011}).

\bibitem{majorelTheoryPlasmonicProperties2019}
\bibinfo{author}{Majorel, C.} \emph{et~al.}
\newblock \bibinfo{title}{Theory of plasmonic properties of hyper-doped silicon
  nanostructures}.
\newblock \emph{\bibinfo{journal}{Optics Communications}}
  \textbf{\bibinfo{volume}{453}}, \bibinfo{pages}{124336}
  (\bibinfo{year}{2019}).

\bibitem{teulleVisibilityPlasmonicParticles2013}
\bibinfo{author}{Teulle, A.}, \bibinfo{author}{Marty, R.},
  \bibinfo{author}{Girard, C.}, \bibinfo{author}{Arbouet, A.} \&
  \bibinfo{author}{Dujardin, E.}
\newblock \bibinfo{title}{Visibility of plasmonic particles embedded in
  transparent materials}.
\newblock \emph{\bibinfo{journal}{Optics Communications}}
  \textbf{\bibinfo{volume}{291}}, \bibinfo{pages}{412--415}
  (\bibinfo{year}{2013}).

\bibitem{mannaSelectiveExcitationEnhancement2020}
\bibinfo{author}{Manna, U.} \emph{et~al.}
\newblock \bibinfo{title}{Selective excitation and enhancement of multipolar
  resonances in dielectric nanospheres using cylindrical vector beams}.
\newblock \emph{\bibinfo{journal}{Journal of Applied Physics}}
  \textbf{\bibinfo{volume}{127}}, \bibinfo{pages}{033101}
  (\bibinfo{year}{2020}).

\bibitem{chaumetGeneralizationCoupledDipole2003}
\bibinfo{author}{Chaumet, P.~C.}, \bibinfo{author}{Rahmani, A.} \&
  \bibinfo{author}{Bryant, G.~W.}
\newblock \bibinfo{title}{Generalization of the coupled dipole method to
  periodic structures}.
\newblock \emph{\bibinfo{journal}{Physical Review B}}
  \textbf{\bibinfo{volume}{67}}, \bibinfo{pages}{165404}
  (\bibinfo{year}{2003}).

\bibitem{paulusGreenTensorTechnique2001}
\bibinfo{author}{Paulus, M.} \& \bibinfo{author}{Martin, O. J.~F.}
\newblock \bibinfo{title}{Green's tensor technique for scattering in
  two-dimensional stratified media}.
\newblock \emph{\bibinfo{journal}{Physical Review E}}
  \textbf{\bibinfo{volume}{63}}, \bibinfo{pages}{066615}
  (\bibinfo{year}{2001}).

\bibitem{baffouMolecularQuenchingRelaxation2008}
\bibinfo{author}{Baffou, G.}, \bibinfo{author}{Girard, C.},
  \bibinfo{author}{Dujardin, E.}, \bibinfo{author}{{Colas des Francs}, G.} \&
  \bibinfo{author}{Martin, O. J.~F.}
\newblock \bibinfo{title}{Molecular quenching and relaxation in a plasmonic
  tunable system}.
\newblock \emph{\bibinfo{journal}{Physical Review B}}
  \textbf{\bibinfo{volume}{77}}, \bibinfo{pages}{121101}
  (\bibinfo{year}{2008}).

\bibitem{arbouetInteractionUltrashortOptical2012}
\bibinfo{author}{Arbouet, A.}, \bibinfo{author}{Houdellier, F.},
  \bibinfo{author}{Marty, R.} \& \bibinfo{author}{Girard, C.}
\newblock \bibinfo{title}{Interaction of an ultrashort optical pulse with a
  metallic nanotip: {{A Green}} dyadic approach}.
\newblock \emph{\bibinfo{journal}{Journal of Applied Physics}}
  \textbf{\bibinfo{volume}{112}}, \bibinfo{pages}{053103}
  (\bibinfo{year}{2012}).

\bibitem{govindarajuHighPerformanceDiscrete2008}
\bibinfo{author}{Govindaraju, N.~K.}, \bibinfo{author}{Lloyd, B.},
  \bibinfo{author}{Dotsenko, Y.}, \bibinfo{author}{Smith, B.} \&
  \bibinfo{author}{Manferdelli, J.}
\newblock \bibinfo{title}{High performance discrete {{Fourier}} transforms on
  graphics processors}.
\newblock In \emph{\bibinfo{booktitle}{{{SC}} '08: {{Proceedings}} of the 2008
  {{ACM}}/{{IEEE Conference}} on {{Supercomputing}}}}, \bibinfo{pages}{1--12}
  (\bibinfo{year}{2008}).

\bibitem{huntemannDiscreteDipoleApproximation2011}
\bibinfo{author}{Huntemann, M.}, \bibinfo{author}{Heygster, G.} \&
  \bibinfo{author}{Hong, G.}
\newblock \bibinfo{title}{Discrete dipole approximation simulations on {{GPUs}}
  using {{OpenCL}}\textemdash{{Application}} on cloud ice particles}.
\newblock \emph{\bibinfo{journal}{Journal of Computational Science}}
  \textbf{\bibinfo{volume}{2}}, \bibinfo{pages}{262--271}
  (\bibinfo{year}{2011}).

\bibitem{wiechaDeepLearningMeets2020}
\bibinfo{author}{Wiecha, P.~R.} \& \bibinfo{author}{Muskens, O.~L.}
\newblock \bibinfo{title}{Deep {{Learning Meets Nanophotonics}}: {{A
  Generalized Accurate Predictor}} for {{Near Fields}} and {{Far Fields}} of
  {{Arbitrary 3D Nanostructures}}}.
\newblock \emph{\bibinfo{journal}{Nano Letters}} \textbf{\bibinfo{volume}{20}},
  \bibinfo{pages}{329--338} (\bibinfo{year}{2020}).
\newblock \eprint{1909.12056}.

\bibitem{colasdesfrancsTheoryNearfieldOptical2002}
\bibinfo{author}{{Colas des Francs}, G.}, \bibinfo{author}{Girard, C.} \&
  \bibinfo{author}{Dereux, A.}
\newblock \bibinfo{title}{Theory of near-field optical imaging with a single
  molecule as light source}.
\newblock \emph{\bibinfo{journal}{The Journal of Chemical Physics}}
  \textbf{\bibinfo{volume}{117}}, \bibinfo{pages}{4659--4666}
  (\bibinfo{year}{2002}).

\bibitem{bornPrinciplesOpticsElectromagnetic1999}
\bibinfo{author}{Born, M.} \& \bibinfo{author}{Wolf, E.}
\newblock \emph{\bibinfo{title}{Principles of {{Optics}}: {{Electromagnetic
  Theory}} of {{Propagation}}, {{Interference}} and {{Diffraction}} of
  {{Light}}}} (\bibinfo{publisher}{{Cambridge University Press}},
  \bibinfo{address}{{Cambridge}}, \bibinfo{year}{1999}),
  \bibinfo{edition}{seventh} edn.

\bibitem{martinElectromagneticScatteringPolarizable1998}
\bibinfo{author}{Martin, O. J.~F.} \& \bibinfo{author}{Piller, N.~B.}
\newblock \bibinfo{title}{Electromagnetic scattering in polarizable
  backgrounds}.
\newblock \emph{\bibinfo{journal}{Physical Review E}}
  \textbf{\bibinfo{volume}{58}}, \bibinfo{pages}{3909--3915}
  (\bibinfo{year}{1998}).

\bibitem{gradshteynTableIntegralsSeries2007}
\bibinfo{author}{Gradshteyn, I.~S.} \& \bibinfo{author}{Ryzhik, I.~M.}
\newblock \emph{\bibinfo{title}{Table of Integrals, Series, and Products}}
  (\bibinfo{publisher}{{Elsevier/Academic Press, Amsterdam}},
  \bibinfo{year}{2007}), \bibinfo{edition}{seventh} edn.

\bibitem{bernasconiWhereDoesEnergy2017}
\bibinfo{author}{Bernasconi, G.~D.} \emph{et~al.}
\newblock \bibinfo{title}{Where {{Does Energy Go}} in {{Electron Energy Loss
  Spectroscopy}} of {{Nanostructures}}?}
\newblock \emph{\bibinfo{journal}{ACS Photonics}} \textbf{\bibinfo{volume}{4}},
  \bibinfo{pages}{156--164} (\bibinfo{year}{2017}).

\end{thebibliography}
